\documentclass[aps,prd,superscriptaddress,nofootinbib,floats,reprint,floatfix]{revtex4-2}
\usepackage{bm}
\usepackage{indentfirst}
\usepackage{amsmath}
\usepackage{graphicx}
\usepackage{float}
\usepackage{amssymb}
\usepackage{subfigure}
\usepackage{psfrag}
\usepackage{hyperref}
\usepackage{tensor}
\usepackage{braket}
\usepackage{multirow}
\usepackage{verbatim}
\usepackage{ulem}
\usepackage{cancel}
\usepackage[dvipsnames, svgnames, x11names]{xcolor}
\hypersetup{
	colorlinks=true,
	linkcolor=blue,
	citecolor=blue,
} 

\allowdisplaybreaks[1]
\usepackage[utf8]{inputenc}

\begin{document}

\title{Scalar induced gravitational waves in metric teleparallel gravity with the Nieh-Yan term}

\author{Fengge Zhang}
\email{zhangfengge@hnas.ac.cn}
\affiliation{Henan Academy of Sciences, Zhengzhou 450046, Henan, China}
\affiliation{School of Physics and Astronomy, Sun Yat-sen University, Zhuhai 519082, China}
\author{Jia-Xi Feng}
\email{fengjx57@mail2.sysu.edu.cn}
\affiliation{School of Physics and Astronomy, Sun Yat-sen University, Zhuhai 519082, China}
\author{Xian Gao}
\email{gaoxian@mail.sysu.edu.cn (Corresponding author)}
\affiliation{School of Physics and Astronomy, Sun Yat-sen University, Zhuhai 519082, China}

\begin{abstract}
We investigate the scalar induced gravitational waves (SIGWs) in metric teleparallel gravity with the Nieh-Yan (NY) term, which results in parity violation during the radiation-dominated era. By solving the equations of motion of linear scalar perturbations from both the metric and the tetrad fields, we obtain the corresponding analytic expressions. Then, we calculate the SIGWs in metric teleparallel gravity with the NY term and evaluate the energy density of SIGWs with a monochromatic power spectrum numerically. We find that the spectrum of the energy density of SIGWs in metric teleparallel gravity with the NY term is significantly different from that in general relativity (GR), which makes metric teleparallel gravity distinguishable from GR.
\end{abstract}

\maketitle


\section{Introduction}

Gravitational waves (GWs) play an important role in exploring the early universe. The successful detection of GWs generated from the merger of compact objects by the Laser Interferometer Gravitational-Wave Observatory (LIGO) scientific collaboration and the Virgo collaboration \cite{Abbott:2016nmj,Abbott:2016blz,Abbott:2017gyy,TheLIGOScientific:2017qsa,Abbott:2017oio,Abbott:2017vtc,LIGOScientific:2018mvr,Abbott:2020khf,Abbott:2020uma,LIGOScientific:2020stg} opens a new window to probe the nature of gravity in the strong gravitational field and nonlinear regime. It also marks the dawn of multimessenger astronomy. The scalar induced gravitational waves (SIGWs) originated during the early universe due to the nonlinear interaction between scalar and tensor perturbation, which contribute to the stochastic gravitational wave background, have attracted much attention recently \cite{Ananda:2006af,Saito:2008jc,Nakama:2016gzw,Wang:2016ana,Kohri:2018awv,Espinosa:2018eve,Kuroyanagi:2018csn,Fumagalli:2020nvq,Braglia:2020taf,Lin:2020goi,Lu:2020diy,Domenech:2021ztg,Zhang:2021vak,Zhang:2021rqs,Zhang:2021rqs,Papanikolaou:2021uhe,Yi:2022ymw,You:2023rmn,Lu:2019sti,Gu:2023mmd,Choudhury:2023hfm,Wang:2024euw,Choudhury:2024one}. The frequency of SIGWs varies widely, and SIGWs can be detected by space-based GW detectors like the Laser Interferometer Space Antenna (LISA) \cite{Danzmann:1997hm,LISA:2017pwj}, Taiji \cite{Hu:2017mde}, TianQin \cite{Luo:2015ght,Gong:2021gvw}, the Deci-hertz Interferometer Gravitational-Wave Observatory (DECIGO) \cite{Kawamura:2011zz}, as well as by the pulsar timing array (PTA) \cite{Kramer:2013kea,Hobbs:2009yy,McLaughlin:2013ira,Hobbs:2013aka} and the Square Kilometer Array \cite{Moore:2014lga}. The recent stochastic GW signal captured by PTA \cite{NANOGrav:2023gor,NANOGrav:2023hde,Zic:2023gta,Reardon:2023gzh,EPTA:2023sfo,EPTA:2023fyk,Xu:2023wog} can also be explained with SIGWs \cite{NANOGrav:2023hvm,EPTA:2023xxk,Yi:2023mbm,Yi:2023tdk,Yi:2023npi,Cai:2023dls,Liu:2023ymk,Jin:2023wri,Chen:2024fir,Liu:2023hpw,Chen:2024twp}.

The gravity theory with parity-violating (PV) terms attracted significant attention recently \cite{Horava:2009uw,Gao:2019liu,Hu:2021bbo,Hu:2021yaq,Takahashi:2009wc,Myung:2009ug,Wang:2012fi,Zhu:2013fja,Cannone:2015rra,Zhao:2019szi,Zhao:2019xmm,Qiao:2019hkz,Qiao:2019wsh,Qiao:2021fwi,Gong:2021jgg,Zhu:2023lhv,Akama:2024bav}. On the one hand, the violation of parity symmetry in weak interaction \cite{Lee:1956qn,Wu:1957my} prompts the investigation of whether such parity violation occurs in gravitational interaction. On the other hand, the recent studies on galaxy trispectrum and the cross-correlation of the $E$ and $B$ mode polarization of the cosmic microwave background (CMB) \cite{Philcox:2022hkh,Hou:2022wfj,Minami:2020odp,Eskilt:2022cff} have hinted at the existence of parity violation in our universe. The PV scalar trispectrum was also studied in \cite{Liu:2019fag,Niu:2022fki,Cabass:2022rhr,Creque-Sarbinowski:2023wmb,Garcia-Saenz:2023zue}.

The simplest PV term in the Riemannian geometry is the Chern-Simons (CS) term, which is quadratic in the Riemann tensor. CS gravity was initially proposed in four-dimensional spacetime in \cite{Jackiw:2003pm}, and has since been extensively studied in cosmology, GWs, and primordial non-Gaussianity \cite{Lue:1998mq,Satoh:2007gn,Saito:2007kt,Satoh:2007gn,Alexander:2009tp,Yunes:2010yf,Gluscevic:2010vv,Yunes:2010yf,Myung:2014jha,Kawai:2017kqt,Nair:2019iur,Nishizawa:2018srh,Bartolo:2017szm,Bartolo:2018elp}. Recently, the SIGWs in CS gravity have also been studied \cite{Zhang:2022xmm,Feng:2023veu}. However, CS gravity suffers from Ostrogradsky instability \cite{Crisostomi:2017ugk} and propagates ghost modes \cite{Bartolo:2017szm,Bartolo:2018elp} due to the presence of higher-derivative field equations. As a result, it can only be treated as a low-energy effective theory. To cure this issue, CS gravity was generalized to ghost-free PV gravity \cite{Crisostomi:2017ugk}.

Considering gravity theory beyond Riemann geometry, various gravity theories based on teleparallel geometry have been proposed \cite{Chatzistavrakidis:2020wum,Wu:2021ndf,Langvik:2020nrs,Rao:2021azn,Li:2021mdp,Battista:2021rlh,Li:2022mti,Li:2022vtn,Hohmann:2020dgy,Bombacigno:2021bpk,Iosifidis:2020dck,Hohmann:2022wrk,Conroy:2019ibo,Iosifidis:2021bad,Pagani:2015ema,Chen:2022wtz,Gialamas:2022xtt,Papanikolaou:2022hkg,Salvio:2022suk,Gialamas:2023emn,Tzerefos:2023mpe,DeFalco:2024ojf}. Interestingly, the symmetric teleparallel gravity with the method of spatially covariant gravity was also studied \cite{Yu:2024drx} recently. Within the framework of metric teleparallel gravity, similar to the CS gravity, the simplest PV term is $T_{A\mu\nu}\mathcal{T}^{A\mu\nu}$ \cite{Li:2020xjt}, where $T_{A\mu\nu}$ represents the torsion tensor and $\mathcal{T}^{A\mu\nu}=\varepsilon^{\mu\nu\rho\sigma}T^{A}_{\ \rho\sigma}/2$ is the dual of the torsion tensor. This term is a generalization of the Nieh-Yan (NY) term, initially proposed in Riemann-Cartan geometry \cite{Nieh:1981ww,Nieh:2008btw}, and the thermal Nieh-Yan anomaly in Weyl superfluids was also studied \cite{Nissinen:2019mkw}. The simplest metric teleparallel gravity with the PV term was constructed by adding the NY term to the metric teleparallel equivalent Einstein-Hilbert action. Furthermore, the linear cosmological perturbations in this theory have also been investigated \cite{Li:2020xjt,Li:2021wij,Cai:2021uup,Li:2023fto}.

Unlike CS gravity, the equations of motion (EOMs) in the above-mentioned gravity model do not involve higher-order derivatives, and thus the Ostrogradsky instability is effectively avoided. In this paper, we concentrate on the nonlinear perturbation in metric teleparallel gravity with the NY term, specifically investigating the behavior of the SIGWs. However, when the perturbations beyond linear order are taken into account, inconsistencies may arise in the simplest PV metric teleparallel gravity mentioned above. This situation is similar to what we encountered in symmetric teleparallel gravity with the PV term when calculating SIGWs \cite{Zhang:2023scq}. Briefly, the theory contains extra scalar degrees of freedom due to the PV term, which, however, do not manifest themselves at the linear order around a homogeneous and isotropic background.
This is reminiscent of the so-called strong coupling problem in the study of Ho\v{r}ava gravity \cite{Blas:2009yd,Charmousis:2009tc,Blas:2009qj,Papazoglou:2009fj,Blas:2009ck} and $f(T)$ gravity \cite{Ferraro:2018tpu,Li:2011rn,Izumi:2012qj,Golovnev:2020zpv,Hu:2023juh,Hu:2023xcf}.

In this paper, we will demonstrate that the simplest metric teleparallel gravity with the NY term also suffers from such a strong coupling problem. Specifically, the scalar perturbations arising from the tetrads do not possess their own linear EOMs, while appearing in the EOM of the SIGWs. To avoid this problem, we replace the metric teleparallel equivalent Einstein-Hilbert action with a general linear combination of quadratic monomials of the torsion tensor. We then derive the EOMs governing the perturbations originating from the tetrads and determine their solutions during the radiation-dominated era. Utilizing these results, we calculate the contribution of both the NY term and the scalar perturbations in the tetrads to the energy density of SIGWs in our model, respectively.

This paper is organized as follows. In Sec. \ref{sec2}, we briefly introduce the metric teleparallel gravity with the NY term. In Sec. \ref{sec3}, we give the EOMs for both the background evolution and the linear scalar perturbations, and we then solve these EOMs during the radiation-dominated era. In Sec. \ref{sec4}, we derive the EOM of SIGWs and calculate the power spectra of the SIGWs. To analyze the feature of SIGWs, we compute the energy density of SIGWs with the monochromatic power spectrum of primordial curvature perturbation. Our results are summarized in Sec. \ref{sec5}. The analytic expressions of the integral kernel are included in Appendix \ref{kernel}.

\section{The metric teleparallel gravity with the Nieh-Yan term}\label{sec2}

In this section, we review the metric teleparallel gravity. In this paper, we use the following conventions: the flat space metric is $\eta_{AB}=\mathrm{diag}(+1,-1,-1,-1)$, and lowercase Latin letters $(i, j, ...)$ denote purely spatial indices, while capital Latin letters $(A, B,...)$ and Greek alphabet letters $(\mu, \nu, ...)$ are used for Lorentz indices and spacetime indices, respectively. The metric tensor is produced by the tetrads $e^A_{\ \mu}$ and their inverses $e^{\ \mu}_A$,
\begin{equation}\label{gtr}
g_{\mu\nu}=\eta_{AB}e^{A}_{\ \mu}e^{B}_{\ \nu} \quad \text{and} \quad g^{\mu\nu}=\eta^{AB}e_{A}^{\ \mu}e_{B}^{\ \nu},
\end{equation}
where these tetrads satisfy  $e^{\ \mu}_A e^B_{\ \mu}=\delta^B_{\ A}$  and $e^{\ \mu}_A e^B_{\ \nu}=\delta^{\mu}_{\ \nu}$ .

In metric teleparallel geometry, the curvature vanishes, 
\begin{equation}
R_{\ \rho \mu \nu}^\sigma=\partial_\mu \Gamma_{\ \nu\rho}^\sigma-\partial_\nu \Gamma_{\ \mu\rho}^\sigma+\Gamma_{\ \mu\alpha}^\sigma \Gamma_{\ \nu\rho}^\alpha-\Gamma_{\ \nu\alpha}^\sigma \Gamma_{\ \mu\rho}^\alpha=0.
\end{equation}
The gravitational effects are described in terms of the torsion tensor, which is defined by the antisymmetric part of the affine connection
\begin{equation}
T^{\rho}_{\ \mu\nu}=\Gamma^{\rho}_{\ \mu\nu}-\Gamma^{\rho}_{\ \nu\mu},
\end{equation}
where the affine connection in the Weitzenb{\"o}ck gauge is \cite{Bahamonde:2021gfp}
\begin{equation}\label{gad}
\Gamma^\rho_{\ \mu\nu}=e^{\ \rho}_{A}\partial_\mu e^A_{\ \nu}.
\end{equation}

Considering the following action:
\begin{equation}\label{mat}
S=S_g+S_{\text{NY}}+S_m,
\end{equation}
where $S_g$ is the gravitational action
\begin{equation}
S_g=\frac{1}{2}\int \mathrm{d}^4 x e\mathbb{T},
\end{equation}
with $e=\mathrm{det}(e^A_{\ \mu})=\sqrt{-g}$,
\begin{equation}\label{TS}
\mathbb{T}=\frac{1}{2}S_{\alpha}^{\ \mu\nu}T^{\alpha}_{\ \mu\nu},
\end{equation}
and the superpotential $S_{\alpha}^{\ \mu\nu}$ is \cite{BeltranJimenez:2017tkd}
\begin{equation}
S_{\alpha}^{\ \mu\nu}=t_1T_{\alpha}^{\ \mu\nu}+t_2T^{[\mu\ \nu]}_{\ \ \alpha}+t_3\delta_{\alpha}^{\ [\mu}T^{\nu]},
\end{equation}
where $T^{\mu}=T^{\nu\mu}_{\ \ \ \nu}$ and $t_1$, $t_2$, and $t_3$ are three constants.

The PV term in the action \eqref{mat} is 
\begin{equation}
S_{\text{NY}}=\int \mathrm{d}^4 x e\frac{g(\theta)}{4}\mathcal{L}_{\text{NY}},
\end{equation}
where 
\begin{equation}
\mathcal{L}_{\text{NY}}=T_{A\mu\nu}\mathcal{T}^{A\mu\nu}-\varepsilon^{\mu\nu\rho\sigma}R_{\mu\nu\rho\sigma},
\end{equation}
is the NY term,
\begin{equation}
 \mathcal{T}^{A\mu\nu}=\varepsilon^{\mu\nu\rho\sigma}T^{A}_{\ \rho\sigma}/2,   
\end{equation}
is the dual of the torsion tensor, and $T^{A}_{\ \mu\nu}=e^A_{\ \rho}T^{\rho}_{\ \mu\nu}$, $\varepsilon^{\mu\nu\rho\sigma}=\epsilon^{\mu\nu\rho\sigma}/\sqrt{-g}$ is the Levi-Civita tensor, with $\epsilon^{\mu\nu\rho\sigma}$ the antisymmetric symbol. 
The NY term is a topological term and was first proposed in Riemann-Cartan geometry \cite{Nieh:1981ww,Nieh:2008btw}. In the framework of teleparallelism, i.e., $R_{\mu\nu\rho\sigma}=0$, the NY term coupled to a scalar field was extended to metric teleparallel geometry and added to the metric teleparallel equivalent Einstein-Hilbert Lagrangian \cite{Li:2020xjt}.

The last term in Eq. \eqref{mat} effectively describes the matter filled in the universe,
\begin{equation}
S_m=\int \mathrm{d}^4 x \sqrt{-g}\left[\frac{1}{2}\nabla_{\mu}\theta\nabla^{\mu}\theta-V(\theta)\right].
\end{equation}

In conclusion, the action we consider in this paper is
\begin{equation}
\begin{split}
S=&\int \mathrm{d}^4 x e\left[\frac{1}{2}\mathbb{T}+\frac{g(\theta)}{4}T_{A\mu\nu}\mathcal{T}^{A\mu\nu}\right]\\&+\int d^4 x \sqrt{-g}\left[\frac{1}{2}\nabla_{\mu}\theta\nabla^{\mu}\theta-V(\theta)\right].
\end{split}
\end{equation}

Note that if we choose the parameters to be
\begin{equation}\label{TEGR}
t_1=1/4,\  t_2=1/2,\  t_3=-1, 
\end{equation}
then the first term in action \eqref{mat} becomes the teleparallel equivalent Einstein-Hilbert Lagrangian up to a surface term
\begin{equation}
\mathbb{T}=-T^{\mu}T_{\mu}+\frac{1}{4}T^{\rho\sigma\mu}T_{\rho\sigma\mu}+\frac{1}{2}T^{\mu\sigma\rho}T_{\rho\sigma\mu}=-\mathring{R}-2\mathring{\nabla}_{\mu}T^{\mu},
\end{equation}
where $\mathring{R}$ is the Ricci scalar corresponding to the Levi-Civita connection and $\mathring{\nabla}$ is the metric-compatible covariant derivative. Linear cosmological perturbations, including linear gravitational waves, were studied \cite{Li:2021mdp,Cai:2021uup,Fu:2023aab} with the parameter set \eqref{TEGR}. However, this model suffers from the strong coupling problem beyond linear orders, which we will show in the next section. Nevertheless, this problem can be avoided by choosing a suitable parameter set instead of \eqref{TEGR}.

\section{The background and linear scalar perturbation}\label{sec3}

In this section, we calculate the evolution of the background and linear cosmological scalar perturbations. To this end, we first provide the perturbed tetrads and metric up to cubic order.

\subsection{The tetrads and the corresponding metric}

We consider the background spacetime to be a spatially flat Friedmann-Robertson-Walker universe. The background tetrads can be parametrized as
\begin{equation}
\bar{e}^{A}_{\ \mu}=\mathrm{diag}(a,a,a,a).
\end{equation}
By substituting the tetrads into \eqref{gtr}, we obtain the background metric
\begin{equation}
ds^2=a^2(\mathrm{d}\tau^2-\delta_{ij}\mathrm{d}x^i\mathrm{d}x^j).
\end{equation}

The parametrization for the linearly perturbed tetrad field is \cite{Izumi:2012qj,Golovnev:2018wbh}
\begin{equation}\label{ptetrad}
\begin{split}
&e^{0}_{\ 0}=a(1+\phi),\ \ e^{0}_{\ i}=a\partial_i\beta,\ \ e^{a}_{\ 0}=a\delta^{ai}\partial_{i}\gamma,\\&
e^{a}_{\ i}=a\delta^{aj}\left[(1-\psi)\delta_{ij}+\partial_i\partial_j E+\epsilon_{ijk}\partial^k\lambda+\frac{1}{2}h_{ij}\right],
\end{split}
\end{equation}
where we consider only the linear scalar and tensor perturbations. Substituting the parametrized tetrads into \eqref{gtr},
we obtain the perturbed metric up to linear order,
\begin{equation}
\begin{split}
&g_{00}=a^2(1+2\phi),\ \ g_{0i}=-a^2\partial_i B,\\&
g_{ij}=-a^2[(1-2\psi)\delta_{ij}+2\partial_i\partial_j E +h_{ij}],
\end{split}
\end{equation}
where $B=\gamma-\beta$. From the expressions of the linearly perturbed metric, we can observe that in the Newtonian gauge, where $B=0$ and $E=0$, the scalar perturbations from tetrads satisfy $\gamma=\beta$ and $E=0$. We will use the Newtonian gauge for the remainder of this paper.

In a given coordinate system, we can always write the tetrad field as $e^{A}_{\ \mu}=\bar{e}^{A}_{\ \nu}e^{\nu}_{\ \mu}$, and thus use the exponential expansion \cite{Li:2018ixg,Hu:2023juh}
\begin{equation}
\begin{split}
e_{\ \mu}^\nu=\text{exp}^{\left(\delta e_{\ \mu}^\nu\right)}=&\delta_{\ \mu}^\nu+\delta e_{\ \mu}^\nu+\frac{1}{2}\delta e^{\nu}_{\ \rho}\delta e^{\rho}_{\ \mu}\\&+\frac{1}{6}\delta e^{\nu}_{\ \rho}\delta e^{\rho}_{\ \sigma}\delta e^{\sigma}_{\ \mu}+...,
\end{split}
\end{equation}
where $\delta e^{\nu}_{\ \mu}$ is the perturbation of $e^{\nu}_{\ \mu}$ and $e^{\nu}_{\ \mu}=\bar{e}_{A}^{\ \nu}e^{A}_{\ \nu}$. Then the perturbed tetrad field up to the cubic order is \footnote{For our purpose of calculating SIGWs, we only keep cubic terms that contain two scalar modes and one tensor mode for notational simplicity.} 
\begin{equation}\label{e00}
e^{0}_{\ 0}=a\left[1+\phi+\frac{1}{2}\phi^2+\frac{1}{2}\partial_i\gamma\partial^i\gamma+\frac{1}{12}h_{ij}\partial^i\gamma\partial^j\gamma\right],
\end{equation}
\begin{equation}\label{e0i}
\begin{split}
e^{0}_{\ i}=&a\left[\partial_i\gamma+\frac{1}{2}(\phi-\psi)\partial_i\gamma-\frac{1}{2}\epsilon_{ijk}\partial^j\lambda\partial^k\gamma\right.\\&\left.+\frac{1}{4}h_{ij}\partial^j\gamma+\frac{1}{12}(\phi-2\psi)h_{ij}\partial^j\gamma\right.\\&\left.-\frac{1}{12}(\epsilon_{lij}h^{l}_{\ k}+\epsilon_{klj}h^{l}_{\ i})\partial^j\lambda\partial^k\gamma\right],
\end{split}
\end{equation}
\begin{equation}\label{ea0}
\begin{split}
e^{a}_{\ 0}=&a\delta^{ai}\left[\partial_i\gamma+\frac{1}{2}(\phi-\psi)\partial_i\gamma-\frac{1}{2}\epsilon_{ijk}\partial^k\lambda\partial^j\gamma\right.\\&\left.+\frac{1}{4}h_{ij}\partial^j\gamma+\frac{1}{12}(\phi-2\psi)h_{ij}\partial^{j}\gamma\right.\\&\left.-\frac{1}{12}(\epsilon_{klj}h^{k}_{\ i}-\epsilon_{ijk}h^{k}_{\ l})\partial^j\lambda\partial^l\gamma\right],
\end{split}
\end{equation}
\begin{equation}\label{eai}
\begin{split}
e^{a}_{\ i}=&a\delta^{aj}\left[(1-\psi)\delta_{ij}+\epsilon_{ijk}\partial^k\lambda+\frac{1}{2}h_{ij}+\frac{1}{2}\delta_{ij}\psi^2\right.\\&\left.+\frac{1}{2}\partial_i\gamma\partial_j\gamma-\epsilon_{ijk}\psi\partial^k\lambda-\frac{1}{2}\left(\delta_{ij}\partial_l\lambda\partial^l\lambda-\partial_i\lambda\partial_j\lambda\right)\right.\\&\left.
-\frac{1}{2}h_{ij}\psi-\frac{1}{4}\epsilon^{k}_{\ il}h_{jk}\partial^l\lambda-\frac{1}{4}\epsilon_{jkl}h^{k}_{\ i}\partial^l\lambda+\frac{1}{8}h_{jk}h^{k}_{\ i}\right.\\&\left.+\frac{1}{4}h_{ij}\psi^2-\frac{1}{4}(\epsilon_{ikl}h^k_{\ j}-\epsilon_{jkl}h^k_{\ i})\psi\partial^l\lambda\right.\\&\left.+\frac{1}{12}(h_{jk}\partial_i\gamma\partial^k\gamma+h_{ik}\partial_j\gamma\partial^k\gamma)\right.\\&\left.-\frac{1}{12}(h_{ij}\partial_k\lambda\partial^k\lambda-h_{kl}\partial^k\lambda\partial^l\lambda\delta_{ij})\right].
\end{split}
\end{equation}

The inverse tetrad field $e^{\ \mu}_{A}$ can be obtained with the relation $e^{\ \mu}_{A}e^{A}_{\ \nu}=\delta^{\mu}_{\nu}$,
\begin{equation}\label{ie00}
e^{\ 0}_{0}=\frac{1}{a}\left[1-\phi+\frac{1}{2}\phi^2+\frac{1}{2}\partial_i\gamma\partial^i\gamma-\frac{1}{12}h_{ij}\partial^i\gamma\partial^j\gamma\right],
\end{equation}
\begin{equation}\label{ie0i}
\begin{split}
e^{\ i}_{0}=&\frac{1}{a}\left[-\partial^i\gamma+\frac{1}{2}(\phi-\psi)\partial^i\gamma-\frac{1}{2}\epsilon^{i}_{\ jk}\partial^k\lambda\partial^j\gamma\right.\\&\left.+\frac{1}{4}h^{ij}\partial_{j}\gamma-\frac{1}{12}(\phi-2\psi)h^{ij}\partial_j\gamma\right.\\&\left.+\frac{1}{12}(\epsilon^i_{\ lj}h^l_{\ k}+\epsilon_{lkj}h^{li})\partial^j\lambda\partial^k\gamma\right],
\end{split}
\end{equation}
\begin{equation}\label{iea0}
\begin{split}
e^{\ 0}_{a}=&\frac{1}{a}\delta^{\ i}_{a}\left[-\partial_i\gamma+\frac{1}{2}(\phi-\psi)\partial_i\gamma+\frac{1}{2}\epsilon_{ijk}\partial^k\lambda\partial^j\gamma\right.\\&\left.+\frac{1}{4}h_{ij}\partial^j\gamma-\frac{1}{12}(\phi-2\psi)h_{ij}\partial^j\gamma\right.\\&\left.-\frac{1}{12}(\epsilon_{klj}h^k_i+\epsilon_{ikj}h^k_l)\partial^j\lambda\partial^l\gamma\right],
\end{split}
\end{equation}
\begin{equation}\label{ieai}
\begin{split}
e^{\ i}_{a}=&\frac{1}{a}\delta^{il}\delta^{\ j}_{a}\left[(1+\psi+\frac{1}{2}\psi^2)\delta_{lj}-\epsilon_{jlk}\partial^k\lambda-\frac{1}{2}h_{jl}\right.\\&\left.+\frac{1}{2}\partial_l\gamma\partial_j\gamma-\epsilon_{jlm}\psi\partial^m\lambda-\frac{1}{2}\left(\partial_k\lambda\partial^k\lambda\delta_{lj}-\partial_l\lambda\partial_j\lambda\right)\right.\\&\left.-\frac{1}{2}\psi h_{jl}+\frac{1}{4}\epsilon_{klm}h^k_{\ j}\partial^m\lambda-\frac{1}{4}\epsilon^{k}_{\ jm}h_{lk}\partial^m\lambda+\frac{1}{8}h_{lk}h^k_{\ j}\right.\\&\left.-\frac{1}{4}h_{jl}\psi^2-\frac{1}{12}(h_{jk}\partial_l\gamma\partial^k\gamma+h_{lk}\partial_j\gamma\partial^k\gamma)\right.\\&\left.+\frac{1}{4}(\epsilon_{klm}h^k_{\ j}+\epsilon_{jkm}h^k_{\ l})\psi\partial^m\lambda\right.\\&\left.+\frac{1}{12}(h_{jl}\partial_k\lambda\partial^k\lambda-h_{km}\partial^k\lambda\partial^m\lambda\delta_{jl})\right].
\end{split}
\end{equation}

The corresponding perturbed metric up to the cubic order is
\begin{equation}\label{g00}
\begin{split}
g_{00}=&a^2\left[1+2\phi+2\phi^2-\frac{1}{3}h_{ij}\partial^i\gamma\partial^j\gamma\right],
\end{split}
\end{equation}
\begin{equation}\label{g0i}
\begin{split}
g_{0i}=&a^2\left[(\phi+\psi)\partial_i\gamma-\frac{1}{2}h_{ij}\partial^j\gamma+\psi h_{ji}\partial^j\gamma\right.\\&\left.+\frac{1}{3}(\epsilon_{kil}h^k_{\ j}-\frac{1}{2}\epsilon_{jkl}h^{k}_{\ i})\partial^j\gamma\partial^l\lambda\right],
\end{split}
\end{equation}
and
\begin{equation}\label{gij}
\begin{split}
g_{ij}=&-a^2\left[(1-2\psi+2\psi^2)\delta_{ij}+h_{ij}-2\psi h_{ij}\right.\\&\left.-\frac{1}{2}(\epsilon_{kil}h^k_{\ j}+\epsilon_{kjl}h^k_{\ i})\partial^l\lambda+\frac{1}{2}h_{ki}h^k_{\ j}\right.\\&\left.+\left(\epsilon_{ilk}h^{k}_{\ j}\psi\partial^l\lambda+\frac{1}{6}h_{ik}\partial^k\gamma\partial_j\gamma+i\leftrightarrow j\right)\right.\\&\left.+2h_{ij}\psi^2-\left(\frac{2}{3}h_{ij}\partial_k\lambda\partial^k\lambda+\frac{1}{3}\delta_{ij}h_{kl}\partial^k\lambda\partial^l\lambda\right.\right.\\&\left.\left.-\frac{1}{2}h_{jk}\partial^k\lambda\partial_i\lambda-\frac{1}{2}h_{ik}\partial^k\lambda\partial_j\lambda\right)\right].
\end{split}
\end{equation}

With the relation $g^{\nu\rho}g_{\rho\mu}=\delta^{\nu}_{\ \mu}$, the inverse metric is
\begin{equation}\label{ig00}
\begin{split}
g^{00}=&\frac{1}{a^2}\left[1-2\phi+2\phi^2+\frac{1}{3}h_{ij}\partial^i\gamma\partial^j\gamma\right],
\end{split}
\end{equation}
\begin{equation}\label{ig0i}
\begin{split}
g^{0i}=&\frac{1}{a^2}\left[(\phi+\psi)\partial^i\gamma-\frac{1}{2}h^{ij}\partial_{j}\gamma-\psi h^{ij}\partial_j\gamma\right.\\&\left.+\frac{1}{3}(\epsilon^{\ i}_{k\ l}h^k_{\ j}-\frac{1}{2}\epsilon_{jkl}h^{ki})\partial^j\gamma\partial^l\lambda\right],
\end{split}
\end{equation}
and
\begin{equation}\label{igij}
\begin{split}
g^{ij}=&-\frac{1}{a^2}\left[(1+2\psi+2\psi^2)\delta^{ij}-h^{ij}-2\psi h^{ij}\right.\\&\left.+\frac{1}{2}(\epsilon^{\ j}_{l\ k}h^{li}+\epsilon^{\ i}_{l\ k}h^{lj})\partial^k\lambda+\frac{1}{2}h_{k}^{\ i}h^{jk}-2\psi^2h^{ij}\right.\\&\left.+\left(\epsilon^{i}_{\ lk}h^{kj}\psi\partial^l\lambda-\frac{1}{6}h^{ik}\partial_k\gamma\partial^j\gamma+i\leftrightarrow j\right)\right.\\&\left.+\left(\frac{2}{3}h^{ij}\partial_k\lambda\partial^k\lambda+\frac{1}{3}\delta^{ij}h_{kl}\partial^k\lambda\partial^l\lambda\right.\right.\\&\left.\left.-\frac{1}{2}h^{jk}\partial_k\lambda\partial^i\lambda-\frac{1}{2}h^{ik}\partial_k\lambda\partial^j\lambda\right)\right].
\end{split}
\end{equation}

We also have
\begin{equation}
\begin{split}
e=\sqrt{-g}=&a^4\left(1+\phi-3\psi+\frac{1}{2}\phi^2-3\phi\psi \right.\\&\left.\quad\quad+\frac{9}{2}\psi^2+\frac{1}{4}h_{ij}h^{ij}\right).
\end{split}
\end{equation}

\subsection{The background equations of motion and linear scalar perturbations}
Expanding action \eqref{mat} to linear order, we obtain
\begin{equation}
\begin{split}
S^{(1)}=&\int \mathrm{d}^3x \mathrm{d}\tau a^2\left[-\frac{3}{2}\mathcal{C}_1\mathcal{H}^2\phi-\left(\frac{1}{2}(\theta')^2+a^2V\right)\phi\right.\\&\left.-\frac{9}{2}\mathcal{C}_1\mathcal{H}^2\psi-\frac{3}{2}\left((\theta')^2-2a^2V\right)\psi\right.\\&\left.-3\mathcal{C}_1\mathcal{H}\psi'-a^2V_{\theta}\delta\theta+\delta\theta'\theta'\right],
\end{split}
\end{equation}
where $\mathcal{C}_1=2t_1+t_2+3t_3$ and the prime represents the derivative with respect to conformal time $\tau$.  

Varying the above action with respect to scalar perturbations, we obtain the background EOMs as follows
\begin{gather}\label{BK1}
-\frac{3}{2}\mathcal{C}_1\mathcal{H}^2=\frac{1}{2}(\theta')^2+a^2V,\\
\label{BK2}
\mathcal{C}_1(\mathcal{H}^2+2\mathcal{H}')=(\theta')^2-2a^2V,\\
\label{BK3}
\theta''+2\mathcal{H}\theta'+a^2V_{\theta}=0.
\end{gather}
It is obvious that when $t_1=1/4$, $t_2=1/2$, and $t_3=-1$, yielding $\mathcal{C}_1=-2$, we recover the results of general relativity (GR).

The quadratic action is
\begin{widetext}
\begin{equation}\label{ts2}
\begin{split}
S^{(2)}_{SS}=&\int \mathrm{d}^3x\mathrm{d}\tau a^2\left[\frac{1}{2}(\delta\theta')^2-\frac{1}{2}a^2V_{\theta\theta}\delta\theta^2-a^2V_{\theta}\delta\theta(\phi-3\psi)-(\phi+3\psi)\delta\theta'\theta'\right.\\&\left.-\frac{1}{2}\partial_i\delta\theta\partial^i\delta\theta+\frac{3}{2}\mathcal{C}_1(\psi')^2+\frac{1}{4}(9\psi^2+\phi^2)\left(3\mathcal{C}_1\mathcal{H}^2+(\theta')^2-2a^2V\right)\right.\\&\left.+3\mathcal{C}_1\mathcal{H}\phi\psi'+9\mathcal{C}_1\mathcal
{H}\psi\psi'-\frac{1}{2}\mathcal{C}_2\partial_i\phi\partial^i\phi-\mathcal{C}_3\partial_i\psi\partial^i\psi+2t_3\partial_i\psi\partial^i\phi\right.\\&\left.-\mathcal{C}_2\psi\partial^2\gamma'-2g_{\theta}\mathcal{H}\delta\theta\partial^2\lambda-2g_{\theta}\theta'\psi\partial^2\lambda+\mathcal{C}_2\partial^i\phi\partial_i\gamma'-\frac{1}{2}\mathcal{C}_2\partial_i\gamma'\partial^i\gamma'\right.\\&\left.+\frac{1}{2}\mathcal{C}_2\partial^2\gamma\partial^2\gamma+\mathcal{C}_4\partial_i\lambda'\partial^i\lambda'-\mathcal{C}_4\partial^2\lambda\partial^2\lambda\right],
\end{split}
\end{equation}
\end{widetext}
where $\mathcal{C}_2=2t_1+t_2+t_3$, $\mathcal{C}_3=2t_1+t_2+2t_3$, $\mathcal{C}_4=2t_1-t_2$, $g_\theta=\mathrm{d}g/\mathrm{d}\theta$, and $\partial^2$ represents $\partial^i\partial_i$.

Then the EOMs for scalar perturbations can be obtained by varying the above quadratic action with respect to the corresponding scalar perturbations
\begin{gather}
\label{tle1}
3\mathcal{C}_1\mathcal{H}(\psi'+\mathcal{H}\phi)-2t_3\partial^2\psi+\mathcal{C}_2\partial^2(\phi-\gamma')\\ \nonumber=-(\theta')^2\phi+\delta\theta'\theta'+a^2V_{\theta}\delta\theta,\\
\label{tle2}
3\mathcal{C}_1\mathcal{H}^2(3\psi-2\phi)-2t_3\partial^2\phi+2\mathcal{C}_3\partial^2\psi-\mathcal{C}_2\partial^2\gamma'\\ \nonumber-3\mathcal{C}_{1}\mathcal{H}'(3\psi+\phi) -3\mathcal{C}_1\mathcal{H}(2\psi'+\phi')-3\mathcal{C}_1\psi''\\ \nonumber-2g_{\theta}\theta'\partial^2\lambda-3\delta\theta'\theta'+3a^2V_{\theta}\delta\theta+9(\theta')^2\psi=0,\\
\label{tle3}
\mathcal{C}_3\psi-t_3\phi=g_{\theta}\theta'\lambda,\\
\label{tle4}
\delta\theta''+2\mathcal{H}\delta\theta'-\partial^2\delta\theta+a^2V_{\theta\theta}\delta\theta-\theta'(\phi'+3\psi')\\ \nonumber+2a^2V_{\theta}\phi=-2g_{\theta}\mathcal{H}\partial^2\lambda,\\
\label{tle5}
g_{\theta}(\mathcal{H}\delta\theta+\theta'\psi)=\mathcal{C}_4(\lambda''+2\mathcal{H}\lambda'-\partial^2\lambda),\\
\label{tle6}
(2\mathcal{H}\phi+\phi'+2\mathcal{H}\psi+\psi')=(\gamma''+2\mathcal{H}\gamma'-\partial^2\gamma), (\mathcal{C}_2\neq0).
\end{gather}

From the above EOMs, we can see that if we choose the parameter set defined in \eqref{TEGR}, which corresponds to the teleparallel equivalent of GR with $\mathcal{C}_2=\mathcal{C}_4=0$, the EOM for the scalar perturbation $\gamma$ disappears. However, $\gamma$ exists in the EOM of SIGWs, which leads to a strong coupling problem. Note that even if we choose the parameter set as \eqref{TEGR}, the linear perturbation $\lambda$ from tetrads exists in the EOMs \eqref{tle3} and \eqref{tle4}, resulting in the EOMs of the perturbations from the metric and scalar field being different from those in GR. Obviously, the extra scalar degrees of freedom exist in this model, whose nature and characterization need to be further studied, we will leave this to our future work.

\subsection{The evolution of background and linear scalar perturbations during the radiation-dominated era}

During the radiation-dominated era, the equation of state is $\bar P/\bar{\rho}=1/3$, with
\begin{equation}\label{BPE}
\bar P=\frac{(\theta')^2}{2a^2}-V,\ \ \ \bar{\rho}=\frac{(\theta')^2}{2a^2}+V,
\end{equation}
being the pressure and energy density of the background universe. We also have $\delta P/\delta \rho=1/3$,
where
\begin{equation}
\begin{split}\label{DPR}
& \delta \rho=-a^{-2}\theta'(\phi \theta'-\delta \theta')+V_{\theta}\delta \theta, \\
& \delta P=-V_{\theta}\delta \theta +a^{-2}\theta'(\delta \theta' - \phi \theta')
\end{split}
\end{equation}
 are the perturbations of the energy density and pressure.

Combining the above two equations \eqref{BPE} and the first two equations for the background evolution \eqref{BK1} and \eqref{BK2} , we obtain 
\begin{equation}\label{bkso}
\mathcal{H}=\frac{1}{\tau}, \ \theta'=\pm\sqrt{-2\mathcal{C}_1}\frac{1}{\tau}.
\end{equation}
From the above equation, we can see that the evolution of the Hubble parameter is independent of $\mathcal{C}_1$.

Obviously, for any choice of parameters $t_1$, $t_2$, and $t_3$, the EOMs for the linear scalar perturbations, \eqref{tle1}-\eqref{tle6}, are very difficult to solve analytically. This poses significant challenges for us in analyzing the behavior of SIGWs in our model.
In this paper, we primarily focus on the contributions from the PV term and the perturbations from the tetrads, denoted as $\lambda$ and $\gamma$, to SIGWs. We expect the background evolution to be the same as that of GR, implying the selection of $\mathcal{C}_1=2t_1+t_2+3t_3=-2$. Furthermore, we aim to minimize the differences between the scalar perturbations from the metric in GR and teleparallel gravity with the NY term. In the case of GR, $\phi=\psi$ without consideration of anisotropic stress. We maintain this assumption in our model, setting $\phi=\psi$. Additionally, in \eqref{tle5}, $\mathcal{H}\delta\theta+\theta'\psi=\mathcal{R}$ is the gauge-invariant curvature perturbation. As introduced in Sec \ref{sec2}, the curvature vanishes in teleparallel gravity, so we set $\mathcal{C}_4=0$ to ensure $\mathcal{R}=0$.

With the above assumptions, $\mathcal{C}_1=-2$, $\mathcal{C}_4=0$, and $\phi=\psi$, the EOMs for the linear perturbations \eqref{tle1}-\eqref{tle6} reduce to
\begin{gather}
\label{ls11}
\psi''+3\mathcal{H}\psi'+(\mathcal{H}^2+2\mathcal{H}')\psi=-\mathcal{H}(\psi'+\mathcal{H}\psi)+\mathbb{C}\partial^2\psi,\\
\label{ls21}
\mathcal{C}_2\psi=g_{\theta}\theta'\lambda,\\
\label{ls31}
\mathcal{H}\delta\theta+\theta'\psi=0,\\
\label{ls41}
2(2\mathcal{H}\psi+\psi')=(\gamma''+2\mathcal{H}\gamma'-\partial^2\gamma), (\mathcal{C}_2\neq0),
\end{gather}
with
\begin{equation}
   \mathbb{C}= \frac{2t_3-\mathcal{C}_2}{3\mathcal{C}_1}.
\end{equation}
Here, Eq. \eqref{ls11} is derived by combining Eqs. \eqref{tle1}-\eqref{tle3}, and \eqref{DPR}. 

Note that in the case of GR, $\mathbb{C}=1/3$, we obtain the solution easily
\begin{equation}
\mathcal{C}_2=0,
\end{equation}
which results in the disappearance of EOM for $\gamma$ \eqref{ls41}, giving rise to the strong coupling problem. Thus we require $\mathbb{C}\neq 1/3$ to avoid the strong coupling problem. Besides, $\mathbb{C}$ relates to the propagating speed of perturbation $\psi$, we assume $\mathbb{C}$ is a real number, and $\mathbb{C}\leq1$.

For late convenience to calculate the SIGWs, we split the perturbations into the primordial  perturbation and the transfer functions as follows:
\begin{gather}\label{trans1}
\psi(\bm k,\tau)=\frac{2}{3}\zeta(\bm k)T_\psi(x),\\
\label{trans2}
\gamma(\bm k,\tau)=\frac{2}{3}\zeta(\bm k)\frac{1}{k}T_\gamma(x),
\end{gather}
where $\zeta$ is the primordial curvature perturbation generated during the inflationary era and $x=k\tau$. It is worth noting that we assume $\mathcal{R}=0$ in teleparallel gravity. However, teleparallel gravity can only be viewed as a low-energy theory. During inflation, gravity may be described by another UV-complete theory. Thus, we expect $\zeta$ to be nonzero, resulting in observable effects in the CMB.

Recalling the evolution of the conformal Hubble parameter, the EOMs for the transfer functions can be written as
\begin{gather}
T^{**}_{\psi}(x)+\frac{4}{x}T^{*}_{\psi}(x)+\mathbb{C}T_{\psi}(x)=0,\\
T^{**}_{\gamma}(x)+\frac{2}{x}T^{*}_{\gamma}(x)+T_{\gamma}(x)=2T^{*}_{\psi}(x)+\frac{4}{x}T_{\psi}(x),
\end{gather}
where ``$*$" represents the derivative with respect to the argument. Then we can solve the above EOMs easily,
\begin{equation}\label{Tpsi}
T_{\psi}(x)=\frac{3}{x^2 \mathbb{C}}\left(\frac{\sin\left(x\sqrt{\mathbb{C}}\right)}{x\sqrt{\mathbb{C}}}-\cos\left(x\sqrt{\mathbb{C}}\right)\right),
\end{equation}
\begin{widetext}
\begin{equation}
\begin{split}\label{tgamma}
T_{\gamma}(x)=&\frac{1}{2\mathbb{C}^{3/2} x^2}\left\{-2 i \mathbb{C}^{3/2} c_1 x \sin (x)+\mathbb{C}^{3/2} c_2 x \sin (x)+2 \mathbb{C}^{3/2} c_1 x \cos (x)\right.\\ &\left.- i \mathbb{C}^{3/2} c_2 x \cos (x)+3(\mathbb{C}+1) x \cos (x) \left[\text{Ci}\left(x+\sqrt{\mathbb{C}}x\right)-\text{Ci}\left(x-\sqrt{\mathbb{C}}x\right)\right]\right.\\ &\left.+3 (\mathbb{C}+1) x \sin (x) \left[\text{Si}\left(\sqrt{\mathbb{C}} x+x\right)-\text{Si}\left(x-\sqrt{\mathbb{C}} x\right)\right]-6 \sin \left(\sqrt{\mathbb{C}} x\right)\right\},
\end{split}
\end{equation}
\end{widetext}
where
\begin{equation}
\text{Si}(x)=\int_0^x \mathrm{d} y\frac{\sin y}{y},\quad \text{Ci}(x)=-\int_x^\infty \mathrm{d} y \frac{\cos y}{y}
\end{equation}
are sine integral and cosine integral, respectively.

If we choose the parameter set \eqref{TEGR}, then $\mathbb{C}=1/3$, and the transfer function $T_{\psi}$ is the same as that in GR. However, from EOM \eqref{ls31}, the fluctuation of the scalar field $\delta\theta$ still differs from that in GR. There are two integral constants, $c_1$ and $c_2$, in the transfer function $T_{\gamma}$. We expect that $T_{\gamma}$ is a real function and finite as $x\rightarrow 0$. Then, we can obtain
\begin{equation}
c_1=\frac{1}{4\mathbb{C}^{3/2}}\left(6\sqrt{\mathbb{C}}+3(1+\mathbb{C})\text{log}\left(\frac{1-\sqrt{\mathbb{C}}}{1+\sqrt{\mathbb{C}}}\right)\right),
\end{equation}
\begin{equation}
c_2=\frac{i}{2\mathbb{C}^{3/2}}\left(6\sqrt{\mathbb{C}}+3(1+\mathbb{C})\text{log}\left(\frac{1-\sqrt{\mathbb{C}}}{1+\sqrt{\mathbb{C}}}\right)\right).
\end{equation}
By substituting $c_1$ and $c_2$ into Eq. \eqref{tgamma},  the transfer function $T_{\gamma}$ becomes 
\begin{widetext}
\begin{equation}
\begin{split}\label{Tga}
T_{\gamma}(x)=&\frac{1}{2 \mathbb{C}^{3/2} x^2}\left\{x\cos(x)\left[6\sqrt{\mathbb{C}}+3(1+\mathbb{C})\text{log}\left(\frac{1-\sqrt{\mathbb{C}}}{1+\sqrt{\mathbb{C}}}\right)\right]-6\sin \left(\sqrt{\mathbb{C}} x\right)\right.\\ &\left.\quad \quad\quad\quad+3(\mathbb{C}+1)x\cos(x)\left[\text{Ci}\left(x+\sqrt{\mathbb{C}}x \right)-\text{Ci}\left(x-\sqrt{\mathbb{C}}x\right)\right]\right.\\ &\left.\quad \quad\quad\quad+3(\mathbb{C}+1)x\sin(x) \left[\text{Si}\left(x+\sqrt{\mathbb{C}}x\right)-\text{Si}\left(x-\sqrt{\mathbb{C}}x\right)\right]\right\}.
\end{split}
\end{equation}
\end{widetext}
It seems that $T_{\gamma}$ is singular when $\mathbb{C}\rightarrow 1$ due to the logarithmic divergence. In fact, 
in the limit $\mathbb{C}\rightarrow 1$, we have
\begin{equation}
\begin{split}
T_{\gamma}(x)|_{\mathbb{C}\rightarrow 1}=&\frac{3}{x^2}\left[-x\cos x\left(E_{\gamma}-1-\text{Ci}(2x)+\log(2x)\right)\right.\\&\left.+\sin x\left(x\text{Si}(2x)-1\right)\right],
\end{split}
\end{equation}
where $E_{\gamma}$ is the Euler Gamma constant.

Recalling the assumptions we made above, $\mathcal{C}_1=-2$ and $\mathcal{C}_4=0$, only $t_1$ is a free parameter, the others can be expressed as
\begin{equation}
\begin{split}
&t_2=2t_1, \ t_3=-\frac{4t_1+2}{3},\ \mathcal{C}_2=\frac{2}{3}(4t_1-1),\\& \mathcal{C}_3=\frac{4}{3}(t_1-1),\ \mathbb{C}=\frac{1}{9}(8t_1+1).
\end{split}
\end{equation}

\section{The scalar induced gravitational waves}\label{sec4}

In this section, we first derive the EOMs for SIGWs, and then we calculate the power spectrum and the energy density of SIGWs. Expanding action \eqref{mat} to cubic order, we obtain
\begin{equation}\label{AGWs}
S_{\text{GW}}=S^{(2)}_{TT}+S^{(3)}_{TT},
\end{equation}
where
\begin{equation}
\begin{split}
S^{(2)}_{TT}=\int \mathrm{d}^3x \mathrm{d}\tau \frac{a^2}{8}&\left[\mathcal{C}_{5}(h^{'}_{ij}h^{'ij}-\partial_kh_{ij}\partial^kh^{ij})\right.\\&\left.+g'\epsilon_{ijk}h^{li}\partial^jh^{\ k}_{l}\right],
\end{split}
\end{equation}
and
\begin{equation}
S^{(3)}_{TT}=\int \mathrm{d}^3x \mathrm{d}\tau a^2 (\mathcal{L}_{ij}+\mathcal{L}^{\mathrm{PV}}_{ij})h^{ij},
\end{equation}
with
\begin{widetext}
\begin{equation}
\begin{split}
\mathcal{L}_{ij}=&\frac{1}{2}\partial_i\delta\theta\partial_j\delta\theta+\mathcal{C}_4\partial_i\partial_j\lambda\partial^2\lambda-\frac{1}{2}g_{\theta\theta}\theta'\delta\theta\partial_i\partial_j\lambda-\frac{1}{2}g_{\theta}\delta\theta'\partial_i\partial_j\lambda -\frac{1}{2}g_{\theta}\delta\theta\partial_i\partial_j\lambda'
\\& +g_{\theta}\theta'\psi\partial_i\partial_j\lambda+\frac{1}{2}\mathcal{C}_2\partial_i\phi\partial_j\phi-2t_3\partial_i\phi\partial_j\psi+\mathcal{C}_3\partial_i\psi\partial_j\psi-\frac{1}{2}\mathcal{C}_2\partial_i\gamma'\partial_j\phi
\\&-t_3\mathcal{H}\partial_i\gamma\partial_j\phi+\frac{1}{4}\mathcal{C}_2\partial_i\gamma\partial_j\phi'+\mathcal{C}_2\partial_i\gamma'\partial_j\psi+\frac{7}{2}\mathcal{C}_2\mathcal{H}\partial_i\gamma\partial_j\psi-\frac{1}{4}(5\mathcal{C}_{5}+t_3)\psi'\partial_i\partial_j\gamma
\\&-\frac{1}{2}\theta'\partial_i\gamma\partial_j\delta\theta-\frac{1}{2}\mathcal{C}_2\mathcal{H}\partial_i\gamma'\partial_j\gamma-\frac{1}{4}\mathcal{C}_2\partial_i\gamma\partial_j\gamma''+\frac{1}{6}\mathcal{C}_1\mathcal{H}^2\partial_i\gamma\partial_j\gamma-\frac{1}{6}\mathcal{C}_1\mathcal{H}'\partial_i\gamma\partial_j\gamma\\&-\frac{1}{4}\mathcal{C}_2\partial_i\partial_j\gamma\partial^2\gamma
+\frac{1}{6}(\theta')^2\partial_i\gamma\partial_j\gamma,
\end{split}
\end{equation}
\end{widetext}
and
\begin{equation}
\begin{split}
\mathcal{L}^{\mathrm{PV}}_{ij}=&\epsilon_{klj}\left[-\frac{1}{2}\mathcal{C}_{5}\partial^k\partial_i\lambda\partial^l\phi+\frac{1}{2}\mathcal{C}_{5}\partial^k\partial_i\lambda\partial^l\psi\right.\\&\left.\quad \quad-\frac{1}{4}g_{\theta}\theta'\partial^k\lambda\partial^l\partial_i\lambda -\mathcal{C}_4\partial^k\lambda'\partial^l\partial_i\gamma\right.\\&\left.-\frac{1}{2}g_{\theta}\partial^k\partial_i\gamma\partial^l\delta\theta-\frac{1}{4}g_{\theta}\theta'\partial^k\gamma\partial^l\partial_i\gamma\right],
\end{split}
\end{equation}
where $\mathcal{C}_5=2t_1+t_2=4t_1$.

\subsection{The EOM for SIGWs}

Varying the cubic action for SIGWs \eqref{AGWs} with respective to $h_{ij}$, the EOM for SIGWs is
\begin{equation}
\begin{split}
&-\frac{\mathcal{C}_{5}}{4}\left(h^{''}_{ij}+2\mathcal{H}h^{'}_{ij}-\nabla^2h_{ij}\right)\\&+\frac{1}{8}g'\left(\epsilon_{lki}\partial^lh^{k}_{\ j}+\epsilon_{lkj}\partial^lh^{k}_{\ i}\right)=\mathcal{T}^{lm}_{\ \ ij}s_{lm},
\end{split}
\end{equation}
where
\begin{equation}
s_{ij}=-\frac{1}{2}(\mathcal{L}_{ij}+\mathcal{L}_{ji}+\mathcal{L}^{\mathrm{PV}}_{ij}+\mathcal{L}^{\mathrm{PV}}_{ji}),
\end{equation}
$\mathcal{T}^{lm}_{\ \ ij}$ is the projection tensor.

We decompose $h_{ij}$ into circularly polarized modes as
\begin{equation}
h_{ij}(\bm{x},\tau)=\sum\limits_{A=R,L}\int \frac{\mathrm{d}^3k}{(2\pi)^{3/2}}e^{i\bm{k}\cdot\bm{x}}p^{A}_{ij}h^A_{\bm{k}}(\tau),
\end{equation}
where the circular polarization tensors are defined as
\begin{equation}\label{cpt}
p^R_{ij}=\frac{1}{\sqrt{2}}(\mathbf e^{+}_{ij}+i\mathbf e^{\times}_{ij}), \ \ p^L_{ij}=\frac{1}{\sqrt{2}}(\mathbf e^{+}_{ij}-i\mathbf e^{\times}_{ij}).
\end{equation}
The plus and cross polarization tensors can be expressed as
\begin{equation}
\label{poltensor1}
\begin{split}
\mathbf e^+_{ij}=&\frac{1}{\sqrt{2}}(\mathbf e_i \mathbf e_j-\bar{\mathbf e}_i \bar{\mathbf e}_j),\\
\mathbf e_{ij}^\times=&\frac{1}{\sqrt{2}}(\mathbf e_i\bar{\mathbf e}_j+\bar{\mathbf e}_i \mathbf e_j),
\end{split}
\end{equation}
where ${\mathbf e_{i}\left(\bm{k}\right)}$ and ${\bar{\mathbf e}_{i}\left(\bm{k}\right)}$ are two basis vectors which are orthogonal to each other and perpendicular to the wave vector ${\bm{k}}$, i.e., satisfying ${\bm k}\cdot {\mathbf e}={\bm k}\cdot \bar{\mathbf e}={\mathbf e}\cdot \bar{\mathbf e}=0$ and $|{\mathbf e}|=|\bar{\mathbf e}|=1$.

The projection tensor  extracts the transverse and trace-free part of the source, of which the definition is
\begin{equation}
\mathcal{T}^{lm}_{\ \ \ ij}s_{lm}(\bm{x},\tau)=\sum\limits_{A=R,L}\int \frac{\mathrm{d}^3\bm k}{(2\pi)^{3/2}}e^{i {\bm k} \cdot {\bm x}}p_{ij}^A p^{Alm}\tilde{s}_{lm}(\bm k,\tau), 
\end{equation}
where $\tilde{s}_{ij}$ is the Fourier transformation of the source $s_{ij}$.

With the above settings, we can now rewrite the EOM for SIGWs in Fourier space as
\begin{equation}\label{eu}
u^{A''}_{\bm k}+\left(\omega^2_A-\frac{a''}{a}\right)u^{A}_{\bm k}=-\frac{4a}{\mathcal{C}_5}S^A_{\bm{k}},
\end{equation}
where  $u^A=ah^A$ and
\begin{equation}\label{omgA}
\omega^2_A=k^2\left(1-\frac{\lambda^A M_{\text{PV}}}{k}\right), \ \ (\lambda^R=1,\ \lambda^L=-1),
\end{equation}
here  $M_{\text{PV}}= g'/\mathcal{C}_5$,  the source
\begin{equation}
S^A_{\bm{k}}=p^{Aij}\tilde{s}_{ij}(\bm{k},\tau).
\end{equation}

The source $S^A_{\bm k}$ can be divided into four parts,
\begin{equation}
S^A_{\bm k}=S^{A(\mathrm{PC}1)}_{\bm k}+S^{A(\mathrm{PC}2)}_{\bm k}+S^{A(\mathrm{PV}1)}_{\bm k}+S^{A(\mathrm{PV}2)}_{\bm k},
\end{equation}
where $S^{A(\mathrm{PC}1)}_{\bm k}$ and $S^{A(\mathrm{PV}1)}_{\bm k}$ do not contain the contribution from $\gamma$, representing the parity-conserved and parity-violating parts, respectively, while $S^{A(\mathrm{PC}2)}_{\bm k}$ and $S^{A(\mathrm{PV}2)}_{\bm k}$ represent the parts that contain the contribution from $\gamma$,
\begin{widetext}
\begin{equation}
\begin{split}
 S^{A(\mathrm{PC}1)}_{\bm k}=&\int \frac{\mathrm{d}^3\bm k'}{(2\pi)^{3/2}}p^{Aij}k^{'}_i k^{'}_j\zeta(\bm k')\zeta(\bm k-\bm k') f_{\mathrm{PC}1}(u,v,x),\\
  S^{A(\mathrm{PC}2)}_{\bm k}=&\int \frac{\mathrm{d}^3\bm k'}{(2\pi)^{3/2}}p^{Aij}k^{'}_i k^{'}_j\zeta(\bm k')\zeta(\bm k-\bm k') f_{\mathrm{PC}2}(u,v,x),\\
 S^{A(\mathrm{PV}1)}_{\bm k}=&\int \frac{\mathrm{d}^3\bm k'}{(2\pi)^{3/2}}p^{Aij}k^{'}_i k^{'}_j\zeta(\bm k')\zeta(\bm k-\bm k') f^A_{\mathrm{PV}1}(k,u,v,x),\\
  S^{A(\mathrm{PV}2)}_{\bm k}=&\int \frac{\mathrm{d}^3\bm k'}{(2\pi)^{3/2}}p^{Aij}k^{'}_i k^{'}_j\zeta(\bm k')\zeta(\bm k-\bm k') f^A_{\mathrm{PV}2}(k,u,v,x),
\end{split}
\end{equation}
\end{widetext}
where $u=k'/k$, $v=|\bm k-\bm{k}'|/k$, and
\begin{equation}
p^{Aij}k^{'}_i k^{'}_j=\frac{1}{2}k^{'2}\sin^2(\vartheta)\text{e}^{2i\lambda^A\ell},
\end{equation}
with $\vartheta$ being the angle between $\bm{k}'$ and $\bm{k}$ and $\ell$ being the azimuthal angle of $\bm{k}'$. The function $f_{\mathrm{PC}1}(u,v,x)$, $f_{\mathrm{PC}2}(u,v,x)$, $f^A_{\mathrm{PV}1}(u,v,x)$, and $f^A_{\mathrm{PV}2}(u,v,x)$ are defined as
\begin{equation}
\begin{split}
f_{\mathrm{PC}1}(u,v,x)=&-\frac{2}{9}\left[\frac{4t_1+8}{3}T_{\psi}(ux)T_{\psi}(vx)\right.\\&\left.-\frac{2}{3}(4t_1-1)\frac{uk}{\mathcal{H}}T^{*}_{\psi}(ux)T_{\psi}(vx)+u\leftrightarrow v\right],
\end{split}
\end{equation}
\begin{equation}
\begin{split}
f_{\mathrm{PC}2}(u,v,x)=&-\frac{2}{9}\left[\frac{4t_1-1}{3}T^{*}_{\gamma}(ux)T_{\psi}(vx)\right.\\&\left.+\frac{16t_1-1}{3}\frac{v}{u}T_{\gamma}(ux)T^{*}_{\psi}(vx)\right.\\&\left.+\frac{32t_1+1}{3}\frac{\mathcal{H}}{uk}T_{\gamma}(ux)T_{\psi}(vx)\right.\\&\left.-\frac{1}{3}(4t_1-1)\frac{\mathcal{H}}{vk}T^{*}_{\gamma}(ux)T_{\gamma}(vx)\right.\\&\left.-\frac{1}{6}(4t_1-1)\frac{v}{u}T_{\gamma}(ux)T^{**}_{\gamma}(vx)\right.\\&\left.-\frac{1}{6}(4t_1-1)\frac{v}{u}T_{\gamma}(ux)T_{\gamma}(vx)+u\leftrightarrow v\right],
\end{split}
\end{equation}
\begin{equation}\label{spv1}
f^A_{\mathrm{PV}1}(u,v,x)=\frac{2}{9}\frac{\lambda^Ak}{M_{\text{PV}}}\frac{\mathcal{C}^2_2}{\mathcal{C}_5}\left(\frac{u}{4}T_{\psi}(ux)T_{\psi}(vx)+u\leftrightarrow v\right),
\end{equation}
\begin{equation}\label{spv2}
\begin{split}
f^A_{\mathrm{PV}2}(u,v,x)=&\frac{2}{9}\frac{\lambda^A M_{\text{PV}}}{k}\mathcal{C}_5\left[\frac{uk}{2v\mathcal{H}}T_{\psi}(ux)T_{\gamma}(vx)\right.\\&\left.+\frac{1}{2v}T_{\gamma}(ux)T_{\gamma}(vx)+u\leftrightarrow v\right].
\end{split}
\end{equation}
We have used the equations for the background and linear perturbations to simplify the above expressions.

Eq. \eqref{eu} can be solved by the method of Green's function,
\begin{equation}\label{Eh}
h^{A}_{\bm k}\left(\tau\right)=-\frac{4}{\mathcal{C}_5a(\tau)}\int^{\tau}\mathrm{d}\bar{\tau}~G^A_{k}\left(\tau,\bar{\tau}\right)
a\left(\bar{\tau}\right)S^A_{\bm k}\left(\bar{\tau}\right),
\end{equation}
where Green's function $G^A_{k}\left(\tau,\bar{\tau}\right)$ satisfies the equation
\begin{equation}\label{GREEN}
G^{A''}_{k}(\tau,\bar{\tau})+\left(\omega_A^2-\frac{a''}{a}\right)G^{A}_{k}(\tau,\bar{\tau})=\delta(\tau-\bar{\tau}).
\end{equation}
The coupling function $g$ characterizes the deviation of Green's function from standard GR. For an arbitrary form of $g$, $\omega_A$ defined in Eq. (\ref{omgA}) is a complex function of both the wave number $k$ and the conformal time $\tau$. Consequently, it is challenging to solve Eq. \eqref{GREEN} and obtain the expression for Green's function analytically. On the one hand, for our purpose of studying the contributions from the scalar perturbations to the SIGWs, we assume that the change in Green's function is as minimal as possible relative to that in GR. On the other hand, since $\omega_A$ is associated with the propagation speed of the GWs, we assume that $\omega_A$ is approximately time-independent and depends only on the wave number during the generation of SIGWs.
We will consider an exponential form of the coupling function
\begin{equation}\label{gvp}
	g(\theta)=g_0\mathrm{e}^{\alpha\theta},
\end{equation}
which renders $\omega_A$ independent of time and allows us to obtain an analytical solution to Eq. \eqref{GREEN}.

Recalling the background equations \eqref{bkso}, the solution for the scalar field is found to be
\begin{equation}\label{evp}
	\theta=2\beta\ln(\tau/\tau_0)+\theta_0,
\end{equation}
where $\theta_0$ is the value of $\theta$ at $\tau_0$ and $\beta=\pm1$, corresponding to $\theta'=\pm 2/\tau$, respectively. Substituting Eqs. \eqref{gvp} and \eqref{evp} into the coupling function $g$, we obtain
\begin{equation}\label{fprime}
	g'=\frac{2\alpha\beta g_0\mathrm{e}^{\alpha\theta_0}}{\tau^{2\alpha\beta}_0}\tau^{2\alpha\beta-1}.
\end{equation}
From Eq. \eqref{fprime}, it is evident that if we set $2\alpha\beta-1=0$, then $g'$ becomes a constant. Consequently, $M_{\text{PV}}$ defined above becomes
\begin{equation}
	M_{\text{PV}}=\frac{g_0\mathrm{e}^{\alpha\theta_0}}{\mathcal{C}_5\tau_0}, \label{calM0}
\end{equation} 
which is independent of time, as is $\omega_A$. With these assumptions, we can analytically solve Eq. \eqref{GREEN} to obtain the expression for Green's function,
\begin{equation}
	G^{A}_{k}(\tau,\bar\tau)=\frac{\sin[\omega_A(\tau-\bar\tau)]}{\omega_A}\Theta(\tau-\bar\tau), \label{gf}
\end{equation}
where $\Theta$ is the Heaviside step function.

The constant $M_{\text{PV}}$ defined in Eq. \eqref{calM0} has the dimension of energy, which can be interpreted as the characteristic energy scale of parity violation in our model. Therefore, it is of interest to estimate $M_{\text{PV}}$ based on current observations.
The recent observations from GW170817 \cite{LIGOScientific:2017vwq} and GRB170817A \cite{LIGOScientific:2017zic} constrain the propagating speed of GWs to be
\begin{equation}\label{cgwc}
-3\times 10^{-15}\leq c_{\rm {gw}}-1\leq 7\times 10^{-16}.
\end{equation}
Using the definition of $\omega_A$ in Eq. \eqref{omgA}, we have
\begin{equation}\label{cpvc}
c_{\rm {gw}}=\frac{\omega_A}{k}=\left(1-\frac{\lambda^A M_{\text{PV}}}{k}\right)^{1/2}\simeq 1-\frac{\lambda^A M_{\text{PV}}}{2k},
\end{equation}
which means
\begin{equation}
\frac{|M_{\text{PV}}|}{k}<1.4\times 10^{-15}.
\end{equation}
Therefore, the typical energy scale of parity violation is much smaller than the wave numbers that we are interested in.

Besides, the constraint on the parity-violating energy scale $M_{\text{PV}}$ from the GW events of binary black hole mergers in the LIGO-Virgo catalogs GWTC-1 and GWTC-2 is $M_{\text{PV}}<6.4 \times 10^{-42}\text{Gev}$ at $90\%$ confidence level \cite{Wu:2021ndf}, which corresponds to $M_{\text{PV}}\sim \mathcal{O}(10^{-3})\ \text{Mpc}^{-1}$. Since SIGWs are generated on small scales where $k\gg1 \ \text{Mpc}^{-1}$, this also implies $M_{\text{PV}}/k\ll 1$.

From Eq. \eqref{spv1}, it might appear that the term $f^A_{\text{PV}1}\propto (k/{M_{\text{PV}}})\mathcal{C}^2_2/\mathcal{C}_5$ could be very large and potentially violate perturbation theory. However, we will show that this term is negligible.

The dimensions of scalar perturbations $\psi$ and $\lambda$ are $[\psi]=k^0$ and $[\lambda]=k^{-1}$, while the coefficient $[\mathcal{C}_2]=k^0$. Furthermore, according to \eqref{ptetrad}, we have $\psi\sim k\lambda$. Taking into account relation \eqref{ls21}, we find that
\begin{equation}
	\frac{\mathcal{C}_2}{g_{\theta}\theta'}k=\frac{k\lambda}{\psi}\sim 1,
\end{equation}
which implies
\begin{equation}
	\mathcal{C}_2\sim \frac{g_{\theta}\theta'}{k}\sim \frac{M_\text{PV}}{k}\mathcal{C}_5\ll\mathcal{C}_5.
\end{equation}
Recalling the relation $\mathcal{C}_2=2/3(\mathcal{C}_5-1)$, we have $\mathcal{C}_5\sim 1$.

Consequently, the coefficient in Eq. \eqref{spv1} can be estimated as
\begin{equation}
	\frac{k}{M_{\text{PV}}}\frac{\mathcal{C}^2_2}{\mathcal{C}_5}\sim \frac{M_{\text{PV}}}{k}\mathcal{C}_5\ll1.
\end{equation}

Considering the above analysis, we can conclude that the contribution from the PV term to SIGWs is negligible.

\subsection{The power spectrum of SIGWs}
The solutions of the circularly polarized modes can be written as
\begin{widetext}
\begin{equation}
\label{hsolution}
h^A_{\bm k}(\tau)=\frac{4}{\mathcal{C}_5} \int\frac{\mathrm{d}^3\bm k'}{(2\pi)^{3/2}} p^{Aij}k^{'}_i k^{'}_j\zeta(\bm k')\zeta(\bm k-\bm k')\frac{1}{k^2}I^{A}(k,u,v,x),
\end{equation}
where
\begin{equation}
\begin{split}
\label{I_int}
I^{A}(k,u,v,x)&=-\int_0^x\mathrm{d}\bar{x}\frac{a(\bar{\tau})}{a(\tau)}k G^A_{k}(\tau,\bar{\tau})\sum_{i=1,2}\left(f_{\mathrm{PC}i}(u,v,\bar x)+f^A_{\mathrm{PV}i}(k,u,v,\bar x)\right)\\&
= \sum_{i=1,2}\left(I^{A}_{\mathrm{PC}i}(k,u,v,x)+I^{A}_{\mathrm{PV}i}(k,u,v,x)\right),
\end{split}
\end{equation}
\end{widetext}
with
\begin{equation}
\label{ISC}
I^{A}_{\mathrm{PC}i}(k,u,v,x)=-\int_0^x\mathrm{d}\bar{x}\frac{a(\bar{\tau})}{a(\tau)}kG^A_{k}(\tau,\bar{\tau})f_{\mathrm{PC}i}(u,v,\bar x),\\
\end{equation}
and
\begin{equation}
\label{IPV}
I^{A}_{\mathrm{PV}i}(k,u,v,x)=-\int_0^x\mathrm{d}\bar{x}\frac{a(\bar{\tau})}{a(\tau)}k G^A_{k}(\tau,\bar{\tau})f^A_{\mathrm{PV}i}(u,v,\bar x).
\end{equation}
The analytic expressions for $I^A_{\mathrm{PC}1}$ and $I^A_{\mathrm{PV}1}$ can be found in Appendix \ref{kernel}. The remaining parts, $I^A_{\mathrm{PC}2}$ and $I^A_{\mathrm{PV}2}$ cannot be calculated analytically, so we will compute them numerically.

The power spectra of the SIGWs $\mathcal{P}_{h}^{A}$ are defined by 
\begin{equation}\label{PTD}
\langle h^A_{\bm{k}} h^C_{\bm{k}'}\rangle =\frac{2\pi^2}{k^3}\delta^3(\bm k+\bm k')\delta^{AC}\mathcal{P}^{A}_{h}(k).
\end{equation}
With the definition of $\mathcal{P}_{h}^{A}$ \eqref{PTD} and the solution of SIGWs, we can obtain the power spectra of the SIGWs \footnote{We have assumed that $\zeta$ is Gaussian to derive Eq. \eqref{PStensor}. For the non-Gaussian contributions, please refer to Refs. \cite{Cai:2018dig,Unal:2018yaa,Adshead:2021hnm,Garcia-Saenz:2022tzu,Garcia-Saenz:2023zue} and references therein.}
\begin{equation}\label{PStensor}
\begin{split}
\mathcal{P}^{A}_h(k,x)=\frac{4}{\mathcal{C}^2_5}\int_{0}^\infty\mathrm{d}u\int_{|1-u|}^{1+u}\mathrm{d}v&\left(
\mathcal{J}(u,v)I^{A}(u,v,x)^2\right.\\&\left.\mathcal{P}_\zeta(uk)\mathcal{P}_\zeta(vk)\right),
\end{split}
\end{equation}
where
\begin{equation}
\mathcal{J}(u,v)=\left[\frac{4u^2-(1+u^2-v^2)^2}{4uv}\right]^2,
\end{equation}
and $\mathcal{P}_\zeta$ is the power spectrum of primordial curvature perturbation.

The fractional energy density of the SIGWs is 
\begin{widetext}
\begin{equation}\label{OGW}
\begin{split}
\Omega_{\mathrm{GW}}(k,x)&=\frac{1}{12}\left(\frac{k}{\mathcal{H}}\right)^2\sum\limits_{A=R,L}\overline{\mathcal{P}^A_h(k,x)}=\frac{x^2}{12}\sum\limits_{A=R,L}\overline{\mathcal{P}^A_h(k,x)}\\
&=\frac{1}{3\mathcal{C}^2_5}\int_{0}^\infty\mathrm{d}u\int_{|1-u|}^{1+u}\mathrm{d}v
\mathcal{J}(u,v)\sum\limits_{A=R,L}\overline{\tilde{I}^{A}(k,u,v,x)^2}\mathcal{P}_\zeta(uk)\mathcal{P}_\zeta(vk),
\end{split}
\end{equation}
\end{widetext}
where the overline represents the time average and $\overline{\tilde{I}^{A}(k,u,v,x)^2}=\overline{I^{A}(k,u,v,x)^2}x^2$. 

The GWs behave as free radiation, thus the fractional energy density of the SIGWs at the present time $\Omega_{\mathrm{GW},0}$ can be expressed as 
\cite{Espinosa:2018eve}
\begin{equation}\label{EGW}
\Omega_{\mathrm{GW},0}\left(k\right)=\Omega_{\mathrm{GW}}\left(k,\eta\rightarrow\infty\right)\Omega_{r,0},
\end{equation}
where $\Omega_{r,0} \simeq 9\times 10^{-5}$ is the current fractional energy density of the radiation \cite{Sato-Polito:2019hws}.

\begin{figure}[htp]
\centering
\includegraphics[width=0.8\linewidth]{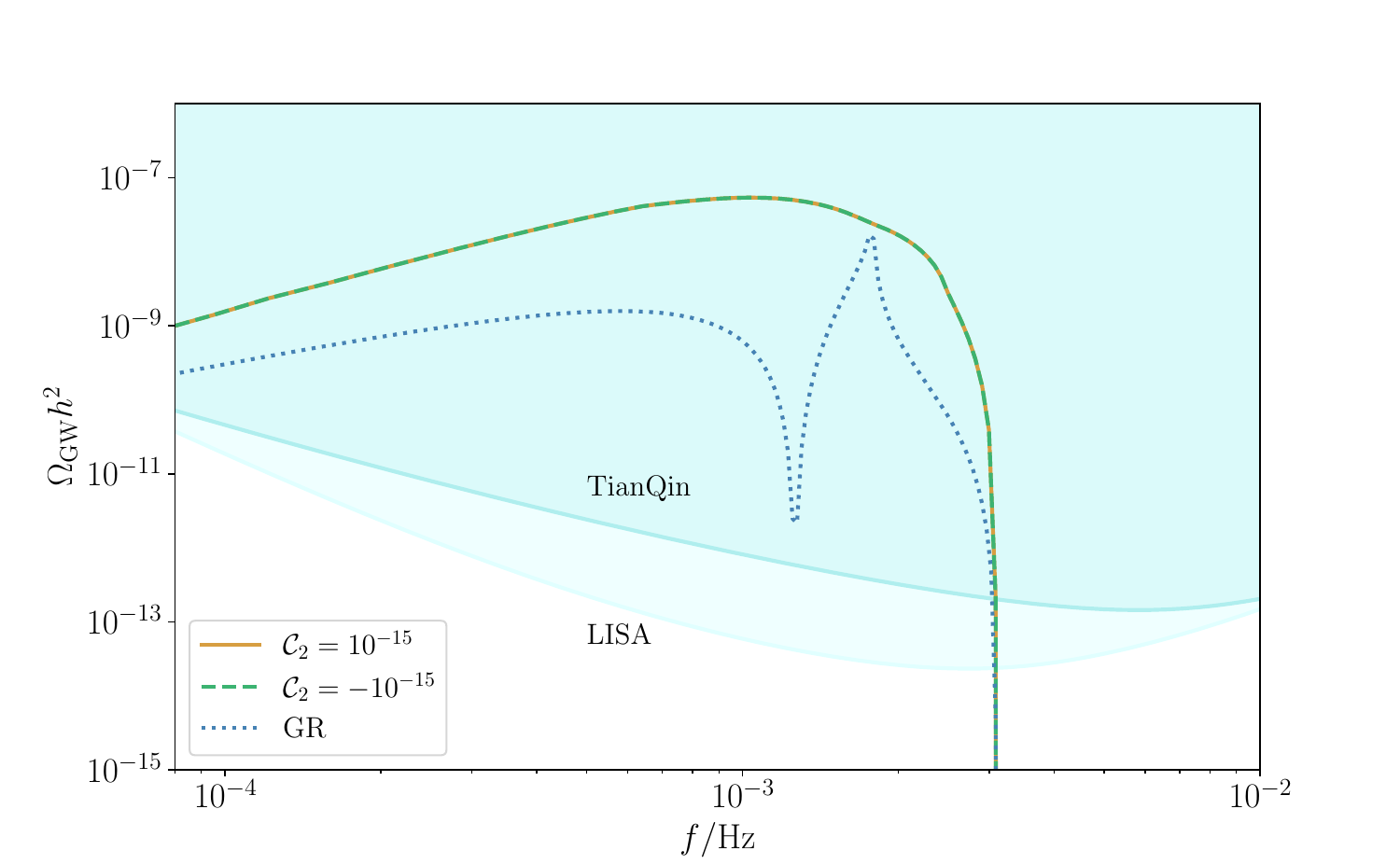}
\caption{The energy density of SIGWs from GR (dotted line) and our model (solid and dashed lines). The peak scale is $k_p=10^{12} \text{Mpc}^{-1}$, which corresponds to the maximum sensitivity of TianQin and LISA. The amplitude of the power spectrum is fixed to be $\mathcal{A}_{\zeta}=10^{-2}$.}
\label{fig:GWs1}
\end{figure}

To analyze the behavior of SIGWs in our model, in the following part of this section, we compute the energy density of SIGWs with a concrete power spectrum of primordial curvature perturbations. Considering the SIGWs induced by the monochromatic power spectrum,
\begin{equation}\label{ps1}
\mathcal{P}_\zeta(k)=\mathcal{A}_\zeta\delta(\ln(k/k_p)),
\end{equation}
then we obtain the energy density of SIGWs at the present time, 
\begin{equation}
\begin{split}
\Omega_{\text{GW},0}(k)=&\frac{1}{3\mathcal{C}^2_5}\Omega_{r,0}\mathcal{A}_{\zeta}^2\tilde{k}^{-2}\mathcal{J}(\tilde{k}^{-1},\tilde{k}^{-1})\\&\sum\limits_{A=R,L}\overline{\tilde{I}^{A}(k,\tilde{k}^{-1},\tilde{k}^{-1},x\rightarrow \infty)^2}\Theta(2-\tilde{k}),
\end{split}
\end{equation}
where $\tilde{k}=k/k_p$. We perform numerical calculations to determine the energy density of SIGWs, and the result is shown in Fig. \ref{fig:GWs1}.

Recalling the discussion in Sec. \ref{sec3}, the energy density of SIGWs may exhibit discrepancies between PV metric teleparallel gravity and GR. This discrepancy arises from the extra scalar perturbation in tetrads $\gamma$, as well as the distinct evolution of scalar perturbation from metric $\psi$ and the fluctuation of the scalar field $\delta\theta$, which deviate from their counterparts in GR. From Fig. \ref{fig:GWs1}, as expected, it is evident that the spectrum of the energy density of SIGWs differs significantly between GR and our model. For the SIGWs from GR, there is a divergence at $\tilde{k}=2/\sqrt{3}$ due to the resonant amplification \cite{Ananda:2006af,Kohri:2018awv}. In contrast, the energy density of SIGWs in our model is regular across all frequencies. This feature makes metric teleparallel gravity distinguishable from GR.

It is difficult to completely analyze the behavior of the spectrum of the energy density of SIGWs due to the absence of analytic expressions for $I^A$. We only consider the analytic part of $I^A$, especially the contribution from $I^A_{\text{PC}1}$, which is analytic. From the expression \eqref{Ipc1} in Appendix \ref{kernel}, the possible divergence in $I^A_{\text{PC}1}$ comes from a logarithmic term, $\log|w-\sqrt{\mathbb{C}}(u+v)|$, which is similar to GR. In the case of SIGWs induced by the monochromatic power spectrum, the term in $I^A_{\text{PC}1}$ that contains this logarithmic divergence is given by
\begin{equation}
\begin{split}
I^A_{\text{PC}1} \supset &\frac{3}{4}\tilde{k}^3 \Bigg(\frac{9\tilde{k}\left(8(5t_1+1)(8t_1+1)-3(14t_1+1)\tilde{k}^2\right)}{(8t_1+1)^3}\\
& \quad \quad-\frac{8(16t_1+5)}{(8 t_1+1)^{3/2}}\Bigg)\log \left(\left|-9\tilde{k}^2+32t_1+4\right|\right).
\end{split}
\end{equation}
For convenience of discussion, we have approximated $w=\omega_k/k \simeq1$, taking into account the observational constraints on the propagating speed of GWs \eqref{cgwc} and \eqref{cpvc}. The above expression vanishes even when $-9\tilde{k}^2+32t_1+4=0$, thus $I^A_{\text{PC}1}$ does not contribute a divergent term.

\section{Conclusion}\label{sec5}

SIGWs are a useful tool to test gravitational theory and probe the early universe. In this paper, we calculate the SIGWs from metric teleparallel gravity with the NY term in teleparallel geometry. By replacing the teleparallel equivalent Einstein-Hilbert Lagrangian with the general torsion scalar $\mathbb{T}$, Eq. \eqref{TS}, the strong coupling problem was avoided effectively only if $\mathcal{C}_2\neq 0$.

In the context of teleparallelism, we assumed $\mathcal{C}_4=0$ to maintain vanishing curvature perturbation. Furthermore, we aimed to minimize deviations in the background evolution and the scalar perturbation from the metric compared to those in GR, so that we can focus on the contribution from scalar perturbations in the tetrad field and the PV source. With this consideration, we solved EOMs for the background and the linear scalar perturbations during the radiation-dominated era and obtained the analytic solutions. We had chosen the coupling function of the PV term to be the exponential form, which ensures that the speed of SIGWs is independent of time. We further derived the analytical expression for Green's function of the tensor perturbations. With these analytic solutions, we calculated the power spectrum and the energy density of SIGWs. Then, we evaluated numerically the energy density of SIGWs with a monochromatic power spectrum of the primordial curvature perturbation. Considering the observational constraints on the propagating speeds of the GWs, we conclude that the effect of the PV term on the SIGWs is negligible. Nevertheless, the spectrum of the energy density of SIGWs in our model still differs from that in GR. Crucially, there is no divergence in the energy density of SIGWs in our model, which makes teleparallel gravity distinguishable from GR.

\begin{acknowledgments}
F.Z. thanks Yumin Hu for helpful discussions. This work was supported by the National Natural Science Foundation of China (NSFC) under the Grants No. 12305075 and No. 11975020. 
\end{acknowledgments}

\appendix

\section{The integral kernel}\label{kernel}
\begin{widetext}
In this appendix, we derive the integral kernel $I^A_{\text{PC}1}$ and $I^A_{\text{PV}1}$. With Green's function \eqref{GREEN} and the definition of $I^A$, we can express the integral kernel as follows
\begin{equation}
I^A_{1}(u,v,x)=\frac{\sin(wx)}{wx}I^A_{1s}(u,v,x)+\frac{\cos(wx)}{wx}I^A_{1c}(u,v,x),
\end{equation}
where the subscripts ``s'' and ``c'' stand for contributions involving the sine and cosine functions, respectively, and $w=\omega_A/k$. We also write
\begin{gather}
 I^A_{1s}(u,v,x)=\mathcal{I}^A_{1s}(u,v,x)- \mathcal{I}^A_{1s}(u,v,0), \nonumber\\
 I^A_{1c}(u,v,x)=\mathcal{I}^A_{1c}(u,v,x)- \mathcal{I}^A_{1c}(u,v,0),
\end{gather}
where $\mathcal{I}^A_{1s}$ and $\mathcal{I}^A_{1c}$ are defined by
\begin{gather}
\mathcal{I}^A_{1s}(u,v,y)=-\int \mathrm{d} y \cos wy f(u,v,y) y, \nonumber\\
\mathcal{I}^A_{1c}(u,v,y)=\int \mathrm{d} y \sin wy f(u,v,y) y.
\end{gather}
After lengthy calculations, we obtain the kernel of PC part
\begin{equation}
\begin{split}
\mathcal{I}^A_{\mathrm{PC}1s}(u,v,y)=&-\frac{1}{9\mathbb{C}^3u^3v^3y^4}\left[6\mathbb{C}(14t_1+1) u v y^2 \cos (w y) \cos \left(\sqrt{\mathbb{C}} u y\right) \cos \left(\sqrt{\mathbb{C}} v y\right)\right.\\&\left.+ \sqrt{\mathbb{C}}vy\left(-2t_1 \left(9\mathbb{C} y^2 \left(u^2-v^2\right)-7 w^2 y^2+42\right)\right.\right.\\& \left.\left.+9\mathbb{C} y^2 \left(v^2-u^2\right)+w^2 y^2-6\right)\cos (w y) \sin \left(\sqrt{\mathbb{C}} u y\right) \cos \left(\sqrt{\mathbb{C}} v y\right)\right.\\&\left.+\sqrt{\mathbb{C}} u y \left(2t_1 \left(9 \mathbb{C} y^2 \left(u^2-v^2\right)+7 \left(w^2 y^2-6\right)\right)\right.\right.\\& \left.\left.+9\mathbb{C} y^2 \left(u^2-v^2\right)+w^2 y^2-6\right)\cos (w y) \cos \left(\sqrt{\mathbb{C}} u y\right) \sin \left(\sqrt{\mathbb{C}} v y\right)\right.\\&\left.+\left(2t_1 \left(9\mathbb{C} y^2 \left(u^2+v^2\right)-7 w^2 y^2+42\right)\right.\right.\\&\left.\left.+y^2 \left(9\mathbb{C}\left(u^2+v^2\right)-w^2\right)+6\right) \cos (w y) \sin \left(\sqrt{\mathbb{C}} u y\right) \sin \left(\sqrt{\mathbb{C}} v y\right) \right.\\&\left.-2\mathbb{C} (14t_1+1) u v w y^3 \sin (w y) \cos \left(\sqrt{\mathbb{C}} u y\right) \cos \left(\sqrt{\mathbb{C}} v y\right)\right.\\&\left.+2\sqrt{\mathbb{C}} (14t_1+1) v w y^2 \sin (w y) \sin \left(\sqrt{\mathbb{C}} u y\right) \cos \left(\sqrt{\mathbb{C}} v y\right)\right.\\&\left.+2\sqrt{\mathbb{C}} (14t_1+1) u w y^2 \sin (w y) \cos \left(\sqrt{\mathbb{C}} u y\right) \sin \left(\sqrt{\mathbb{C}} v y\right)\right.\\&\left.+wy(-2-11\mathbb{C}(u^2+v^2)y^2+w^2y^2\right.\\&\left.-2t_1(14+23\mathbb{C}(u^2+v^2)y^2-7 w^2 y^2))\sin (w y) \sin \left(\sqrt{\mathbb{C}}uy\right)\sin\left(\sqrt{\mathbb{C}}vy\right)\right]\\&-
\frac{1}{36\mathbb{C}^3u^3v^3}\left[B_1\left(\text{Ci}\left(A_3y\right)+\text{Ci}\left(A_4y\right)-\text{Ci}\left(|A_2|y\right)-\text{Ci}\left(A_1y\right)\right)\right.\\&\left.+B_2\left(\text{Ci}\left(A_3y\right)-\text{Ci}\left(A_4y\right)+\text{Ci}\left(|A_2|y\right)-\text{Ci}\left(A_1y\right)\right)\right.\\&\left.+B_3\left(\text{Ci}\left(|A_2|y\right)-\text{Ci}\left(A_1y\right)-\text{Ci}\left(A_3y\right)+\text{Ci}\left(A_4y\right)\right)\right],
\end{split}
\end{equation}
where
\begin{equation}
A_1=w+\sqrt{\mathbb{C}}(u+v),\ A_2=w-\sqrt{\mathbb{C}}(u+v), \ A_3=w+\sqrt{\mathbb{C}}(u-v), \ A_4=w-\sqrt{\mathbb{C}}(u-v),
\end{equation}
and
\begin{gather}
B_1=9\mathbb{C}^2 (2t_1+1) \left(u^2-v^2\right)^2+12\mathbb{C}(5t_1+1) w^2 \left(u^2+v^2\right)-(14t_1+1)w^4,\\
B_2=4\mathbb{C}^{3/2}(5+16t_1)u^3w,\ B_3=4\mathbb{C}^{3/2}(5+16t_1)v^3w.
\end{gather}

We also get
\begin{equation}
\begin{split}
\mathcal{I}^A_{\mathrm{PC}1c}(u,v,y)=&\frac{1}{9\mathbb{C}^3u^3v^3y^4}\left[6\mathbb{C}(14t_1+1) u v y^2 \sin (w y) \cos \left(\sqrt{\mathbb{C}} u y\right) \cos \left(\sqrt{\mathbb{C}} v y\right)\right.\\&\left.+ \sqrt{\mathbb{C}}vy\left(-2t_1 \left(9\mathbb{C} y^2 \left(u^2-v^2\right)-7 w^2 y^2+42\right)\right.\right.\\& \left.\left.+9\mathbb{C} y^2 \left(v^2-u^2\right)+w^2 y^2-6\right)\sin (w y) \sin \left(\sqrt{\mathbb{C}} u y\right) \cos \left(\sqrt{\mathbb{C}} v y\right)\right.\\&\left.+\sqrt{\mathbb{C}} u y \left(2t_1 \left(9 \mathbb{C} y^2 \left(u^2-v^2\right)+7 \left(w^2 y^2-6\right)\right)\right.\right.\\& \left.\left.+9\mathbb{C} y^2 \left(u^2-v^2\right)+w^2 y^2-6\right)\sin (w y) \cos \left(\sqrt{\mathbb{C}} u y\right) \sin \left(\sqrt{\mathbb{C}} v y\right)\right.\\&\left.+\left(2t_1 \left(9\mathbb{C} y^2 \left(u^2+v^2\right)-7 w^2 y^2+42\right)\right.\right.\\&\left.\left.+y^2 \left(9\mathbb{C}\left(u^2+v^2\right)-w^2\right)+6\right) \cos (w y) \sin \left(\sqrt{\mathbb{C}} u y\right) \sin \left(\sqrt{\mathbb{C}} v y\right) \right.\\&\left.+2\mathbb{C} (14t_1+1) u v w y^3 \cos (w y) \cos \left(\sqrt{\mathbb{C}} u y\right) \cos \left(\sqrt{\mathbb{C}} v y\right)\right.\\&\left.+2\sqrt{\mathbb{C}} (14t_1+1) v w y^2 \cos (w y) \sin \left(\sqrt{\mathbb{C}} u y\right) \cos \left(\sqrt{\mathbb{C}} v y\right)\right.\\&\left.+2\sqrt{\mathbb{C}} (14t_1+1) u w y^2 \cos (w y) \cos \left(\sqrt{\mathbb{C}} u y\right) \sin \left(\sqrt{\mathbb{C}} v y\right)\right.\\&\left.+wy(2+11\mathbb{C}(u^2+v^2)y^2-w^2y^2\right.\\&\left.+2t_1(14+23\mathbb{C}(u^2+v^2)y^2-7 w^2 y^2))\cos(w y) \sin \left(\sqrt{\mathbb{C}}uy\right)\sin\left(\sqrt{\mathbb{C}}vy\right)\right]\\&-
\frac{1}{36\mathbb{C}^3u^3v^3}\left[B_1\left(\text{Si}\left(A_1y\right)+\text{Si}\left(A_2y\right)-\text{Si}\left(A_3y\right)-\text{Si}\left(A_4y\right)\right)\right.\\&\left.-B_2\left(\text{Si}\left(A_3y\right)+\text{Si}\left(A_2y\right)-\text{Si}\left(A_1y\right)-\text{Si}\left(A_4y\right)\right)\right.\\&\left.+B_3\left(\text{Si}\left(A_1y\right)+\text{Si}\left(A_3y\right)-\text{Ci}\left(A_2y\right)-\text{Ci}\left(A_4y\right)\right)\right].
\end{split}
\end{equation}

We have the following limit:
\begin{equation}
  \mathcal{I}^A_{\mathrm{PC}1s}(u,v,x\rightarrow \infty)=0,  
\end{equation}
and
\begin{equation}
\begin{split}
\mathcal{I}^A_{\mathrm{PC}1s}(u,v,x\rightarrow 0)=&-\frac{9\mathbb{C}(2t_1+1) \left(u^2+v^2\right)+(14t_1+1) w^2}{9\mathbb{C}^2 u^2 v^2}\\&+\frac{1}{36\mathbb{C}^3u^3v^3}\left(B_1\log\left|\frac{A_1A_2}{A_3A_4}\right|-B_2\log\left|\frac{A_2A_3}{A_1A_4}\right|-B_3\log\left|\frac{A_2A_4}{A_1A_3}\right|\right),
\end{split}
\end{equation}
thus
\begin{equation}\label{Ipc1}
\begin{split}
I^A_{\mathrm{PC}1s}(u,v,x\rightarrow \infty)=&\frac{9\mathbb{C}(2t_1+1) \left(u^2+v^2\right)+(14t_1+1) w^2}{9\mathbb{C}^2 u^2 v^2}\\&-\frac{1}{36\mathbb{C}^3u^3v^3}\left(B_1\log\left|\frac{A_1A_2}{A_3A_4}\right|-B_2\log\left|\frac{A_2A_3}{A_1A_4}\right|-B_3\log\left|\frac{A_2A_4}{A_1A_3}\right|\right).
\end{split}
\end{equation}
We also have
\begin{equation}
  \mathcal{I}^A_{\mathrm{PC}1c}(u,v,x\rightarrow \infty)=
\frac{B_1-B_2-B_3}{36\mathbb{C}^3u^3v^3}\pi\Theta(-A_2),  
\end{equation}
and
\begin{equation}
  \mathcal{I}^A_{\mathrm{PC}1c}(u,v,0)=0,  
\end{equation}
thus
\begin{equation}
I^A_{\mathrm{PC}1c}(u,v,x\rightarrow \infty)=
\frac{B_1-B_2-B_3}{36\mathbb{C}^3u^3v^3}\pi\Theta(-A_2).
\end{equation}

As for the PV part, we have the following expressions:
\begin{equation}
\begin{split}
\mathcal{I}^A_{\mathrm{PV}1s}(u,v,y)=&\frac{\lambda^Ak}{M_{\mathrm{PV}}}\frac{\mathcal{C}^2_2}{\mathcal{C}_5}\frac{u+v}{12\mathbb{C}^3 u^3 v^3 y^4}\left[6\mathbb{C}u v y^2 \cos (w y) \cos \left(\sqrt{\mathbb{C}} u y\right) \cos \left(\sqrt{\mathbb{C}} v y\right)\right.\\&\left.-2\mathbb{C} u v w y^3 \sin (w y) \cos \left(\sqrt{\mathbb{C}} u y\right) \cos \left(\sqrt{\mathbb{C}} v y\right)\right.\\&\left.+2 \sqrt{\mathbb{C}} v w y^2 \sin (w y) \sin \left(\sqrt{\mathbb{C}} u y\right) \cos \left(\sqrt{\mathbb{C}} v y\right) \right.\\&\left.+2 \sqrt{\mathbb{C}} u w y^2 \sin (w y) \sin \left(\sqrt{\mathbb{C}} v y\right) \cos \left(\sqrt{\mathbb{C}} u y\right)\right.\\&\left.
+\sqrt{\mathbb{C}}v y \left(3\mathbb{C} y^2 \left(v^2-u^2\right)+w^2 y^2-6\right)\cos (w y) \sin \left(\sqrt{\mathbb{C}} u y\right) \cos \left(\sqrt{\mathbb{C}} v y\right)\right.\\&\left.
+\sqrt{\mathbb{C}} u y \left(3\mathbb{C} y^2 \left(u^2-v^2\right)+w^2 y^2-6\right)\cos (w y) \cos \left(\sqrt{\mathbb{C}} u y\right) \sin \left(\sqrt{\mathbb{C}} v y\right)\right.\\&\left. +\left(3\mathbb{C} y^2 \left(u^2+v^2\right)-w^2 y^2+6\right)\cos (w y) \sin \left(\sqrt{\mathbb{C}} u y\right) \sin \left(\sqrt{\mathbb{C}} v y\right)\right.\\&\left.+w y \left(w^2 y^2-5\mathbb{C} y^2 \left(u^2+v^2\right)-2\right)\sin (w y) \sin \left(\sqrt{\mathbb{C}} u y\right) \sin \left(\sqrt{\mathbb{C}} v y\right)\right]\\&-
\frac{\lambda^Ak}{M_{\mathrm{PV}}}\frac{\mathcal{C}^2_2}{\mathcal{C}_5}\frac{u+v}{48\mathbb{C}^3 u^3 v^3}\left[D_1\left(\text{Ci}\left(A_1y\right)+\text{Ci}\left(|A_2|y\right)-\text{Ci}\left(A_3y\right)-\text{Ci}\left(A_4y\right)\right)\right.\\&\left.-D_2\left(\text{Ci}\left(A_3y\right)+\text{Ci}\left(|A_2|y\right)-\text{Ci}\left(A_1y\right)-\text{Ci}\left(A_4y\right)\right)\right.\\&\left.-D_3\left(\text{Ci}\left(|A_2|y\right)+\text{Ci}\left(A_4y\right)-\text{Ci}\left(A_1y\right)-\text{Ci}\left(A_3y\right)\right)\right],
\end{split}
\end{equation}
where
\begin{equation}
D_1=3\mathbb{C}^2\left(u^2-v^2\right)^2+6\mathbb{C}w^2 \left(u^2+v^2\right)-w^4, \ D_2=8\mathbb{C}^{3/2}u^3w,\ D_3=8\mathbb{C}^{3/2}v^3w.
\end{equation}

We have the following limit:
\begin{equation}
\mathcal{I}^A_{\mathrm{PV}1s}(u,v,x\rightarrow \infty)=0,
\end{equation}
and
\begin{equation}
\begin{split}
\mathcal{I}^A_{\mathrm{PV}1s}(u,v,y\rightarrow 0)=&\frac{\lambda^Ak}{M_{\mathrm{PV}}}\frac{\mathcal{C}^2_2}{\mathcal{C}_5}\frac{(u+v) \left(3\mathbb{C} \left(u^2+v^2\right)+w^2\right)}{12\mathbb{C}^2 u^2 v^2}\\&-\frac{\lambda^Ak}{M_{\mathrm{PV}}}\frac{\mathcal{C}^2_2}{\mathcal{C}_5}
\frac{u+v}{48\mathbb{C}^3 u^3 v^3}\left(D_1\log\left|\frac{A_1A_2}{A_3A_4}\right|-D_2\log\left|\frac{A_2A_3}{A_1A_4}\right|-D_3\log\left|\frac{A_2A_4}{A_1A_3}\right|\right),
\end{split}
\end{equation}
thus
\begin{equation}
\begin{split}
I^A_{\mathrm{PV}1s}(u,v,x\rightarrow \infty)=&-\frac{\lambda^Ak}{M_{\mathrm{PV}}}\frac{\mathcal{C}^2_2}{\mathcal{C}_5}\frac{(u+v) \left(3\mathbb{C} \left(u^2+v^2\right)+w^2\right)}{12\mathbb{C}^2 u^2 v^2}\\&+\frac{\lambda^Ak}{M_{\mathrm{PV}}}\frac{\mathcal{C}^2_2}{\mathcal{C}_5}
\frac{u+v}{48\mathbb{C}^3 u^3 v^3}\left(D_1\log\left|\frac{A_1A_2}{A_3A_4}\right|-D_2\log\left|\frac{A_2A_3}{A_1A_4}\right|-D_3\log\left|\frac{A_2A_4}{A_1A_3}\right|\right).
\end{split}
\end{equation}

We also have
\begin{equation}
\begin{split}
\mathcal{I}^A_{\mathrm{PV}1c}(u,v,y)=&\frac{\lambda^Ak}{M_{\mathrm{PV}}}\frac{\mathcal{C}^2_2}{\mathcal{C}_5}\frac{u+v}{12\mathbb{C}^3 u^3 v^3 y^4}\left[6\mathbb{C}u v y^2 \cos (w y) \cos \left(\sqrt{\mathbb{C}} u y\right) \cos \left(\sqrt{\mathbb{C}} v y\right)\right.\\&\left.-2\mathbb{C} u v w y^3 \sin (w y) \cos \left(\sqrt{\mathbb{C}} u y\right) \cos \left(\sqrt{\mathbb{C}} v y\right)\right.\\&\left.+2 \sqrt{\mathbb{C}} v w y^2 \sin (w y) \sin \left(\sqrt{\mathbb{C}} u y\right) \cos \left(\sqrt{\mathbb{C}} v y\right) \right.\\&\left.+2 \sqrt{\mathbb{C}} u w y^2 \sin (w y) \sin \left(\sqrt{\mathbb{C}} v y\right) \cos \left(\sqrt{\mathbb{C}} u y\right)\right.\\&\left.
+\sqrt{\mathbb{C}}v y \left(3\mathbb{C} y^2 \left(v^2-u^2\right)+w^2 y^2-6\right)\cos (w y) \sin \left(\sqrt{\mathbb{C}} u y\right) \cos \left(\sqrt{\mathbb{C}} v y\right)\right.\\&\left.
+\sqrt{\mathbb{C}} u y \left(3\mathbb{C} y^2 \left(u^2-v^2\right)+w^2 y^2-6\right)\cos (w y) \cos \left(\sqrt{\mathbb{C}} u y\right) \sin \left(\sqrt{\mathbb{C}} v y\right)\right.\\&\left. +\left(3\mathbb{C} y^2 \left(u^2+v^2\right)-w^2 y^2+6\right)\cos (w y) \sin \left(\sqrt{\mathbb{C}} u y\right) \sin \left(\sqrt{\mathbb{C}} v y\right)\right.\\&\left.+w y \left(w^2 y^2-5\mathbb{C} y^2 \left(u^2+v^2\right)-2\right)\sin (w y) \sin \left(\sqrt{\mathbb{C}} u y\right) \sin \left(\sqrt{\mathbb{C}} v y\right)\right]\\&+
\frac{\lambda^Ak}{M_{\mathrm{PV}}}\frac{\mathcal{C}^2_2}{\mathcal{C}_5}\frac{u+v}{48\mathbb{C}^3 u^3 v^3}\left[D_1\left(\text{Si}\left(A_1y\right)+\text{Si}\left(A_2y\right)-\text{Si}\left(A_3y\right)-\text{Si}\left(A_4y\right)\right)\right.\\&\left.+D_2\left(\text{Si}\left(A_1y\right)-\text{Si}\left(A_2y\right)-\text{Si}\left(A_3y\right)+\text{Si}\left(A_4y\right)\right)\right.\\&\left.+D_3\left(\text{Si}\left(A_1y\right)-\text{Si}\left(A_2y\right)+\text{Si}\left(A_3y\right)-\text{Si}\left(A_4y\right)\right)\right].
\end{split}
\end{equation}

We have the following limit:
\begin{equation}
\mathcal{I}^A_{\mathrm{PV}1c}(u,v,x\rightarrow \infty)=-
\frac{\lambda^Ak}{M_{\mathrm{PV}}}\frac{\mathcal{C}^2_2}{\mathcal{C}_5}\frac{u+v}{48\mathbb{C}^3 u^3 v^3}(D_1-D_2-D_3)\pi\Theta(-A_2),
\end{equation}
and
\begin{equation}
\begin{split}
\mathcal{I}^A_{\mathrm{PV}1c}(u,v,x\rightarrow 0)=0,
\end{split}
\end{equation}
thus
\begin{equation}
\begin{split}
I^A_{\mathrm{PV}1c}(u,v,x\rightarrow \infty)=-
\frac{\lambda^Ak}{M_{\mathrm{PV}}}\frac{\mathcal{C}^2_2}{\mathcal{C}_5}\frac{u+v}{48\mathbb{C}^3 u^3 v^3}(D_1-D_2-D_3)\pi\Theta(-A_2).
\end{split}
\end{equation}
\end{widetext}


\begin{thebibliography}{159}%
\makeatletter
\providecommand \@ifxundefined [1]{%
 \@ifx{#1\undefined}
}%
\providecommand \@ifnum [1]{%
 \ifnum #1\expandafter \@firstoftwo
 \else \expandafter \@secondoftwo
 \fi
}%
\providecommand \@ifx [1]{%
 \ifx #1\expandafter \@firstoftwo
 \else \expandafter \@secondoftwo
 \fi
}%
\providecommand \natexlab [1]{#1}%
\providecommand \enquote  [1]{``#1''}%
\providecommand \bibnamefont  [1]{#1}%
\providecommand \bibfnamefont [1]{#1}%
\providecommand \citenamefont [1]{#1}%
\providecommand \href@noop [0]{\@secondoftwo}%
\providecommand \href [0]{\begingroup \@sanitize@url \@href}%
\providecommand \@href[1]{\@@startlink{#1}\@@href}%
\providecommand \@@href[1]{\endgroup#1\@@endlink}%
\providecommand \@sanitize@url [0]{\catcode `\\12\catcode `\$12\catcode
  `\&12\catcode `\#12\catcode `\^12\catcode `\_12\catcode `\%12\relax}%
\providecommand \@@startlink[1]{}%
\providecommand \@@endlink[0]{}%
\providecommand \url  [0]{\begingroup\@sanitize@url \@url }%
\providecommand \@url [1]{\endgroup\@href {#1}{\urlprefix }}%
\providecommand \urlprefix  [0]{URL }%
\providecommand \Eprint [0]{\href }%
\providecommand \doibase [0]{https://doi.org/}%
\providecommand \selectlanguage [0]{\@gobble}%
\providecommand \bibinfo  [0]{\@secondoftwo}%
\providecommand \bibfield  [0]{\@secondoftwo}%
\providecommand \translation [1]{[#1]}%
\providecommand \BibitemOpen [0]{}%
\providecommand \bibitemStop [0]{}%
\providecommand \bibitemNoStop [0]{.\EOS\space}%
\providecommand \EOS [0]{\spacefactor3000\relax}%
\providecommand \BibitemShut  [1]{\csname bibitem#1\endcsname}%
\let\auto@bib@innerbib\@empty
\bibitem [{\citenamefont {Abbott}\ {\it
  et~al.}(2016{\natexlab{a}})\citenamefont {Abbott} {\it
  et~al.}}]{Abbott:2016nmj}%
  \BibitemOpen
  \bibfield  {author} {\bibinfo {author} {\bibfnamefont {B.~P.}\ \bibnamefont
  {Abbott}} {\it et~al.} (\bibinfo {collaboration} {LIGO Scientific and Virgo
  Collaborations}),\ }\bibinfo {title} {{GW151226: Observation of Gravitational
  Waves from a 22-Solar-Mass Binary Black Hole Coalescence}},\ \href
  {https://doi.org/10.1103/PhysRevLett.116.241103} {\bibfield  {journal} {Phys.
  Rev. Lett.\ }\textbf {\bibinfo {volume} {116}},\ \bibinfo {pages} {241103}
  (\bibinfo {year} {2016}{\natexlab{a}})}\BibitemShut {NoStop}%
\bibitem [{\citenamefont {Abbott}\ {\it
  et~al.}(2016{\natexlab{b}})\citenamefont {Abbott} {\it
  et~al.}}]{Abbott:2016blz}%
  \BibitemOpen
  \bibfield  {author} {\bibinfo {author} {\bibfnamefont {B.~P.}\ \bibnamefont
  {Abbott}} {\it et~al.} (\bibinfo {collaboration} {LIGO Scientific and Virgo
  Collaborations}),\ }\bibinfo {title} {{Observation of Gravitational Waves
  from a Binary Black Hole Merger}},\ \href
  {https://doi.org/10.1103/PhysRevLett.116.061102} {\bibfield  {journal} {Phys.
  Rev. Lett.\ }\textbf {\bibinfo {volume} {116}},\ \bibinfo {pages} {061102}
  (\bibinfo {year} {2016}{\natexlab{b}})}\BibitemShut {NoStop}%
\bibitem [{\citenamefont {Abbott}\ {\it
  et~al.}(2017{\natexlab{a}})\citenamefont {Abbott} {\it
  et~al.}}]{Abbott:2017gyy}%
  \BibitemOpen
  \bibfield  {author} {\bibinfo {author} {\bibfnamefont {B.~P.}\ \bibnamefont
  {Abbott}} {\it et~al.} (\bibinfo {collaboration} {LIGO Scientific and Virgo
  Collaborations}),\ }\bibinfo {title} {{GW170608: Observation of a
  19-solar-mass Binary Black Hole Coalescence}},\ \href
  {https://doi.org/10.3847/2041-8213/aa9f0c} {\bibfield  {journal} {Astrophys.
  J. Lett.\ }\textbf {\bibinfo {volume} {851}},\ \bibinfo {pages} {L35}
  (\bibinfo {year} {2017}{\natexlab{a}})}\BibitemShut {NoStop}%
\bibitem [{\citenamefont {Abbott}\ {\it
  et~al.}(2017{\natexlab{b}})\citenamefont {Abbott} {\it
  et~al.}}]{TheLIGOScientific:2017qsa}%
  \BibitemOpen
  \bibfield  {author} {\bibinfo {author} {\bibfnamefont {B.~P.}\ \bibnamefont
  {Abbott}} {\it et~al.} (\bibinfo {collaboration} {LIGO Scientific and Virgo
  Collaborations}),\ }\bibinfo {title} {{GW170817: Observation of Gravitational
  Waves from a Binary Neutron Star Inspiral}},\ \href
  {https://doi.org/10.1103/PhysRevLett.119.161101} {\bibfield  {journal} {Phys.
  Rev. Lett.\ }\textbf {\bibinfo {volume} {119}},\ \bibinfo {pages} {161101}
  (\bibinfo {year} {2017}{\natexlab{b}})}\BibitemShut {NoStop}%
\bibitem [{\citenamefont {Abbott}\ {\it
  et~al.}(2017{\natexlab{c}})\citenamefont {Abbott} {\it
  et~al.}}]{Abbott:2017oio}%
  \BibitemOpen
  \bibfield  {author} {\bibinfo {author} {\bibfnamefont {B.~P.}\ \bibnamefont
  {Abbott}} {\it et~al.} (\bibinfo {collaboration} {LIGO Scientific and Virgo
  Collaborations}),\ }\bibinfo {title} {{GW170814: A Three-Detector Observation
  of Gravitational Waves from a Binary Black Hole Coalescence}},\ \href
  {https://doi.org/10.1103/PhysRevLett.119.141101} {\bibfield  {journal} {Phys.
  Rev. Lett.\ }\textbf {\bibinfo {volume} {119}},\ \bibinfo {pages} {141101}
  (\bibinfo {year} {2017}{\natexlab{c}})}\BibitemShut {NoStop}%
\bibitem [{\citenamefont {Abbott}\ {\it
  et~al.}(2017{\natexlab{d}})\citenamefont {Abbott} {\it
  et~al.}}]{Abbott:2017vtc}%
  \BibitemOpen
  \bibfield  {author} {\bibinfo {author} {\bibfnamefont {B.~P.}\ \bibnamefont
  {Abbott}} {\it et~al.} (\bibinfo {collaboration} {LIGO Scientific and Virgo
  Collaborations}),\ }\bibinfo {title} {{GW170104: Observation of a
  50-Solar-Mass Binary Black Hole Coalescence at Redshift 0.2}},\ \href
  {https://doi.org/10.1103/PhysRevLett.118.221101} {\bibfield  {journal} {Phys.
  Rev. Lett.\ }\textbf {\bibinfo {volume} {118}},\ \bibinfo {pages} {221101}
  (\bibinfo {year} {2017}{\natexlab{d}})},\ \bibinfo {note} {[Erratum:
  Phys.Rev.Lett. 121, 129901 (2018)]}\BibitemShut {NoStop}%
\bibitem [{\citenamefont {Abbott}\ {\it et~al.}(2019)\citenamefont {Abbott}
  {\it et~al.}}]{LIGOScientific:2018mvr}%
  \BibitemOpen
  \bibfield  {author} {\bibinfo {author} {\bibfnamefont {B.~P.}\ \bibnamefont
  {Abbott}} {\it et~al.} (\bibinfo {collaboration} {LIGO Scientific and Virgo
  Collaborations}),\ }\bibinfo {title} {{GWTC-1: A Gravitational-Wave Transient
  Catalog of Compact Binary Mergers Observed by LIGO and Virgo during the First
  and Second Observing Runs}},\ \href
  {https://doi.org/10.1103/PhysRevX.9.031040} {\bibfield  {journal} {Phys. Rev.
  X\ }\textbf {\bibinfo {volume} {9}},\ \bibinfo {pages} {031040} (\bibinfo
  {year} {2019})}\BibitemShut {NoStop}%
\bibitem [{\citenamefont {Abbott}\ {\it
  et~al.}(2020{\natexlab{a}})\citenamefont {Abbott} {\it
  et~al.}}]{Abbott:2020khf}%
  \BibitemOpen
  \bibfield  {author} {\bibinfo {author} {\bibfnamefont {R.}~\bibnamefont
  {Abbott}} {\it et~al.} (\bibinfo {collaboration} {LIGO Scientific and Virgo
  Collaborations}),\ }\bibinfo {title} {{GW190814: Gravitational Waves from the
  Coalescence of a 23 Solar Mass Black Hole with a 2.6 Solar Mass Compact
  Object}},\ \href {https://doi.org/10.3847/2041-8213/ab960f} {\bibfield
  {journal} {Astrophys. J. Lett.\ }\textbf {\bibinfo {volume} {896}},\ \bibinfo
  {pages} {L44} (\bibinfo {year} {2020}{\natexlab{a}})}\BibitemShut {NoStop}%
\bibitem [{\citenamefont {Abbott}\ {\it
  et~al.}(2020{\natexlab{b}})\citenamefont {Abbott} {\it
  et~al.}}]{Abbott:2020uma}%
  \BibitemOpen
  \bibfield  {author} {\bibinfo {author} {\bibfnamefont {B.~P.}\ \bibnamefont
  {Abbott}} {\it et~al.} (\bibinfo {collaboration} {LIGO Scientific and Virgo
  Collaborations}),\ }\bibinfo {title} {{GW190425: Observation of a Compact
  Binary Coalescence with Total Mass $\sim 3.4 M_{\odot}$}},\ \href
  {https://doi.org/10.3847/2041-8213/ab75f5} {\bibfield  {journal} {Astrophys.
  J. Lett.\ }\textbf {\bibinfo {volume} {892}},\ \bibinfo {pages} {L3}
  (\bibinfo {year} {2020}{\natexlab{b}})}\BibitemShut {NoStop}%
\bibitem [{\citenamefont {Abbott}\ {\it
  et~al.}(2020{\natexlab{c}})\citenamefont {Abbott} {\it
  et~al.}}]{LIGOScientific:2020stg}%
  \BibitemOpen
  \bibfield  {author} {\bibinfo {author} {\bibfnamefont {R.}~\bibnamefont
  {Abbott}} {\it et~al.} (\bibinfo {collaboration} {LIGO Scientific and Virgo
  Collaborations}),\ }\bibinfo {title} {{GW190412: Observation of a
  Binary-Black-Hole Coalescence with Asymmetric Masses}},\ \href
  {https://doi.org/10.1103/PhysRevD.102.043015} {\bibfield  {journal} {Phys.
  Rev. D\ }\textbf {\bibinfo {volume} {102}},\ \bibinfo {pages} {043015}
  (\bibinfo {year} {2020}{\natexlab{c}})}\BibitemShut {NoStop}%
\bibitem [{\citenamefont {Ananda}\ {\it et~al.}(2007)\citenamefont {Ananda},
  \citenamefont {Clarkson},\ and\ \citenamefont {Wands}}]{Ananda:2006af}%
  \BibitemOpen
  \bibfield  {author} {\bibinfo {author} {\bibfnamefont {K.~N.}\ \bibnamefont
  {Ananda}}, \bibinfo {author} {\bibfnamefont {C.}~\bibnamefont {Clarkson}},\
  and\ \bibinfo {author} {\bibfnamefont {D.}~\bibnamefont {Wands}},\ }\bibinfo
  {title} {{The Cosmological gravitational wave background from primordial
  density perturbations}},\ \href {https://doi.org/10.1103/PhysRevD.75.123518}
  {\bibfield  {journal} {Phys. Rev. D\ }\textbf {\bibinfo {volume} {75}},\
  \bibinfo {pages} {123518} (\bibinfo {year} {2007})}\BibitemShut {NoStop}%
\bibitem [{\citenamefont {Saito}\ and\ \citenamefont
  {Yokoyama}(2009)}]{Saito:2008jc}%
  \BibitemOpen
  \bibfield  {author} {\bibinfo {author} {\bibfnamefont {R.}~\bibnamefont
  {Saito}}\ and\ \bibinfo {author} {\bibfnamefont {J.}~\bibnamefont
  {Yokoyama}},\ }\bibinfo {title} {{Gravitational wave background as a probe of
  the primordial black hole abundance}},\ \href
  {https://doi.org/10.1103/PhysRevLett.102.161101} {\bibfield  {journal} {Phys.
  Rev. Lett.\ }\textbf {\bibinfo {volume} {102}},\ \bibinfo {pages} {161101}
  (\bibinfo {year} {2009})},\ \bibinfo {note} {[Erratum: Phys.Rev.Lett. 107,
  069901 (2011)]}\BibitemShut {NoStop}%
\bibitem [{\citenamefont {Nakama}\ {\it et~al.}(2017)\citenamefont {Nakama},
  \citenamefont {Silk},\ and\ \citenamefont {Kamionkowski}}]{Nakama:2016gzw}%
  \BibitemOpen
  \bibfield  {author} {\bibinfo {author} {\bibfnamefont {T.}~\bibnamefont
  {Nakama}}, \bibinfo {author} {\bibfnamefont {J.}~\bibnamefont {Silk}},\ and\
  \bibinfo {author} {\bibfnamefont {M.}~\bibnamefont {Kamionkowski}},\
  }\bibinfo {title} {{Stochastic gravitational waves associated with the
  formation of primordial black holes}},\ \href
  {https://doi.org/10.1103/PhysRevD.95.043511} {\bibfield  {journal} {Phys.
  Rev. D\ }\textbf {\bibinfo {volume} {95}},\ \bibinfo {pages} {043511}
  (\bibinfo {year} {2017})}\BibitemShut {NoStop}%
\bibitem [{\citenamefont {Wang}\ {\it et~al.}(2018)\citenamefont {Wang},
  \citenamefont {Wang}, \citenamefont {Huang},\ and\ \citenamefont
  {Li}}]{Wang:2016ana}%
  \BibitemOpen
  \bibfield  {author} {\bibinfo {author} {\bibfnamefont {S.}~\bibnamefont
  {Wang}}, \bibinfo {author} {\bibfnamefont {Y.-F.}\ \bibnamefont {Wang}},
  \bibinfo {author} {\bibfnamefont {Q.-G.}\ \bibnamefont {Huang}},\ and\
  \bibinfo {author} {\bibfnamefont {T.~G.~F.}\ \bibnamefont {Li}},\ }\bibinfo
  {title} {{Constraints on the Primordial Black Hole Abundance from the First
  Advanced LIGO Observation Run Using the Stochastic Gravitational-Wave
  Background}},\ \href {https://doi.org/10.1103/PhysRevLett.120.191102}
  {\bibfield  {journal} {Phys. Rev. Lett.\ }\textbf {\bibinfo {volume} {120}},\
  \bibinfo {pages} {191102} (\bibinfo {year} {2018})}\BibitemShut {NoStop}%
\bibitem [{\citenamefont {Kohri}\ and\ \citenamefont
  {Terada}(2018)}]{Kohri:2018awv}%
  \BibitemOpen
  \bibfield  {author} {\bibinfo {author} {\bibfnamefont {K.}~\bibnamefont
  {Kohri}}\ and\ \bibinfo {author} {\bibfnamefont {T.}~\bibnamefont {Terada}},\
  }\bibinfo {title} {{Semianalytic calculation of gravitational wave spectrum
  nonlinearly induced from primordial curvature perturbations}},\ \href
  {https://doi.org/10.1103/PhysRevD.97.123532} {\bibfield  {journal} {Phys.
  Rev. D\ }\textbf {\bibinfo {volume} {97}},\ \bibinfo {pages} {123532}
  (\bibinfo {year} {2018})}\BibitemShut {NoStop}%
\bibitem [{\citenamefont {Espinosa}\ {\it et~al.}(2018)\citenamefont
  {Espinosa}, \citenamefont {Racco},\ and\ \citenamefont
  {Riotto}}]{Espinosa:2018eve}%
  \BibitemOpen
  \bibfield  {author} {\bibinfo {author} {\bibfnamefont {J.~R.}\ \bibnamefont
  {Espinosa}}, \bibinfo {author} {\bibfnamefont {D.}~\bibnamefont {Racco}},\
  and\ \bibinfo {author} {\bibfnamefont {A.}~\bibnamefont {Riotto}},\ }\bibinfo
  {title} {{A Cosmological Signature of the SM Higgs Instability: Gravitational
  Waves}},\ \href {https://doi.org/10.1088/1475-7516/2018/09/012} {J. Cosmol.
  Astropart. Phys.\ \bibinfo {volume} {09}\bibfield  {year} {\bibinfo  {year} {
  (\textbf {2018})}\ }\bibinfo  {pages} {012}}\BibitemShut {NoStop}%
\bibitem [{\citenamefont {Kuroyanagi}\ {\it et~al.}(2018)\citenamefont
  {Kuroyanagi}, \citenamefont {Chiba},\ and\ \citenamefont
  {Takahashi}}]{Kuroyanagi:2018csn}%
  \BibitemOpen
\bibfield  {pages} {  }\bibfield  {author} {\bibinfo {author} {\bibfnamefont
  {S.}~\bibnamefont {Kuroyanagi}}, \bibinfo {author} {\bibfnamefont
  {T.}~\bibnamefont {Chiba}},\ and\ \bibinfo {author} {\bibfnamefont
  {T.}~\bibnamefont {Takahashi}},\ }\bibinfo {title} {{Probing the Universe
  through the Stochastic Gravitational Wave Background}},\ \href
  {https://doi.org/10.1088/1475-7516/2018/11/038} {J. Cosmol. Astropart. Phys.\
  \bibinfo {volume} {11}\bibfield  {year} {\bibinfo  {year} { (\textbf
  {2018})}\ }\bibinfo  {pages} {038}}\BibitemShut {NoStop}%
\bibitem [{\citenamefont {Fumagalli}\ {\it et~al.}(2021)\citenamefont
  {Fumagalli}, \citenamefont {Renaux-Petel},\ and\ \citenamefont
  {Witkowski}}]{Fumagalli:2020nvq}%
  \BibitemOpen
\bibfield  {pages} {  }\bibfield  {author} {\bibinfo {author} {\bibfnamefont
  {J.}~\bibnamefont {Fumagalli}}, \bibinfo {author} {\bibfnamefont
  {S.}~\bibnamefont {Renaux-Petel}},\ and\ \bibinfo {author} {\bibfnamefont
  {L.~T.}\ \bibnamefont {Witkowski}},\ }\bibinfo {title} {{Oscillations in the
  stochastic gravitational wave background from sharp features and particle
  production during inflation}},\ \href
  {https://doi.org/10.1088/1475-7516/2021/08/030} {J. Cosmol. Astropart. Phys.\
  \bibinfo {volume} {08}\bibfield  {year} {\bibinfo  {year} { (\textbf
  {2021})}\ }\bibinfo  {pages} {030}}\BibitemShut {NoStop}%
\bibitem [{\citenamefont {Braglia}\ {\it et~al.}(2021)\citenamefont {Braglia},
  \citenamefont {Chen},\ and\ \citenamefont {Hazra}}]{Braglia:2020taf}%
  \BibitemOpen
\bibfield  {pages} {  }\bibfield  {author} {\bibinfo {author} {\bibfnamefont
  {M.}~\bibnamefont {Braglia}}, \bibinfo {author} {\bibfnamefont
  {X.}~\bibnamefont {Chen}},\ and\ \bibinfo {author} {\bibfnamefont {D.~K.}\
  \bibnamefont {Hazra}},\ }\bibinfo {title} {{Probing Primordial Features with
  the Stochastic Gravitational Wave Background}},\ \href
  {https://doi.org/10.1088/1475-7516/2021/03/005} {J. Cosmol. Astropart. Phys.\
  \bibinfo {volume} {03}\bibfield  {year} {\bibinfo  {year} { (\textbf
  {2021})}\ }\bibinfo  {pages} {005}}\BibitemShut {NoStop}%
\bibitem [{\citenamefont {Lin}\ {\it et~al.}(2020)\citenamefont {Lin},
  \citenamefont {Gao}, \citenamefont {Gong}, \citenamefont {Lu}, \citenamefont
  {Zhang},\ and\ \citenamefont {Zhang}}]{Lin:2020goi}%
  \BibitemOpen
\bibfield  {pages} {  }\bibfield  {author} {\bibinfo {author} {\bibfnamefont
  {J.}~\bibnamefont {Lin}}, \bibinfo {author} {\bibfnamefont {Q.}~\bibnamefont
  {Gao}}, \bibinfo {author} {\bibfnamefont {Y.}~\bibnamefont {Gong}}, \bibinfo
  {author} {\bibfnamefont {Y.}~\bibnamefont {Lu}}, \bibinfo {author}
  {\bibfnamefont {C.}~\bibnamefont {Zhang}},\ and\ \bibinfo {author}
  {\bibfnamefont {F.}~\bibnamefont {Zhang}},\ }\bibinfo {title} {{Primordial
  black holes and secondary gravitational waves from $k$ and $G$ inflation}},\
  \href {https://doi.org/10.1103/PhysRevD.101.103515} {\bibfield  {journal}
  {Phys. Rev. D\ }\textbf {\bibinfo {volume} {101}},\ \bibinfo {pages} {103515}
  (\bibinfo {year} {2020})}\BibitemShut {NoStop}%
\bibitem [{\citenamefont {Lu}\ {\it et~al.}(2020)\citenamefont {Lu},
  \citenamefont {Ali}, \citenamefont {Gong}, \citenamefont {Lin},\ and\
  \citenamefont {Zhang}}]{Lu:2020diy}%
  \BibitemOpen
  \bibfield  {author} {\bibinfo {author} {\bibfnamefont {Y.}~\bibnamefont
  {Lu}}, \bibinfo {author} {\bibfnamefont {A.}~\bibnamefont {Ali}}, \bibinfo
  {author} {\bibfnamefont {Y.}~\bibnamefont {Gong}}, \bibinfo {author}
  {\bibfnamefont {J.}~\bibnamefont {Lin}},\ and\ \bibinfo {author}
  {\bibfnamefont {F.}~\bibnamefont {Zhang}},\ }\bibinfo {title} {{Gauge
  transformation of scalar induced gravitational waves}},\ \href
  {https://doi.org/10.1103/PhysRevD.102.083503(2020)} {\bibfield  {journal}
  {Phys. Rev. D\ }\textbf {\bibinfo {volume} {102}},\ \bibinfo {pages} {083503}
  (\bibinfo {year} {2020})}\BibitemShut {NoStop}%
\bibitem [{\citenamefont {Dom\`enech}(2021)}]{Domenech:2021ztg}%
  \BibitemOpen
  \bibfield  {author} {\bibinfo {author} {\bibfnamefont {G.}~\bibnamefont
  {Dom\`enech}},\ }\bibinfo {title} {{Scalar Induced Gravitational Waves
  Review}},\ \href {https://doi.org/10.3390/universe7110398} {\bibfield
  {journal} {Universe\ }\textbf {\bibinfo {volume} {7}},\ \bibinfo {pages}
  {398} (\bibinfo {year} {2021})}\BibitemShut {NoStop}%
\bibitem [{\citenamefont {Zhang}\ {\it et~al.}(2021)\citenamefont {Zhang},
  \citenamefont {Lin},\ and\ \citenamefont {Lu}}]{Zhang:2021vak}%
  \BibitemOpen
  \bibfield  {author} {\bibinfo {author} {\bibfnamefont {F.}~\bibnamefont
  {Zhang}}, \bibinfo {author} {\bibfnamefont {J.}~\bibnamefont {Lin}},\ and\
  \bibinfo {author} {\bibfnamefont {Y.}~\bibnamefont {Lu}},\ }\bibinfo {title}
  {{Double-peaked inflation model: Scalar induced gravitational waves and
  primordial-black-hole suppression from primordial non-Gaussianity}},\ \href
  {https://doi.org/10.1103/PhysRevD.104.063515} {\bibfield  {journal} {Phys.
  Rev. D\ }\textbf {\bibinfo {volume} {104}},\ \bibinfo {pages} {063515}
  (\bibinfo {year} {2021})},\ \bibinfo {note} {[Erratum: Phys.Rev.D 104, 129902
  (2021)]}\BibitemShut {NoStop}%
\bibitem [{\citenamefont {Zhang}(2022)}]{Zhang:2021rqs}%
  \BibitemOpen
  \bibfield  {author} {\bibinfo {author} {\bibfnamefont {F.}~\bibnamefont
  {Zhang}},\ }\bibinfo {title} {{Primordial black holes and scalar induced
  gravitational waves from the E model with a Gauss-Bonnet term}},\ \href
  {https://doi.org/10.1103/PhysRevD.105.063539} {\bibfield  {journal} {Phys.
  Rev. D\ }\textbf {\bibinfo {volume} {105}},\ \bibinfo {pages} {063539}
  (\bibinfo {year} {2022})}\BibitemShut {NoStop}%
\bibitem [{\citenamefont {Papanikolaou}\ {\it et~al.}(2022)\citenamefont
  {Papanikolaou}, \citenamefont {Tzerefos}, \citenamefont {Basilakos},\ and\
  \citenamefont {Saridakis}}]{Papanikolaou:2021uhe}%
  \BibitemOpen
  \bibfield  {author} {\bibinfo {author} {\bibfnamefont {T.}~\bibnamefont
  {Papanikolaou}}, \bibinfo {author} {\bibfnamefont {C.}~\bibnamefont
  {Tzerefos}}, \bibinfo {author} {\bibfnamefont {S.}~\bibnamefont
  {Basilakos}},\ and\ \bibinfo {author} {\bibfnamefont {E.~N.}\ \bibnamefont
  {Saridakis}},\ }\bibinfo {title} {{Scalar induced gravitational waves from
  primordial black hole Poisson fluctuations in f(R) gravity}},\ \href
  {https://doi.org/10.1088/1475-7516/2022/10/013} {J. Cosmol. Astropart. Phys.\
  \bibinfo {volume} {10}\bibfield  {year} {\bibinfo  {year} { (\textbf
  {2022})}\ }\bibinfo  {pages} {013}}\BibitemShut {NoStop}%
\bibitem [{\citenamefont {Yi}\ and\ \citenamefont {Fei}(2023)}]{Yi:2022ymw}%
  \BibitemOpen
\bibfield  {pages} {  }\bibfield  {author} {\bibinfo {author} {\bibfnamefont
  {Z.}~\bibnamefont {Yi}}\ and\ \bibinfo {author} {\bibfnamefont
  {Q.}~\bibnamefont {Fei}},\ }\bibinfo {title} {{Constraints on primordial
  curvature spectrum from primordial black holes and scalar-induced
  gravitational waves}},\ \href
  {https://doi.org/10.1140/epjc/s10052-023-11233-3} {\bibfield  {journal} {Eur.
  Phys. J. C\ }\textbf {\bibinfo {volume} {83}},\ \bibinfo {pages} {82}
  (\bibinfo {year} {2023})}\BibitemShut {NoStop}%
\bibitem [{\citenamefont {You}\ {\it et~al.}(2023)\citenamefont {You},
  \citenamefont {Yi},\ and\ \citenamefont {Wu}}]{You:2023rmn}%
  \BibitemOpen
  \bibfield  {author} {\bibinfo {author} {\bibfnamefont {Z.-Q.}\ \bibnamefont
  {You}}, \bibinfo {author} {\bibfnamefont {Z.}~\bibnamefont {Yi}},\ and\
  \bibinfo {author} {\bibfnamefont {Y.}~\bibnamefont {Wu}},\ }\bibinfo {title}
  {{Constraints on primordial curvature power spectrum with pulsar timing
  arrays}},\ \href {https://doi.org/10.1088/1475-7516/2023/11/065} {J. Cosmol.
  Astropart. Phys.\ \bibinfo {volume} {11}\bibfield  {year} {\bibinfo  {year} {
  (\textbf {2023})}\ }\bibinfo  {pages} {065}}\BibitemShut {NoStop}%
\bibitem [{\citenamefont {Lu}\ {\it et~al.}(2019)\citenamefont {Lu},
  \citenamefont {Gong}, \citenamefont {Yi},\ and\ \citenamefont
  {Zhang}}]{Lu:2019sti}%
  \BibitemOpen
\bibfield  {pages} {  }\bibfield  {author} {\bibinfo {author} {\bibfnamefont
  {Y.}~\bibnamefont {Lu}}, \bibinfo {author} {\bibfnamefont {Y.}~\bibnamefont
  {Gong}}, \bibinfo {author} {\bibfnamefont {Z.}~\bibnamefont {Yi}},\ and\
  \bibinfo {author} {\bibfnamefont {F.}~\bibnamefont {Zhang}},\ }\bibinfo
  {title} {{Constraints on primordial curvature perturbations from primordial
  black hole dark matter and secondary gravitational waves}},\ \href
  {https://doi.org/10.1088/1475-7516/2019/12/031} {J. Cosmol. Astropart. Phys.\
  \bibinfo {volume} {12}\bibfield  {year} {\bibinfo  {year} { (\textbf
  {2019})}\ }\bibinfo  {pages} {031}}\BibitemShut {NoStop}%
\bibitem [{\citenamefont {Gu}\ {\it et~al.}(2023)\citenamefont {Gu},
  \citenamefont {Shu},\ and\ \citenamefont {Yang}}]{Gu:2023mmd}%
  \BibitemOpen
\bibfield  {pages} {  }\bibfield  {author} {\bibinfo {author} {\bibfnamefont
  {B.-M.}\ \bibnamefont {Gu}}, \bibinfo {author} {\bibfnamefont {F.-W.}\
  \bibnamefont {Shu}},\ and\ \bibinfo {author} {\bibfnamefont {K.}~\bibnamefont
  {Yang}},\ }\bibinfo {title} {{Inflation with shallow dip and primordial black
  holes}},\ \Eprint {https://arxiv.org/abs/2307.00510} {arXiv:2307.00510}
  \BibitemShut {NoStop}%
\bibitem [{\citenamefont {Choudhury}\ {\it et~al.}(2023)\citenamefont
  {Choudhury}, \citenamefont {Karde}, \citenamefont {Panda},\ and\
  \citenamefont {Sami}}]{Choudhury:2023hfm}%
  \BibitemOpen
  \bibfield  {author} {\bibinfo {author} {\bibfnamefont {S.}~\bibnamefont
  {Choudhury}}, \bibinfo {author} {\bibfnamefont {A.}~\bibnamefont {Karde}},
  \bibinfo {author} {\bibfnamefont {S.}~\bibnamefont {Panda}},\ and\ \bibinfo
  {author} {\bibfnamefont {M.}~\bibnamefont {Sami}},\ }\bibinfo {title}
  {{Scalar induced gravity waves from ultra slow-roll Galileon inflation}},\
  \Eprint {https://arxiv.org/abs/2308.09273} {arXiv:2308.09273} \BibitemShut
  {NoStop}%
\bibitem [{\citenamefont {Wang}\ {\it et~al.}(2024)\citenamefont {Wang},
  \citenamefont {Gao}, \citenamefont {Gong},\ and\ \citenamefont
  {Wang}}]{Wang:2024euw}%
  \BibitemOpen
  \bibfield  {author} {\bibinfo {author} {\bibfnamefont {Z.}~\bibnamefont
  {Wang}}, \bibinfo {author} {\bibfnamefont {S.}~\bibnamefont {Gao}}, \bibinfo
  {author} {\bibfnamefont {Y.}~\bibnamefont {Gong}},\ and\ \bibinfo {author}
  {\bibfnamefont {Y.}~\bibnamefont {Wang}},\ }\bibinfo {title} {{Primordial
  black holes and scalar-induced gravitational waves from the polynomial
  attractor model}},\ \href {https://doi.org/10.1103/PhysRevD.109.103532}
  {\bibfield  {journal} {Phys. Rev. D\ }\textbf {\bibinfo {volume} {109}},\
  \bibinfo {pages} {103532} (\bibinfo {year} {2024})}\BibitemShut {NoStop}%
\bibitem [{\citenamefont {Choudhury}\ {\it et~al.}(2024)\citenamefont
  {Choudhury}, \citenamefont {Karde}, \citenamefont {Panda},\ and\
  \citenamefont {Sami}}]{Choudhury:2024one}%
  \BibitemOpen
  \bibfield  {author} {\bibinfo {author} {\bibfnamefont {S.}~\bibnamefont
  {Choudhury}}, \bibinfo {author} {\bibfnamefont {A.}~\bibnamefont {Karde}},
  \bibinfo {author} {\bibfnamefont {S.}~\bibnamefont {Panda}},\ and\ \bibinfo
  {author} {\bibfnamefont {M.}~\bibnamefont {Sami}},\ }\bibinfo {title}
  {{Realisation of the ultra-slow roll phase in Galileon inflation and PBH
  overproduction}},\ \Eprint {https://arxiv.org/abs/2401.10925}
  {arXiv:2401.10925} \BibitemShut {NoStop}%
\bibitem [{\citenamefont {Danzmann}(1997)}]{Danzmann:1997hm}%
  \BibitemOpen
  \bibfield  {author} {\bibinfo {author} {\bibfnamefont {K.}~\bibnamefont
  {Danzmann}},\ }\bibinfo {title} {{LISA: An ESA cornerstone mission for a
  gravitational wave observatory}},\ \href
  {https://doi.org/10.1088/0264-9381/14/6/002} {\bibfield  {journal} {Classical
  Quantum Gravity\ }\textbf {\bibinfo {volume} {14}},\ \bibinfo {pages} {1399}
  (\bibinfo {year} {1997})}\BibitemShut {NoStop}%
\bibitem [{\citenamefont {Amaro-Seoane}\ {\it et~al.}(2017)\citenamefont
  {Amaro-Seoane} {\it et~al.}}]{LISA:2017pwj}%
  \BibitemOpen
  \bibfield  {author} {\bibinfo {author} {\bibfnamefont {P.}~\bibnamefont
  {Amaro-Seoane}} {\it et~al.} (\bibinfo {collaboration} {LISA}),\ }\bibinfo
  {title} {{Laser Interferometer Space Antenna}},\ \Eprint
  {https://arxiv.org/abs/1702.00786} {arXiv:1702.00786} \BibitemShut {NoStop}%
\bibitem [{\citenamefont {Hu}\ and\ \citenamefont {Wu}(2017)}]{Hu:2017mde}%
  \BibitemOpen
  \bibfield  {author} {\bibinfo {author} {\bibfnamefont {W.-R.}\ \bibnamefont
  {Hu}}\ and\ \bibinfo {author} {\bibfnamefont {Y.-L.}\ \bibnamefont {Wu}},\
  }\bibinfo {title} {{The Taiji Program in Space for gravitational wave physics
  and the nature of gravity}},\ \href {https://doi.org/10.1093/nsr/nwx116}
  {\bibfield  {journal} {Natl. Sci. Rev.\ }\textbf {\bibinfo {volume} {4}},\
  \bibinfo {pages} {685} (\bibinfo {year} {2017})}\BibitemShut {NoStop}%
\bibitem [{\citenamefont {Luo}\ {\it et~al.}(2016)\citenamefont {Luo} {\it
  et~al.}}]{Luo:2015ght}%
  \BibitemOpen
  \bibfield  {author} {\bibinfo {author} {\bibfnamefont {J.}~\bibnamefont
  {Luo}} {\it et~al.} (\bibinfo {collaboration} {TianQin}),\ }\bibinfo {title}
  {{TianQin: a space-borne gravitational wave detector}},\ \href
  {https://doi.org/10.1088/0264-9381/33/3/035010} {\bibfield  {journal}
  {Classical Quantum Gravity\ }\textbf {\bibinfo {volume} {33}},\ \bibinfo
  {pages} {035010} (\bibinfo {year} {2016})}\BibitemShut {NoStop}%
\bibitem [{\citenamefont {Gong}\ {\it et~al.}(2021)\citenamefont {Gong},
  \citenamefont {Luo},\ and\ \citenamefont {Wang}}]{Gong:2021gvw}%
  \BibitemOpen
  \bibfield  {author} {\bibinfo {author} {\bibfnamefont {Y.}~\bibnamefont
  {Gong}}, \bibinfo {author} {\bibfnamefont {J.}~\bibnamefont {Luo}},\ and\
  \bibinfo {author} {\bibfnamefont {B.}~\bibnamefont {Wang}},\ }\bibinfo
  {title} {{Concepts and status of Chinese space gravitational wave detection
  projects}},\ \href {https://doi.org/10.1038/s41550-021-01480-3} {\bibfield
  {journal} {Nat. Astron.\ }\textbf {\bibinfo {volume} {5}},\ \bibinfo {pages}
  {881} (\bibinfo {year} {2021})}\BibitemShut {NoStop}%
\bibitem [{\citenamefont {Kawamura}\ {\it et~al.}(2011)\citenamefont {Kawamura}
  {\it et~al.}}]{Kawamura:2011zz}%
  \BibitemOpen
  \bibfield  {author} {\bibinfo {author} {\bibfnamefont {S.}~\bibnamefont
  {Kawamura}} {\it et~al.},\ }\bibinfo {title} {{The Japanese space
  gravitational wave antenna: DECIGO}},\ \href
  {https://doi.org/10.1088/0264-9381/28/9/094011} {\bibfield  {journal}
  {Classical Quantum Gravity\ }\textbf {\bibinfo {volume} {28}},\ \bibinfo
  {pages} {094011} (\bibinfo {year} {2011})}\BibitemShut {NoStop}%
\bibitem [{\citenamefont {Kramer}\ and\ \citenamefont
  {Champion}(2013)}]{Kramer:2013kea}%
  \BibitemOpen
  \bibfield  {author} {\bibinfo {author} {\bibfnamefont {M.}~\bibnamefont
  {Kramer}}\ and\ \bibinfo {author} {\bibfnamefont {D.~J.}\ \bibnamefont
  {Champion}} (\bibinfo {collaboration} {EPTA}),\ }\bibinfo {title} {{The
  European Pulsar Timing Array and the Large European Array for Pulsars}},\
  \href {https://doi.org/10.1088/0264-9381/30/22/224009} {\bibfield  {journal}
  {Classical Quantum Gravity\ }\textbf {\bibinfo {volume} {30}},\ \bibinfo
  {pages} {224009} (\bibinfo {year} {2013})}\BibitemShut {NoStop}%
\bibitem [{\citenamefont {Hobbs}\ {\it et~al.}(2010)\citenamefont {Hobbs} {\it
  et~al.}}]{Hobbs:2009yy}%
  \BibitemOpen
  \bibfield  {author} {\bibinfo {author} {\bibfnamefont {G.}~\bibnamefont
  {Hobbs}} {\it et~al.},\ }\bibinfo {title} {{The international pulsar timing
  array project: using pulsars as a gravitational wave detector}},\ \href
  {https://doi.org/10.1088/0264-9381/27/8/084013} {\bibfield  {journal}
  {Classical Quantum Gravity\ }\textbf {\bibinfo {volume} {27}},\ \bibinfo
  {pages} {084013} (\bibinfo {year} {2010})}\BibitemShut {NoStop}%
\bibitem [{\citenamefont {McLaughlin}(2013)}]{McLaughlin:2013ira}%
  \BibitemOpen
  \bibfield  {author} {\bibinfo {author} {\bibfnamefont {M.~A.}\ \bibnamefont
  {McLaughlin}},\ }\bibinfo {title} {{The North American Nanohertz Observatory
  for Gravitational Waves}},\ \href
  {https://doi.org/10.1088/0264-9381/30/22/224008} {\bibfield  {journal}
  {Classical Quantum Gravity\ }\textbf {\bibinfo {volume} {30}},\ \bibinfo
  {pages} {224008} (\bibinfo {year} {2013})}\BibitemShut {NoStop}%
\bibitem [{\citenamefont {Hobbs}(2013)}]{Hobbs:2013aka}%
  \BibitemOpen
  \bibfield  {author} {\bibinfo {author} {\bibfnamefont {G.}~\bibnamefont
  {Hobbs}},\ }\bibinfo {title} {{The Parkes Pulsar Timing Array}},\ \href
  {https://doi.org/10.1088/0264-9381/30/22/224007} {\bibfield  {journal}
  {Classical Quantum Gravity\ }\textbf {\bibinfo {volume} {30}},\ \bibinfo
  {pages} {224007} (\bibinfo {year} {2013})}\BibitemShut {NoStop}%
\bibitem [{\citenamefont {Moore}\ {\it et~al.}(2015)\citenamefont {Moore},
  \citenamefont {Cole},\ and\ \citenamefont {Berry}}]{Moore:2014lga}%
  \BibitemOpen
  \bibfield  {author} {\bibinfo {author} {\bibfnamefont {C.~J.}\ \bibnamefont
  {Moore}}, \bibinfo {author} {\bibfnamefont {R.~H.}\ \bibnamefont {Cole}},\
  and\ \bibinfo {author} {\bibfnamefont {C.~P.~L.}\ \bibnamefont {Berry}},\
  }\bibinfo {title} {{Gravitational-wave sensitivity curves}},\ \href
  {https://doi.org/10.1088/0264-9381/32/1/015014} {\bibfield  {journal}
  {Classical Quantum Gravity\ }\textbf {\bibinfo {volume} {32}},\ \bibinfo
  {pages} {015014} (\bibinfo {year} {2015})}\BibitemShut {NoStop}%
\bibitem [{\citenamefont {Agazie}\ {\it
  et~al.}(2023{\natexlab{a}})\citenamefont {Agazie} {\it
  et~al.}}]{NANOGrav:2023gor}%
  \BibitemOpen
  \bibfield  {author} {\bibinfo {author} {\bibfnamefont {G.}~\bibnamefont
  {Agazie}} {\it et~al.} (\bibinfo {collaboration} {NANOGrav}),\ }\bibinfo
  {title} {{The NANOGrav 15 yr Data Set: Evidence for a Gravitational-wave
  Background}},\ \href {https://doi.org/10.3847/2041-8213/acdac6} {\bibfield
  {journal} {Astrophys. J. Lett.\ }\textbf {\bibinfo {volume} {951}},\ \bibinfo
  {pages} {L8} (\bibinfo {year} {2023}{\natexlab{a}})}\BibitemShut {NoStop}%
\bibitem [{\citenamefont {Agazie}\ {\it
  et~al.}(2023{\natexlab{b}})\citenamefont {Agazie} {\it
  et~al.}}]{NANOGrav:2023hde}%
  \BibitemOpen
  \bibfield  {author} {\bibinfo {author} {\bibfnamefont {G.}~\bibnamefont
  {Agazie}} {\it et~al.} (\bibinfo {collaboration} {NANOGrav}),\ }\bibinfo
  {title} {{The NANOGrav 15 yr Data Set: Observations and Timing of 68
  Millisecond Pulsars}},\ \href {https://doi.org/10.3847/2041-8213/acda9a}
  {\bibfield  {journal} {Astrophys. J. Lett.\ }\textbf {\bibinfo {volume}
  {951}},\ \bibinfo {pages} {L9} (\bibinfo {year}
  {2023}{\natexlab{b}})}\BibitemShut {NoStop}%
\bibitem [{\citenamefont {Zic}\ {\it et~al.}(2023)\citenamefont {Zic} {\it
  et~al.}}]{Zic:2023gta}%
  \BibitemOpen
  \bibfield  {author} {\bibinfo {author} {\bibfnamefont {A.}~\bibnamefont
  {Zic}} {\it et~al.},\ }\bibinfo {title} {{The Parkes Pulsar Timing Array
  third data release}},\ \href {https://doi.org/10.1017/pasa.2023.36}
  {\bibfield  {journal} {Publ. Astron. Soc. Austral.\ }\textbf {\bibinfo
  {volume} {40}},\ \bibinfo {pages} {e049} (\bibinfo {year}
  {2023})}\BibitemShut {NoStop}%
\bibitem [{\citenamefont {Reardon}\ {\it et~al.}(2023)\citenamefont {Reardon}
  {\it et~al.}}]{Reardon:2023gzh}%
  \BibitemOpen
  \bibfield  {author} {\bibinfo {author} {\bibfnamefont {D.~J.}\ \bibnamefont
  {Reardon}} {\it et~al.},\ }\bibinfo {title} {{Search for an Isotropic
  Gravitational-wave Background with the Parkes Pulsar Timing Array}},\ \href
  {https://doi.org/10.3847/2041-8213/acdd02} {\bibfield  {journal} {Astrophys.
  J. Lett.\ }\textbf {\bibinfo {volume} {951}},\ \bibinfo {pages} {L6}
  (\bibinfo {year} {2023})}\BibitemShut {NoStop}%
\bibitem [{\citenamefont {Antoniadis}\ {\it
  et~al.}(2023{\natexlab{a}})\citenamefont {Antoniadis} {\it
  et~al.}}]{EPTA:2023sfo}%
  \BibitemOpen
  \bibfield  {author} {\bibinfo {author} {\bibfnamefont {J.}~\bibnamefont
  {Antoniadis}} {\it et~al.} (\bibinfo {collaboration} {EPTA}),\ }\bibinfo
  {title} {{The second data release from the European Pulsar Timing Array - I.
  The dataset and timing analysis}},\ \href
  {https://doi.org/10.1051/0004-6361/202346841} {\bibfield  {journal} {Astron.
  Astrophys.\ }\textbf {\bibinfo {volume} {678}},\ \bibinfo {pages} {A48}
  (\bibinfo {year} {2023}{\natexlab{a}})}\BibitemShut {NoStop}%
\bibitem [{\citenamefont {Antoniadis}\ {\it
  et~al.}(2023{\natexlab{b}})\citenamefont {Antoniadis} {\it
  et~al.}}]{EPTA:2023fyk}%
  \BibitemOpen
  \bibfield  {author} {\bibinfo {author} {\bibfnamefont {J.}~\bibnamefont
  {Antoniadis}} {\it et~al.} (\bibinfo {collaboration} {EPTA, InPTA:}),\
  }\bibinfo {title} {{The second data release from the European Pulsar Timing
  Array - III. Search for gravitational wave signals}},\ \href
  {https://doi.org/10.1051/0004-6361/202346844} {\bibfield  {journal} {Astron.
  Astrophys.\ }\textbf {\bibinfo {volume} {678}},\ \bibinfo {pages} {A50}
  (\bibinfo {year} {2023}{\natexlab{b}})}\BibitemShut {NoStop}%
\bibitem [{\citenamefont {Xu}\ {\it et~al.}(2023)\citenamefont {Xu} {\it
  et~al.}}]{Xu:2023wog}%
  \BibitemOpen
  \bibfield  {author} {\bibinfo {author} {\bibfnamefont {H.}~\bibnamefont {Xu}}
  {\it et~al.},\ }\bibinfo {title} {{Searching for the Nano-Hertz Stochastic
  Gravitational Wave Background with the Chinese Pulsar Timing Array Data
  Release I}},\ \href {https://doi.org/10.1088/1674-4527/acdfa5} {\bibfield
  {journal} {Res. Astron. Astrophys.\ }\textbf {\bibinfo {volume} {23}},\
  \bibinfo {pages} {075024} (\bibinfo {year} {2023})}\BibitemShut {NoStop}%
\bibitem [{\citenamefont {Afzal}\ {\it et~al.}(2023)\citenamefont {Afzal} {\it
  et~al.}}]{NANOGrav:2023hvm}%
  \BibitemOpen
  \bibfield  {author} {\bibinfo {author} {\bibfnamefont {A.}~\bibnamefont
  {Afzal}} {\it et~al.} (\bibinfo {collaboration} {NANOGrav}),\ }\bibinfo
  {title} {{The NANOGrav 15 yr Data Set: Search for Signals from New
  Physics}},\ \href {https://doi.org/10.3847/2041-8213/acdc91} {\bibfield
  {journal} {Astrophys. J. Lett.\ }\textbf {\bibinfo {volume} {951}},\ \bibinfo
  {pages} {L11} (\bibinfo {year} {2023})}\BibitemShut {NoStop}%
\bibitem [{\citenamefont {Antoniadis}\ {\it
  et~al.}(2023{\natexlab{c}})\citenamefont {Antoniadis} {\it
  et~al.}}]{EPTA:2023xxk}%
  \BibitemOpen
  \bibfield  {author} {\bibinfo {author} {\bibfnamefont {J.}~\bibnamefont
  {Antoniadis}} {\it et~al.} (\bibinfo {collaboration} {EPTA}),\ }\bibinfo
  {title} {{The second data release from the European Pulsar Timing Array: IV.
  Implications for massive black holes, dark matter and the early Universe}},\
  \Eprint {https://arxiv.org/abs/2306.16227} {arXiv:2306.16227} \BibitemShut
  {NoStop}%
\bibitem [{\citenamefont {Yi}\ {\it et~al.}(2023{\natexlab{a}})\citenamefont
  {Yi}, \citenamefont {Gao}, \citenamefont {Gong}, \citenamefont {Wang},\ and\
  \citenamefont {Zhang}}]{Yi:2023mbm}%
  \BibitemOpen
  \bibfield  {author} {\bibinfo {author} {\bibfnamefont {Z.}~\bibnamefont
  {Yi}}, \bibinfo {author} {\bibfnamefont {Q.}~\bibnamefont {Gao}}, \bibinfo
  {author} {\bibfnamefont {Y.}~\bibnamefont {Gong}}, \bibinfo {author}
  {\bibfnamefont {Y.}~\bibnamefont {Wang}},\ and\ \bibinfo {author}
  {\bibfnamefont {F.}~\bibnamefont {Zhang}},\ }\bibinfo {title} {{Scalar
  induced gravitational waves in light of Pulsar Timing Array data}},\ \href
  {https://doi.org/10.1007/s11433-023-2266-1} {\bibfield  {journal} {Sci. China
  Phys. Mech. Astron.\ }\textbf {\bibinfo {volume} {66}},\ \bibinfo {pages}
  {120404} (\bibinfo {year} {2023}{\natexlab{a}})}\BibitemShut {NoStop}%
\bibitem [{\citenamefont {Yi}\ {\it et~al.}(2024)\citenamefont {Yi},
  \citenamefont {You},\ and\ \citenamefont {Wu}}]{Yi:2023tdk}%
  \BibitemOpen
  \bibfield  {author} {\bibinfo {author} {\bibfnamefont {Z.}~\bibnamefont
  {Yi}}, \bibinfo {author} {\bibfnamefont {Z.-Q.}\ \bibnamefont {You}},\ and\
  \bibinfo {author} {\bibfnamefont {Y.}~\bibnamefont {Wu}},\ }\bibinfo {title}
  {{Model-independent reconstruction of the primordial curvature power spectrum
  from PTA data}},\ \href {https://doi.org/10.1088/1475-7516/2024/01/066} {J.
  Cosmol. Astropart. Phys.\ \bibinfo {volume} {01}\bibfield  {year} {\bibinfo
  {year} { (\textbf {2024})}\ }\bibinfo  {pages} {066}}\BibitemShut {NoStop}%
\bibitem [{\citenamefont {Yi}\ {\it et~al.}(2023{\natexlab{b}})\citenamefont
  {Yi}, \citenamefont {You}, \citenamefont {Wu}, \citenamefont {Chen},\ and\
  \citenamefont {Liu}}]{Yi:2023npi}%
  \BibitemOpen
\bibfield  {pages} {  }\bibfield  {author} {\bibinfo {author} {\bibfnamefont
  {Z.}~\bibnamefont {Yi}}, \bibinfo {author} {\bibfnamefont {Z.-Q.}\
  \bibnamefont {You}}, \bibinfo {author} {\bibfnamefont {Y.}~\bibnamefont
  {Wu}}, \bibinfo {author} {\bibfnamefont {Z.-C.}\ \bibnamefont {Chen}},\ and\
  \bibinfo {author} {\bibfnamefont {L.}~\bibnamefont {Liu}},\ }\bibinfo {title}
  {{Exploring the NANOGrav Signal and Planet-mass Primordial Black Holes
  through Higgs Inflation}},\ \Eprint {https://arxiv.org/abs/2308.14688}
  {arXiv:2308.14688} \BibitemShut {NoStop}%
\bibitem [{\citenamefont {Cai}\ {\it et~al.}(2023)\citenamefont {Cai},
  \citenamefont {He}, \citenamefont {Ma}, \citenamefont {Yan},\ and\
  \citenamefont {Yuan}}]{Cai:2023dls}%
  \BibitemOpen
  \bibfield  {author} {\bibinfo {author} {\bibfnamefont {Y.-F.}\ \bibnamefont
  {Cai}}, \bibinfo {author} {\bibfnamefont {X.-C.}\ \bibnamefont {He}},
  \bibinfo {author} {\bibfnamefont {X.-H.}\ \bibnamefont {Ma}}, \bibinfo
  {author} {\bibfnamefont {S.-F.}\ \bibnamefont {Yan}},\ and\ \bibinfo {author}
  {\bibfnamefont {G.-W.}\ \bibnamefont {Yuan}},\ }\bibinfo {title} {{Limits on
  scalar-induced gravitational waves from the stochastic background by pulsar
  timing array observations}},\ \href
  {https://doi.org/10.1016/j.scib.2023.10.027} {\bibfield  {journal} {Sci.
  Bull.\ }\textbf {\bibinfo {volume} {68}},\ \bibinfo {pages} {2929} (\bibinfo
  {year} {2023})}\BibitemShut {NoStop}%
\bibitem [{\citenamefont {Liu}\ {\it et~al.}(2024{\natexlab{a}})\citenamefont
  {Liu}, \citenamefont {Chen},\ and\ \citenamefont {Huang}}]{Liu:2023ymk}%
  \BibitemOpen
  \bibfield  {author} {\bibinfo {author} {\bibfnamefont {L.}~\bibnamefont
  {Liu}}, \bibinfo {author} {\bibfnamefont {Z.-C.}\ \bibnamefont {Chen}},\ and\
  \bibinfo {author} {\bibfnamefont {Q.-G.}\ \bibnamefont {Huang}},\ }\bibinfo
  {title} {{Implications for the non-Gaussianity of curvature perturbation from
  pulsar timing arrays}},\ \href {https://doi.org/10.1103/PhysRevD.109.L061301}
  {\bibfield  {journal} {Phys. Rev. D\ }\textbf {\bibinfo {volume} {109}},\
  \bibinfo {pages} {L061301} (\bibinfo {year}
  {2024}{\natexlab{a}})}\BibitemShut {NoStop}%
\bibitem [{\citenamefont {Jin}\ {\it et~al.}(2023)\citenamefont {Jin},
  \citenamefont {Chen}, \citenamefont {Yi}, \citenamefont {You}, \citenamefont
  {Liu},\ and\ \citenamefont {Wu}}]{Jin:2023wri}%
  \BibitemOpen
  \bibfield  {author} {\bibinfo {author} {\bibfnamefont {J.-H.}\ \bibnamefont
  {Jin}}, \bibinfo {author} {\bibfnamefont {Z.-C.}\ \bibnamefont {Chen}},
  \bibinfo {author} {\bibfnamefont {Z.}~\bibnamefont {Yi}}, \bibinfo {author}
  {\bibfnamefont {Z.-Q.}\ \bibnamefont {You}}, \bibinfo {author} {\bibfnamefont
  {L.}~\bibnamefont {Liu}},\ and\ \bibinfo {author} {\bibfnamefont
  {Y.}~\bibnamefont {Wu}},\ }\bibinfo {title} {{Confronting sound speed
  resonance with pulsar timing arrays}},\ \href
  {https://doi.org/10.1088/1475-7516/2023/09/016} {J. Cosmol. Astropart. Phys.\
  \bibinfo {volume} {09}\bibfield  {year} {\bibinfo  {year} { (\textbf
  {2023})}\ }\bibinfo  {pages} {016}}\BibitemShut {NoStop}%
\bibitem [{\citenamefont {Chen}\ {\it et~al.}(2024)\citenamefont {Chen},
  \citenamefont {Li}, \citenamefont {Liu},\ and\ \citenamefont
  {Yi}}]{Chen:2024fir}%
  \BibitemOpen
\bibfield  {pages} {  }\bibfield  {author} {\bibinfo {author} {\bibfnamefont
  {Z.-C.}\ \bibnamefont {Chen}}, \bibinfo {author} {\bibfnamefont
  {J.}~\bibnamefont {Li}}, \bibinfo {author} {\bibfnamefont {L.}~\bibnamefont
  {Liu}},\ and\ \bibinfo {author} {\bibfnamefont {Z.}~\bibnamefont {Yi}},\
  }\bibinfo {title} {{Probing the speed of scalar-induced gravitational waves
  with pulsar timing arrays}},\ \href
  {https://doi.org/10.1103/PhysRevD.109.L101302} {\bibfield  {journal} {Phys.
  Rev. D\ }\textbf {\bibinfo {volume} {109}},\ \bibinfo {pages} {L101302}
  (\bibinfo {year} {2024})}\BibitemShut {NoStop}%
\bibitem [{\citenamefont {Liu}\ {\it et~al.}(2024{\natexlab{b}})\citenamefont
  {Liu}, \citenamefont {Wu},\ and\ \citenamefont {Chen}}]{Liu:2023hpw}%
  \BibitemOpen
  \bibfield  {author} {\bibinfo {author} {\bibfnamefont {L.}~\bibnamefont
  {Liu}}, \bibinfo {author} {\bibfnamefont {Y.}~\bibnamefont {Wu}},\ and\
  \bibinfo {author} {\bibfnamefont {Z.-C.}\ \bibnamefont {Chen}},\ }\bibinfo
  {title} {{Simultaneously probing the sound speed and equation of state of the
  early Universe with pulsar timing arrays}},\ \href
  {https://doi.org/10.1088/1475-7516/2024/04/011} {J. Cosmol. Astropart. Phys.\
  \bibinfo {volume} {04}\bibfield  {year} {\bibinfo  {year} { (\textbf
  {2024})}\ }\bibinfo  {pages} {011}}\BibitemShut {NoStop}%
\bibitem [{\citenamefont {Chen}\ and\ \citenamefont
  {Liu}(2024)}]{Chen:2024twp}%
  \BibitemOpen
\bibfield  {pages} {  }\bibfield  {author} {\bibinfo {author} {\bibfnamefont
  {Z.-C.}\ \bibnamefont {Chen}}\ and\ \bibinfo {author} {\bibfnamefont
  {L.}~\bibnamefont {Liu}},\ }\bibinfo {title} {{Can we distinguish the
  adiabatic fluctuations and isocurvature fluctuations with pulsar timing
  arrays?}},\ \Eprint {https://arxiv.org/abs/2402.16781} {arXiv:2402.16781}
  \BibitemShut {NoStop}%
\bibitem [{\citenamefont {Horava}(2009)}]{Horava:2009uw}%
  \BibitemOpen
  \bibfield  {author} {\bibinfo {author} {\bibfnamefont {P.}~\bibnamefont
  {Horava}},\ }\bibinfo {title} {{Quantum Gravity at a Lifshitz Point}},\ \href
  {https://doi.org/10.1103/PhysRevD.79.084008} {\bibfield  {journal} {Phys.
  Rev. D\ }\textbf {\bibinfo {volume} {79}},\ \bibinfo {pages} {084008}
  (\bibinfo {year} {2009})}\BibitemShut {NoStop}%
\bibitem [{\citenamefont {Gao}\ and\ \citenamefont {Hong}(2020)}]{Gao:2019liu}%
  \BibitemOpen
  \bibfield  {author} {\bibinfo {author} {\bibfnamefont {X.}~\bibnamefont
  {Gao}}\ and\ \bibinfo {author} {\bibfnamefont {X.-Y.}\ \bibnamefont {Hong}},\
  }\bibinfo {title} {{Propagation of gravitational waves in a cosmological
  background}},\ \href {https://doi.org/10.1103/PhysRevD.101.064057} {\bibfield
   {journal} {Phys. Rev. D\ }\textbf {\bibinfo {volume} {101}},\ \bibinfo
  {pages} {064057} (\bibinfo {year} {2020})}\BibitemShut {NoStop}%
\bibitem [{\citenamefont {Hu}\ and\ \citenamefont {Gao}(2022)}]{Hu:2021bbo}%
  \BibitemOpen
  \bibfield  {author} {\bibinfo {author} {\bibfnamefont {Y.-M.}\ \bibnamefont
  {Hu}}\ and\ \bibinfo {author} {\bibfnamefont {X.}~\bibnamefont {Gao}},\
  }\bibinfo {title} {{Covariant 3+1 correspondence of the spatially covariant
  gravity and the degeneracy conditions}},\ \href
  {https://doi.org/10.1103/PhysRevD.105.044023} {\bibfield  {journal} {Phys.
  Rev. D\ }\textbf {\bibinfo {volume} {105}},\ \bibinfo {pages} {044023}
  (\bibinfo {year} {2022})}\BibitemShut {NoStop}%
\bibitem [{\citenamefont {Hu}\ and\ \citenamefont {Gao}(2021)}]{Hu:2021yaq}%
  \BibitemOpen
  \bibfield  {author} {\bibinfo {author} {\bibfnamefont {Y.-M.}\ \bibnamefont
  {Hu}}\ and\ \bibinfo {author} {\bibfnamefont {X.}~\bibnamefont {Gao}},\
  }\bibinfo {title} {{Spatially covariant gravity with 2 degrees of freedom:
  Perturbative analysis}},\ \href {https://doi.org/10.1103/PhysRevD.104.104007}
  {\bibfield  {journal} {Phys. Rev. D\ }\textbf {\bibinfo {volume} {104}},\
  \bibinfo {pages} {104007} (\bibinfo {year} {2021})}\BibitemShut {NoStop}%
\bibitem [{\citenamefont {Takahashi}\ and\ \citenamefont
  {Soda}(2009)}]{Takahashi:2009wc}%
  \BibitemOpen
  \bibfield  {author} {\bibinfo {author} {\bibfnamefont {T.}~\bibnamefont
  {Takahashi}}\ and\ \bibinfo {author} {\bibfnamefont {J.}~\bibnamefont
  {Soda}},\ }\bibinfo {title} {{Chiral Primordial Gravitational Waves from a
  Lifshitz Point}},\ \href {https://doi.org/10.1103/PhysRevLett.102.231301}
  {\bibfield  {journal} {Phys. Rev. Lett.\ }\textbf {\bibinfo {volume} {102}},\
  \bibinfo {pages} {231301} (\bibinfo {year} {2009})}\BibitemShut {NoStop}%
\bibitem [{\citenamefont {Myung}(2010)}]{Myung:2009ug}%
  \BibitemOpen
  \bibfield  {author} {\bibinfo {author} {\bibfnamefont {Y.~S.}\ \bibnamefont
  {Myung}},\ }\bibinfo {title} {{Chiral gravitational waves from z=2
  Ho\v{r}ava-Lifshitz gravity}},\ \href
  {https://doi.org/10.1016/j.physletb.2009.12.059} {\bibfield  {journal} {Phys.
  Lett. B\ }\textbf {\bibinfo {volume} {684}},\ \bibinfo {pages} {1} (\bibinfo
  {year} {2010})}\BibitemShut {NoStop}%
\bibitem [{\citenamefont {Wang}\ {\it et~al.}(2013)\citenamefont {Wang},
  \citenamefont {Wu}, \citenamefont {Zhao},\ and\ \citenamefont
  {Zhu}}]{Wang:2012fi}%
  \BibitemOpen
  \bibfield  {author} {\bibinfo {author} {\bibfnamefont {A.}~\bibnamefont
  {Wang}}, \bibinfo {author} {\bibfnamefont {Q.}~\bibnamefont {Wu}}, \bibinfo
  {author} {\bibfnamefont {W.}~\bibnamefont {Zhao}},\ and\ \bibinfo {author}
  {\bibfnamefont {T.}~\bibnamefont {Zhu}},\ }\bibinfo {title} {{Polarizing
  primordial gravitational waves by parity violation}},\ \href
  {https://doi.org/10.1103/PhysRevD.87.103512} {\bibfield  {journal} {Phys.
  Rev. D\ }\textbf {\bibinfo {volume} {87}},\ \bibinfo {pages} {103512}
  (\bibinfo {year} {2013})}\BibitemShut {NoStop}%
\bibitem [{\citenamefont {Zhu}\ {\it et~al.}(2013)\citenamefont {Zhu},
  \citenamefont {Zhao}, \citenamefont {Huang}, \citenamefont {Wang},\ and\
  \citenamefont {Wu}}]{Zhu:2013fja}%
  \BibitemOpen
  \bibfield  {author} {\bibinfo {author} {\bibfnamefont {T.}~\bibnamefont
  {Zhu}}, \bibinfo {author} {\bibfnamefont {W.}~\bibnamefont {Zhao}}, \bibinfo
  {author} {\bibfnamefont {Y.}~\bibnamefont {Huang}}, \bibinfo {author}
  {\bibfnamefont {A.}~\bibnamefont {Wang}},\ and\ \bibinfo {author}
  {\bibfnamefont {Q.}~\bibnamefont {Wu}},\ }\bibinfo {title} {{Effects of
  parity violation on non-gaussianity of primordial gravitational waves in
  Ho\v{r}ava-Lifshitz gravity}},\ \href
  {https://doi.org/10.1103/PhysRevD.88.063508} {\bibfield  {journal} {Phys.
  Rev. D\ }\textbf {\bibinfo {volume} {88}},\ \bibinfo {pages} {063508}
  (\bibinfo {year} {2013})}\BibitemShut {NoStop}%
\bibitem [{\citenamefont {Cannone}\ {\it et~al.}(2015)\citenamefont {Cannone},
  \citenamefont {Gong},\ and\ \citenamefont {Tasinato}}]{Cannone:2015rra}%
  \BibitemOpen
  \bibfield  {author} {\bibinfo {author} {\bibfnamefont {D.}~\bibnamefont
  {Cannone}}, \bibinfo {author} {\bibfnamefont {J.-O.}\ \bibnamefont {Gong}},\
  and\ \bibinfo {author} {\bibfnamefont {G.}~\bibnamefont {Tasinato}},\
  }\bibinfo {title} {{Breaking discrete symmetries in the effective field
  theory of inflation}},\ \href {https://doi.org/10.1088/1475-7516/2015/08/003}
  {J. Cosmol. Astropart. Phys.\ \bibinfo {volume} {08}\bibfield  {year}
  {\bibinfo  {year} { (\textbf {2015})}\ }\bibinfo  {pages} {003}}\BibitemShut
  {NoStop}%
\bibitem [{\citenamefont {Zhao}\ {\it et~al.}(2020{\natexlab{a}})\citenamefont
  {Zhao}, \citenamefont {Liu}, \citenamefont {Wen}, \citenamefont {Zhu},
  \citenamefont {Wang}, \citenamefont {Hu},\ and\ \citenamefont
  {Zhou}}]{Zhao:2019szi}%
  \BibitemOpen
\bibfield  {pages} {  }\bibfield  {author} {\bibinfo {author} {\bibfnamefont
  {W.}~\bibnamefont {Zhao}}, \bibinfo {author} {\bibfnamefont {T.}~\bibnamefont
  {Liu}}, \bibinfo {author} {\bibfnamefont {L.}~\bibnamefont {Wen}}, \bibinfo
  {author} {\bibfnamefont {T.}~\bibnamefont {Zhu}}, \bibinfo {author}
  {\bibfnamefont {A.}~\bibnamefont {Wang}}, \bibinfo {author} {\bibfnamefont
  {Q.}~\bibnamefont {Hu}},\ and\ \bibinfo {author} {\bibfnamefont
  {C.}~\bibnamefont {Zhou}},\ }\bibinfo {title} {{Model-independent test of the
  parity symmetry of gravity with gravitational waves}},\ \href
  {https://doi.org/10.1140/epjc/s10052-020-8211-4} {\bibfield  {journal} {Eur.
  Phys. J. C\ }\textbf {\bibinfo {volume} {80}},\ \bibinfo {pages} {630}
  (\bibinfo {year} {2020}{\natexlab{a}})}\BibitemShut {NoStop}%
\bibitem [{\citenamefont {Zhao}\ {\it et~al.}(2020{\natexlab{b}})\citenamefont
  {Zhao}, \citenamefont {Zhu}, \citenamefont {Qiao},\ and\ \citenamefont
  {Wang}}]{Zhao:2019xmm}%
  \BibitemOpen
  \bibfield  {author} {\bibinfo {author} {\bibfnamefont {W.}~\bibnamefont
  {Zhao}}, \bibinfo {author} {\bibfnamefont {T.}~\bibnamefont {Zhu}}, \bibinfo
  {author} {\bibfnamefont {J.}~\bibnamefont {Qiao}},\ and\ \bibinfo {author}
  {\bibfnamefont {A.}~\bibnamefont {Wang}},\ }\bibinfo {title} {{Waveform of
  gravitational waves in the general parity-violating gravities}},\ \href
  {https://doi.org/10.1103/PhysRevD.101.024002} {\bibfield  {journal} {Phys.
  Rev. D\ }\textbf {\bibinfo {volume} {101}},\ \bibinfo {pages} {024002}
  (\bibinfo {year} {2020}{\natexlab{b}})}\BibitemShut {NoStop}%
\bibitem [{\citenamefont {Qiao}\ {\it et~al.}(2020)\citenamefont {Qiao},
  \citenamefont {Zhu}, \citenamefont {Zhao},\ and\ \citenamefont
  {Wang}}]{Qiao:2019hkz}%
  \BibitemOpen
  \bibfield  {author} {\bibinfo {author} {\bibfnamefont {J.}~\bibnamefont
  {Qiao}}, \bibinfo {author} {\bibfnamefont {T.}~\bibnamefont {Zhu}}, \bibinfo
  {author} {\bibfnamefont {W.}~\bibnamefont {Zhao}},\ and\ \bibinfo {author}
  {\bibfnamefont {A.}~\bibnamefont {Wang}},\ }\bibinfo {title} {{Polarized
  primordial gravitational waves in the ghost-free parity-violating gravity}},\
  \href {https://doi.org/10.1103/PhysRevD.101.043528} {\bibfield  {journal}
  {Phys. Rev. D\ }\textbf {\bibinfo {volume} {101}},\ \bibinfo {pages} {043528}
  (\bibinfo {year} {2020})}\BibitemShut {NoStop}%
\bibitem [{\citenamefont {Qiao}\ {\it et~al.}(2019)\citenamefont {Qiao},
  \citenamefont {Zhu}, \citenamefont {Zhao},\ and\ \citenamefont
  {Wang}}]{Qiao:2019wsh}%
  \BibitemOpen
  \bibfield  {author} {\bibinfo {author} {\bibfnamefont {J.}~\bibnamefont
  {Qiao}}, \bibinfo {author} {\bibfnamefont {T.}~\bibnamefont {Zhu}}, \bibinfo
  {author} {\bibfnamefont {W.}~\bibnamefont {Zhao}},\ and\ \bibinfo {author}
  {\bibfnamefont {A.}~\bibnamefont {Wang}},\ }\bibinfo {title} {{Waveform of
  gravitational waves in the ghost-free parity-violating gravities}},\ \href
  {https://doi.org/10.1103/PhysRevD.100.124058} {\bibfield  {journal} {Phys.
  Rev. D\ }\textbf {\bibinfo {volume} {100}},\ \bibinfo {pages} {124058}
  (\bibinfo {year} {2019})}\BibitemShut {NoStop}%
\bibitem [{\citenamefont {Qiao}\ {\it et~al.}(2022)\citenamefont {Qiao},
  \citenamefont {Zhu}, \citenamefont {Li},\ and\ \citenamefont
  {Zhao}}]{Qiao:2021fwi}%
  \BibitemOpen
  \bibfield  {author} {\bibinfo {author} {\bibfnamefont {J.}~\bibnamefont
  {Qiao}}, \bibinfo {author} {\bibfnamefont {T.}~\bibnamefont {Zhu}}, \bibinfo
  {author} {\bibfnamefont {G.}~\bibnamefont {Li}},\ and\ \bibinfo {author}
  {\bibfnamefont {W.}~\bibnamefont {Zhao}},\ }\bibinfo {title} {{Post-Newtonian
  parameters of ghost-free parity-violating gravities}},\ \href
  {https://doi.org/10.1088/1475-7516/2022/04/054} {J. Cosmol. Astropart. Phys.\
  \bibinfo {volume} {04}\bibfield  {year} {\bibinfo  {year} { (\textbf
  {2022})}\ }\bibinfo  {pages} {054}}\BibitemShut {NoStop}%
\bibitem [{\citenamefont {Gong}\ {\it et~al.}(2022)\citenamefont {Gong},
  \citenamefont {Zhu}, \citenamefont {Niu}, \citenamefont {Wu}, \citenamefont
  {Cui}, \citenamefont {Zhang}, \citenamefont {Zhao},\ and\ \citenamefont
  {Wang}}]{Gong:2021jgg}%
  \BibitemOpen
\bibfield  {pages} {  }\bibfield  {author} {\bibinfo {author} {\bibfnamefont
  {C.}~\bibnamefont {Gong}}, \bibinfo {author} {\bibfnamefont {T.}~\bibnamefont
  {Zhu}}, \bibinfo {author} {\bibfnamefont {R.}~\bibnamefont {Niu}}, \bibinfo
  {author} {\bibfnamefont {Q.}~\bibnamefont {Wu}}, \bibinfo {author}
  {\bibfnamefont {J.-L.}\ \bibnamefont {Cui}}, \bibinfo {author} {\bibfnamefont
  {X.}~\bibnamefont {Zhang}}, \bibinfo {author} {\bibfnamefont
  {W.}~\bibnamefont {Zhao}},\ and\ \bibinfo {author} {\bibfnamefont
  {A.}~\bibnamefont {Wang}},\ }\bibinfo {title} {{Gravitational wave
  constraints on Lorentz and parity violations in gravity: High-order spatial
  derivative cases}},\ \href {https://doi.org/10.1103/PhysRevD.105.044034}
  {\bibfield  {journal} {Phys. Rev. D\ }\textbf {\bibinfo {volume} {105}},\
  \bibinfo {pages} {044034} (\bibinfo {year} {2022})}\BibitemShut {NoStop}%
\bibitem [{\citenamefont {Zhu}\ and\ \citenamefont {Cai}(2023)}]{Zhu:2023lhv}%
  \BibitemOpen
  \bibfield  {author} {\bibinfo {author} {\bibfnamefont {M.}~\bibnamefont
  {Zhu}}\ and\ \bibinfo {author} {\bibfnamefont {Y.}~\bibnamefont {Cai}},\
  }\bibinfo {title} {{Parity-violation in bouncing cosmology}},\ \href
  {https://doi.org/10.1007/JHEP04(2023)095} {J. High Energ. Phys.\ \bibinfo
  {volume} {04}\bibfield  {year} {\bibinfo  {year} { (\textbf {2023})}\
  }\bibinfo  {pages} {095}}\BibitemShut {NoStop}%
\bibitem [{\citenamefont {Akama}\ and\ \citenamefont
  {Zhu}(2024)}]{Akama:2024bav}%
  \BibitemOpen
\bibfield  {pages} {  }\bibfield  {author} {\bibinfo {author} {\bibfnamefont
  {S.}~\bibnamefont {Akama}}\ and\ \bibinfo {author} {\bibfnamefont
  {M.}~\bibnamefont {Zhu}},\ }\bibinfo {title} {{Parity violation in primordial
  tensor non-Gaussianities from matter bounce cosmology}},\ \Eprint
  {https://arxiv.org/abs/2404.05464} {arXiv:2404.05464} \BibitemShut {NoStop}%
\bibitem [{\citenamefont {Lee}\ and\ \citenamefont {Yang}(1956)}]{Lee:1956qn}%
  \BibitemOpen
  \bibfield  {author} {\bibinfo {author} {\bibfnamefont {T.~D.}\ \bibnamefont
  {Lee}}\ and\ \bibinfo {author} {\bibfnamefont {C.-N.}\ \bibnamefont {Yang}},\
  }\bibinfo {title} {{Question of Parity Conservation in Weak Interactions}},\
  \href {https://doi.org/10.1103/PhysRev.104.254} {\bibfield  {journal} {Phys.
  Rev.\ }\textbf {\bibinfo {volume} {104}},\ \bibinfo {pages} {254} (\bibinfo
  {year} {1956})}\BibitemShut {NoStop}%
\bibitem [{\citenamefont {Wu}\ {\it et~al.}(1957)\citenamefont {Wu},
  \citenamefont {Ambler}, \citenamefont {Hayward}, \citenamefont {Hoppes},\
  and\ \citenamefont {Hudson}}]{Wu:1957my}%
  \BibitemOpen
  \bibfield  {author} {\bibinfo {author} {\bibfnamefont {C.~S.}\ \bibnamefont
  {Wu}}, \bibinfo {author} {\bibfnamefont {E.}~\bibnamefont {Ambler}}, \bibinfo
  {author} {\bibfnamefont {R.~W.}\ \bibnamefont {Hayward}}, \bibinfo {author}
  {\bibfnamefont {D.~D.}\ \bibnamefont {Hoppes}},\ and\ \bibinfo {author}
  {\bibfnamefont {R.~P.}\ \bibnamefont {Hudson}},\ }\bibinfo {title}
  {{Experimental Test of Parity Conservation in $\beta$ Decay}},\ \href
  {https://doi.org/10.1103/PhysRev.105.1413} {\bibfield  {journal} {Phys. Rev.\
  }\textbf {\bibinfo {volume} {105}},\ \bibinfo {pages} {1413} (\bibinfo {year}
  {1957})}\BibitemShut {NoStop}%
\bibitem [{\citenamefont {Philcox}(2022)}]{Philcox:2022hkh}%
  \BibitemOpen
  \bibfield  {author} {\bibinfo {author} {\bibfnamefont {O.~H.~E.}\
  \bibnamefont {Philcox}},\ }\bibinfo {title} {{Probing parity violation with
  the four-point correlation function of BOSS galaxies}},\ \href
  {https://doi.org/10.1103/PhysRevD.106.063501} {\bibfield  {journal} {Phys.
  Rev. D\ }\textbf {\bibinfo {volume} {106}},\ \bibinfo {pages} {063501}
  (\bibinfo {year} {2022})}\BibitemShut {NoStop}%
\bibitem [{\citenamefont {Hou}\ {\it et~al.}(2023)\citenamefont {Hou},
  \citenamefont {Slepian},\ and\ \citenamefont {Cahn}}]{Hou:2022wfj}%
  \BibitemOpen
  \bibfield  {author} {\bibinfo {author} {\bibfnamefont {J.}~\bibnamefont
  {Hou}}, \bibinfo {author} {\bibfnamefont {Z.}~\bibnamefont {Slepian}},\ and\
  \bibinfo {author} {\bibfnamefont {R.~N.}\ \bibnamefont {Cahn}},\ }\bibinfo
  {title} {{Measurement of parity-odd modes in the large-scale 4-point
  correlation function of Sloan Digital Sky Survey Baryon Oscillation
  Spectroscopic Survey twelfth data release CMASS and LOWZ galaxies}},\ \href
  {https://doi.org/10.1093/mnras/stad1062} {\bibfield  {journal} {Mon. Not. R.
  Astron. Soc.\ }\textbf {\bibinfo {volume} {522}},\ \bibinfo {pages} {5701}
  (\bibinfo {year} {2023})}\BibitemShut {NoStop}%
\bibitem [{\citenamefont {Minami}\ and\ \citenamefont
  {Komatsu}(2020)}]{Minami:2020odp}%
  \BibitemOpen
  \bibfield  {author} {\bibinfo {author} {\bibfnamefont {Y.}~\bibnamefont
  {Minami}}\ and\ \bibinfo {author} {\bibfnamefont {E.}~\bibnamefont
  {Komatsu}},\ }\bibinfo {title} {{New Extraction of the Cosmic Birefringence
  from the Planck 2018 Polarization Data}},\ \href
  {https://doi.org/10.1103/PhysRevLett.125.221301} {\bibfield  {journal} {Phys.
  Rev. Lett.\ }\textbf {\bibinfo {volume} {125}},\ \bibinfo {pages} {221301}
  (\bibinfo {year} {2020})}\BibitemShut {NoStop}%
\bibitem [{\citenamefont {Eskilt}\ and\ \citenamefont
  {Komatsu}(2022)}]{Eskilt:2022cff}%
  \BibitemOpen
  \bibfield  {author} {\bibinfo {author} {\bibfnamefont {J.~R.}\ \bibnamefont
  {Eskilt}}\ and\ \bibinfo {author} {\bibfnamefont {E.}~\bibnamefont
  {Komatsu}},\ }\bibinfo {title} {{Improved constraints on cosmic birefringence
  from the WMAP and Planck cosmic microwave background polarization data}},\
  \href {https://doi.org/10.1103/PhysRevD.106.063503} {\bibfield  {journal}
  {Phys. Rev. D\ }\textbf {\bibinfo {volume} {106}},\ \bibinfo {pages} {063503}
  (\bibinfo {year} {2022})}\BibitemShut {NoStop}%
\bibitem [{\citenamefont {Liu}\ {\it et~al.}(2020)\citenamefont {Liu},
  \citenamefont {Tong}, \citenamefont {Wang},\ and\ \citenamefont
  {Xianyu}}]{Liu:2019fag}%
  \BibitemOpen
  \bibfield  {author} {\bibinfo {author} {\bibfnamefont {T.}~\bibnamefont
  {Liu}}, \bibinfo {author} {\bibfnamefont {X.}~\bibnamefont {Tong}}, \bibinfo
  {author} {\bibfnamefont {Y.}~\bibnamefont {Wang}},\ and\ \bibinfo {author}
  {\bibfnamefont {Z.-Z.}\ \bibnamefont {Xianyu}},\ }\bibinfo {title} {{Probing
  P and CP Violations on the Cosmological Collider}},\ \href
  {https://doi.org/10.1007/JHEP04(2020)189} {J. High Energ. Phys.\ \bibinfo
  {volume} {04}\bibfield  {year} {\bibinfo  {year} { (\textbf {2020})}\
  }\bibinfo  {pages} {189}}\BibitemShut {NoStop}%
\bibitem [{\citenamefont {Niu}\ {\it et~al.}(2023)\citenamefont {Niu},
  \citenamefont {Rahat}, \citenamefont {Srinivasan},\ and\ \citenamefont
  {Xue}}]{Niu:2022fki}%
  \BibitemOpen
\bibfield  {pages} {  }\bibfield  {author} {\bibinfo {author} {\bibfnamefont
  {X.}~\bibnamefont {Niu}}, \bibinfo {author} {\bibfnamefont {M.~H.}\
  \bibnamefont {Rahat}}, \bibinfo {author} {\bibfnamefont {K.}~\bibnamefont
  {Srinivasan}},\ and\ \bibinfo {author} {\bibfnamefont {W.}~\bibnamefont
  {Xue}},\ }\bibinfo {title} {{Parity-odd and even trispectrum from axion
  inflation}},\ \href {https://doi.org/10.1088/1475-7516/2023/05/018} {J.
  Cosmol. Astropart. Phys.\ \bibinfo {volume} {05}\bibfield  {year} {\bibinfo
  {year} { (\textbf {2023})}\ }\bibinfo  {pages} {018}}\BibitemShut {NoStop}%
\bibitem [{\citenamefont {Cabass}\ {\it et~al.}(2023)\citenamefont {Cabass},
  \citenamefont {Jazayeri}, \citenamefont {Pajer},\ and\ \citenamefont
  {Stefanyszyn}}]{Cabass:2022rhr}%
  \BibitemOpen
\bibfield  {pages} {  }\bibfield  {author} {\bibinfo {author} {\bibfnamefont
  {G.}~\bibnamefont {Cabass}}, \bibinfo {author} {\bibfnamefont
  {S.}~\bibnamefont {Jazayeri}}, \bibinfo {author} {\bibfnamefont
  {E.}~\bibnamefont {Pajer}},\ and\ \bibinfo {author} {\bibfnamefont
  {D.}~\bibnamefont {Stefanyszyn}},\ }\bibinfo {title} {{Parity violation in
  the scalar trispectrum: no-go theorems and yes-go examples}},\ \href
  {https://doi.org/10.1007/JHEP02(2023)021} {J. High Energ. Phys.\ \bibinfo
  {volume} {02}\bibfield  {year} {\bibinfo  {year} { (\textbf {2023})}\
  }\bibinfo  {pages} {021}}\BibitemShut {NoStop}%
\bibitem [{\citenamefont {Creque-Sarbinowski}\ {\it et~al.}(2023)\citenamefont
  {Creque-Sarbinowski}, \citenamefont {Alexander}, \citenamefont
  {Kamionkowski},\ and\ \citenamefont {Philcox}}]{Creque-Sarbinowski:2023wmb}%
  \BibitemOpen
\bibfield  {pages} {  }\bibfield  {author} {\bibinfo {author} {\bibfnamefont
  {C.}~\bibnamefont {Creque-Sarbinowski}}, \bibinfo {author} {\bibfnamefont
  {S.}~\bibnamefont {Alexander}}, \bibinfo {author} {\bibfnamefont
  {M.}~\bibnamefont {Kamionkowski}},\ and\ \bibinfo {author} {\bibfnamefont
  {O.}~\bibnamefont {Philcox}},\ }\bibinfo {title} {{Parity-violating
  trispectrum from Chern-Simons gravity}},\ \href
  {https://doi.org/10.1088/1475-7516/2023/11/029} {J. Cosmol. Astropart. Phys.\
  \bibinfo {volume} {11}\bibfield  {year} {\bibinfo  {year} { (\textbf
  {2023})}\ }\bibinfo  {pages} {029}}\BibitemShut {NoStop}%
\bibitem [{\citenamefont {Garcia-Saenz}\ {\it
  et~al.}(2023{\natexlab{a}})\citenamefont {Garcia-Saenz}, \citenamefont {Lu},\
  and\ \citenamefont {Shuai}}]{Garcia-Saenz:2023zue}%
  \BibitemOpen
\bibfield  {pages} {  }\bibfield  {author} {\bibinfo {author} {\bibfnamefont
  {S.}~\bibnamefont {Garcia-Saenz}}, \bibinfo {author} {\bibfnamefont
  {Y.}~\bibnamefont {Lu}},\ and\ \bibinfo {author} {\bibfnamefont
  {Z.}~\bibnamefont {Shuai}},\ }\bibinfo {title} {{Scalar-induced gravitational
  waves from ghost inflation and parity violation}},\ \href
  {https://doi.org/10.1103/PhysRevD.108.123507} {\bibfield  {journal} {Phys.
  Rev. D\ }\textbf {\bibinfo {volume} {108}},\ \bibinfo {pages} {123507}
  (\bibinfo {year} {2023}{\natexlab{a}})}\BibitemShut {NoStop}%
\bibitem [{\citenamefont {Jackiw}\ and\ \citenamefont
  {Pi}(2003)}]{Jackiw:2003pm}%
  \BibitemOpen
  \bibfield  {author} {\bibinfo {author} {\bibfnamefont {R.}~\bibnamefont
  {Jackiw}}\ and\ \bibinfo {author} {\bibfnamefont {S.~Y.}\ \bibnamefont
  {Pi}},\ }\bibinfo {title} {{Chern-Simons modification of general
  relativity}},\ \href {https://doi.org/10.1103/PhysRevD.68.104012} {\bibfield
  {journal} {Phys. Rev. D\ }\textbf {\bibinfo {volume} {68}},\ \bibinfo {pages}
  {104012} (\bibinfo {year} {2003})}\BibitemShut {NoStop}%
\bibitem [{\citenamefont {Lue}\ {\it et~al.}(1999)\citenamefont {Lue},
  \citenamefont {Wang},\ and\ \citenamefont {Kamionkowski}}]{Lue:1998mq}%
  \BibitemOpen
  \bibfield  {author} {\bibinfo {author} {\bibfnamefont {A.}~\bibnamefont
  {Lue}}, \bibinfo {author} {\bibfnamefont {L.-M.}\ \bibnamefont {Wang}},\ and\
  \bibinfo {author} {\bibfnamefont {M.}~\bibnamefont {Kamionkowski}},\
  }\bibinfo {title} {{Cosmological signature of new parity violating
  interactions}},\ \href {https://doi.org/10.1103/PhysRevLett.83.1506}
  {\bibfield  {journal} {Phys. Rev. Lett.\ }\textbf {\bibinfo {volume} {83}},\
  \bibinfo {pages} {1506} (\bibinfo {year} {1999})}\BibitemShut {NoStop}%
\bibitem [{\citenamefont {Satoh}\ {\it et~al.}(2008)\citenamefont {Satoh},
  \citenamefont {Kanno},\ and\ \citenamefont {Soda}}]{Satoh:2007gn}%
  \BibitemOpen
  \bibfield  {author} {\bibinfo {author} {\bibfnamefont {M.}~\bibnamefont
  {Satoh}}, \bibinfo {author} {\bibfnamefont {S.}~\bibnamefont {Kanno}},\ and\
  \bibinfo {author} {\bibfnamefont {J.}~\bibnamefont {Soda}},\ }\bibinfo
  {title} {{Circular Polarization of Primordial Gravitational Waves in
  String-inspired Inflationary Cosmology}},\ \href
  {https://doi.org/10.1103/PhysRevD.77.023526} {\bibfield  {journal} {Phys.
  Rev. D\ }\textbf {\bibinfo {volume} {77}},\ \bibinfo {pages} {023526}
  (\bibinfo {year} {2008})}\BibitemShut {NoStop}%
\bibitem [{\citenamefont {Saito}\ {\it et~al.}(2007)\citenamefont {Saito},
  \citenamefont {Ichiki},\ and\ \citenamefont {Taruya}}]{Saito:2007kt}%
  \BibitemOpen
  \bibfield  {author} {\bibinfo {author} {\bibfnamefont {S.}~\bibnamefont
  {Saito}}, \bibinfo {author} {\bibfnamefont {K.}~\bibnamefont {Ichiki}},\ and\
  \bibinfo {author} {\bibfnamefont {A.}~\bibnamefont {Taruya}},\ }\bibinfo
  {title} {{Probing polarization states of primordial gravitational waves with
  CMB anisotropies}},\ \href {https://doi.org/10.1088/1475-7516/2007/09/002}
  {J. Cosmol. Astropart. Phys.\ \bibinfo {volume} {09}\bibfield  {year}
  {\bibinfo  {year} { (\textbf {2007})}\ }\bibinfo  {pages} {002}}\BibitemShut
  {NoStop}%
\bibitem [{\citenamefont {Alexander}\ and\ \citenamefont
  {Yunes}(2009)}]{Alexander:2009tp}%
  \BibitemOpen
\bibfield  {pages} {  }\bibfield  {author} {\bibinfo {author} {\bibfnamefont
  {S.}~\bibnamefont {Alexander}}\ and\ \bibinfo {author} {\bibfnamefont
  {N.}~\bibnamefont {Yunes}},\ }\bibinfo {title} {{Chern-Simons Modified
  General Relativity}},\ \href {https://doi.org/10.1016/j.physrep.2009.07.002}
  {\bibfield  {journal} {Phys. Rep.\ }\textbf {\bibinfo {volume} {480}},\
  \bibinfo {pages} {1} (\bibinfo {year} {2009})}\BibitemShut {NoStop}%
\bibitem [{\citenamefont {Yunes}\ {\it et~al.}(2010)\citenamefont {Yunes},
  \citenamefont {O'Shaughnessy}, \citenamefont {Owen},\ and\ \citenamefont
  {Alexander}}]{Yunes:2010yf}%
  \BibitemOpen
  \bibfield  {author} {\bibinfo {author} {\bibfnamefont {N.}~\bibnamefont
  {Yunes}}, \bibinfo {author} {\bibfnamefont {R.}~\bibnamefont
  {O'Shaughnessy}}, \bibinfo {author} {\bibfnamefont {B.~J.}\ \bibnamefont
  {Owen}},\ and\ \bibinfo {author} {\bibfnamefont {S.}~\bibnamefont
  {Alexander}},\ }\bibinfo {title} {{Testing gravitational parity violation
  with coincident gravitational waves and short gamma-ray bursts}},\ \href
  {https://doi.org/10.1103/PhysRevD.82.064017} {\bibfield  {journal} {Phys.
  Rev. D\ }\textbf {\bibinfo {volume} {82}},\ \bibinfo {pages} {064017}
  (\bibinfo {year} {2010})}\BibitemShut {NoStop}%
\bibitem [{\citenamefont {Gluscevic}\ and\ \citenamefont
  {Kamionkowski}(2010)}]{Gluscevic:2010vv}%
  \BibitemOpen
  \bibfield  {author} {\bibinfo {author} {\bibfnamefont {V.}~\bibnamefont
  {Gluscevic}}\ and\ \bibinfo {author} {\bibfnamefont {M.}~\bibnamefont
  {Kamionkowski}},\ }\bibinfo {title} {{Testing Parity-Violating Mechanisms
  with Cosmic Microwave Background Experiments}},\ \href
  {https://doi.org/10.1103/PhysRevD.81.123529} {\bibfield  {journal} {Phys.
  Rev. D\ }\textbf {\bibinfo {volume} {81}},\ \bibinfo {pages} {123529}
  (\bibinfo {year} {2010})}\BibitemShut {NoStop}%
\bibitem [{\citenamefont {Myung}\ and\ \citenamefont
  {Moon}(2014)}]{Myung:2014jha}%
  \BibitemOpen
  \bibfield  {author} {\bibinfo {author} {\bibfnamefont {Y.~S.}\ \bibnamefont
  {Myung}}\ and\ \bibinfo {author} {\bibfnamefont {T.}~\bibnamefont {Moon}},\
  }\bibinfo {title} {{Primordial massive gravitational waves from
  Einstein-Chern-Simons-Weyl gravity}},\ \href
  {https://doi.org/10.1088/1475-7516/2014/08/061} {J. Cosmol. Astropart. Phys.\
  \bibinfo {volume} {08}\bibfield  {year} {\bibinfo  {year} { (\textbf
  {2014})}\ }\bibinfo  {pages} {061}}\BibitemShut {NoStop}%
\bibitem [{\citenamefont {Kawai}\ and\ \citenamefont
  {Kim}(2019)}]{Kawai:2017kqt}%
  \BibitemOpen
\bibfield  {pages} {  }\bibfield  {author} {\bibinfo {author} {\bibfnamefont
  {S.}~\bibnamefont {Kawai}}\ and\ \bibinfo {author} {\bibfnamefont
  {J.}~\bibnamefont {Kim}},\ }\bibinfo {title} {{Gauss\textendash{}Bonnet
  Chern\textendash{}Simons gravitational wave leptogenesis}},\ \href
  {https://doi.org/10.1016/j.physletb.2018.12.019} {\bibfield  {journal} {Phys.
  Lett. B\ }\textbf {\bibinfo {volume} {789}},\ \bibinfo {pages} {145}
  (\bibinfo {year} {2019})}\BibitemShut {NoStop}%
\bibitem [{\citenamefont {Nair}\ {\it et~al.}(2019)\citenamefont {Nair},
  \citenamefont {Perkins}, \citenamefont {Silva},\ and\ \citenamefont
  {Yunes}}]{Nair:2019iur}%
  \BibitemOpen
  \bibfield  {author} {\bibinfo {author} {\bibfnamefont {R.}~\bibnamefont
  {Nair}}, \bibinfo {author} {\bibfnamefont {S.}~\bibnamefont {Perkins}},
  \bibinfo {author} {\bibfnamefont {H.~O.}\ \bibnamefont {Silva}},\ and\
  \bibinfo {author} {\bibfnamefont {N.}~\bibnamefont {Yunes}},\ }\bibinfo
  {title} {{Fundamental Physics Implications for Higher-Curvature Theories from
  Binary Black Hole Signals in the LIGO-Virgo Catalog GWTC-1}},\ \href
  {https://doi.org/10.1103/PhysRevLett.123.191101} {\bibfield  {journal} {Phys.
  Rev. Lett.\ }\textbf {\bibinfo {volume} {123}},\ \bibinfo {pages} {191101}
  (\bibinfo {year} {2019})}\BibitemShut {NoStop}%
\bibitem [{\citenamefont {Nishizawa}\ and\ \citenamefont
  {Kobayashi}(2018)}]{Nishizawa:2018srh}%
  \BibitemOpen
  \bibfield  {author} {\bibinfo {author} {\bibfnamefont {A.}~\bibnamefont
  {Nishizawa}}\ and\ \bibinfo {author} {\bibfnamefont {T.}~\bibnamefont
  {Kobayashi}},\ }\bibinfo {title} {{Parity-violating gravity and GW170817}},\
  \href {https://doi.org/10.1103/PhysRevD.98.124018} {\bibfield  {journal}
  {Phys. Rev. D\ }\textbf {\bibinfo {volume} {98}},\ \bibinfo {pages} {124018}
  (\bibinfo {year} {2018})}\BibitemShut {NoStop}%
\bibitem [{\citenamefont {Bartolo}\ and\ \citenamefont
  {Orlando}(2017)}]{Bartolo:2017szm}%
  \BibitemOpen
  \bibfield  {author} {\bibinfo {author} {\bibfnamefont {N.}~\bibnamefont
  {Bartolo}}\ and\ \bibinfo {author} {\bibfnamefont {G.}~\bibnamefont
  {Orlando}},\ }\bibinfo {title} {{Parity breaking signatures from a
  Chern-Simons coupling during inflation: the case of non-Gaussian
  gravitational waves}},\ \href {https://doi.org/10.1088/1475-7516/2017/07/034}
  {J. Cosmol. Astropart. Phys.\ \bibinfo {volume} {07}\bibfield  {year}
  {\bibinfo  {year} { (\textbf {2017})}\ }\bibinfo  {pages} {034}}\BibitemShut
  {NoStop}%
\bibitem [{\citenamefont {Bartolo}\ {\it et~al.}(2019)\citenamefont {Bartolo},
  \citenamefont {Orlando},\ and\ \citenamefont {Shiraishi}}]{Bartolo:2018elp}%
  \BibitemOpen
\bibfield  {pages} {  }\bibfield  {author} {\bibinfo {author} {\bibfnamefont
  {N.}~\bibnamefont {Bartolo}}, \bibinfo {author} {\bibfnamefont
  {G.}~\bibnamefont {Orlando}},\ and\ \bibinfo {author} {\bibfnamefont
  {M.}~\bibnamefont {Shiraishi}},\ }\bibinfo {title} {{Measuring chiral
  gravitational waves in Chern-Simons gravity with CMB bispectra}},\ \href
  {https://doi.org/10.1088/1475-7516/2019/01/050} {J. Cosmol. Astropart. Phys.\
  \bibinfo {volume} {01}\bibfield  {year} {\bibinfo  {year} { (\textbf
  {2019})}\ }\bibinfo  {pages} {050}}\BibitemShut {NoStop}%
\bibitem [{\citenamefont {Zhang}\ {\it et~al.}(2022)\citenamefont {Zhang},
  \citenamefont {Feng},\ and\ \citenamefont {Gao}}]{Zhang:2022xmm}%
  \BibitemOpen
\bibfield  {pages} {  }\bibfield  {author} {\bibinfo {author} {\bibfnamefont
  {F.}~\bibnamefont {Zhang}}, \bibinfo {author} {\bibfnamefont {J.-X.}\
  \bibnamefont {Feng}},\ and\ \bibinfo {author} {\bibfnamefont
  {X.}~\bibnamefont {Gao}},\ }\bibinfo {title} {{Circularly polarized scalar
  induced gravitational waves from the Chern-Simons modified gravity}},\ \href
  {https://doi.org/10.1088/1475-7516/2022/10/054} {J. Cosmol. Astropart. Phys.\
  \bibinfo {volume} {10}\bibfield  {year} {\bibinfo  {year} { (\textbf
  {2022})}\ }\bibinfo  {pages} {054}}\BibitemShut {NoStop}%
\bibitem [{\citenamefont {Feng}\ {\it et~al.}(2023)\citenamefont {Feng},
  \citenamefont {Zhang},\ and\ \citenamefont {Gao}}]{Feng:2023veu}%
  \BibitemOpen
\bibfield  {pages} {  }\bibfield  {author} {\bibinfo {author} {\bibfnamefont
  {J.-X.}\ \bibnamefont {Feng}}, \bibinfo {author} {\bibfnamefont
  {F.}~\bibnamefont {Zhang}},\ and\ \bibinfo {author} {\bibfnamefont
  {X.}~\bibnamefont {Gao}},\ }\bibinfo {title} {{Scalar induced gravitational
  waves from Chern-Simons gravity during inflation era}},\ \href
  {https://doi.org/10.1088/1475-7516/2023/07/047} {J. Cosmol. Astropart. Phys.\
  \bibinfo {volume} {07}\bibfield  {year} {\bibinfo  {year} { (\textbf
  {2023})}\ }\bibinfo  {pages} {047}}\BibitemShut {NoStop}%
\bibitem [{\citenamefont {Crisostomi}\ {\it et~al.}(2018)\citenamefont
  {Crisostomi}, \citenamefont {Noui}, \citenamefont {Charmousis},\ and\
  \citenamefont {Langlois}}]{Crisostomi:2017ugk}%
  \BibitemOpen
\bibfield  {pages} {  }\bibfield  {author} {\bibinfo {author} {\bibfnamefont
  {M.}~\bibnamefont {Crisostomi}}, \bibinfo {author} {\bibfnamefont
  {K.}~\bibnamefont {Noui}}, \bibinfo {author} {\bibfnamefont {C.}~\bibnamefont
  {Charmousis}},\ and\ \bibinfo {author} {\bibfnamefont {D.}~\bibnamefont
  {Langlois}},\ }\bibinfo {title} {{Beyond Lovelock gravity: Higher derivative
  metric theories}},\ \href {https://doi.org/10.1103/PhysRevD.97.044034}
  {\bibfield  {journal} {Phys. Rev. D\ }\textbf {\bibinfo {volume} {97}},\
  \bibinfo {pages} {044034} (\bibinfo {year} {2018})}\BibitemShut {NoStop}%
\bibitem [{\citenamefont {Chatzistavrakidis}\ {\it et~al.}(2020)\citenamefont
  {Chatzistavrakidis}, \citenamefont {Karagiannis},\ and\ \citenamefont
  {Schupp}}]{Chatzistavrakidis:2020wum}%
  \BibitemOpen
  \bibfield  {author} {\bibinfo {author} {\bibfnamefont {A.}~\bibnamefont
  {Chatzistavrakidis}}, \bibinfo {author} {\bibfnamefont {G.}~\bibnamefont
  {Karagiannis}},\ and\ \bibinfo {author} {\bibfnamefont {P.}~\bibnamefont
  {Schupp}},\ }\bibinfo {title} {{Torsion-induced gravitational $\theta$ term
  and gravitoelectromagnetism}},\ \href
  {https://doi.org/10.1140/epjc/s10052-020-08600-9} {\bibfield  {journal} {Eur.
  Phys. J. C\ }\textbf {\bibinfo {volume} {80}},\ \bibinfo {pages} {1034}
  (\bibinfo {year} {2020})}\BibitemShut {NoStop}%
\bibitem [{\citenamefont {Wu}\ {\it et~al.}(2022)\citenamefont {Wu},
  \citenamefont {Zhu}, \citenamefont {Niu}, \citenamefont {Zhao},\ and\
  \citenamefont {Wang}}]{Wu:2021ndf}%
  \BibitemOpen
  \bibfield  {author} {\bibinfo {author} {\bibfnamefont {Q.}~\bibnamefont
  {Wu}}, \bibinfo {author} {\bibfnamefont {T.}~\bibnamefont {Zhu}}, \bibinfo
  {author} {\bibfnamefont {R.}~\bibnamefont {Niu}}, \bibinfo {author}
  {\bibfnamefont {W.}~\bibnamefont {Zhao}},\ and\ \bibinfo {author}
  {\bibfnamefont {A.}~\bibnamefont {Wang}},\ }\bibinfo {title} {{Constraints on
  the Nieh-Yan modified teleparallel gravity with gravitational waves}},\ \href
  {https://doi.org/10.1103/PhysRevD.105.024035} {\bibfield  {journal} {Phys.
  Rev. D\ }\textbf {\bibinfo {volume} {105}},\ \bibinfo {pages} {024035}
  (\bibinfo {year} {2022})}\BibitemShut {NoStop}%
\bibitem [{\citenamefont {L\r{a}ngvik}\ {\it et~al.}(2021)\citenamefont
  {L\r{a}ngvik}, \citenamefont {Ojanper\"a}, \citenamefont {Raatikainen},\ and\
  \citenamefont {Rasanen}}]{Langvik:2020nrs}%
  \BibitemOpen
  \bibfield  {author} {\bibinfo {author} {\bibfnamefont {M.}~\bibnamefont
  {L\r{a}ngvik}}, \bibinfo {author} {\bibfnamefont {J.-M.}\ \bibnamefont
  {Ojanper\"a}}, \bibinfo {author} {\bibfnamefont {S.}~\bibnamefont
  {Raatikainen}},\ and\ \bibinfo {author} {\bibfnamefont {S.}~\bibnamefont
  {Rasanen}},\ }\bibinfo {title} {{Higgs inflation with the Holst and the
  Nieh\textendash{}Yan term}},\ \href
  {https://doi.org/10.1103/PhysRevD.103.083514} {\bibfield  {journal} {Phys.
  Rev. D\ }\textbf {\bibinfo {volume} {103}},\ \bibinfo {pages} {083514}
  (\bibinfo {year} {2021})}\BibitemShut {NoStop}%
\bibitem [{\citenamefont {Rao}(2021)}]{Rao:2021azn}%
  \BibitemOpen
  \bibfield  {author} {\bibinfo {author} {\bibfnamefont {H.}~\bibnamefont
  {Rao}},\ }\bibinfo {title} {{Parametrized post-Newtonian limit of the
  Nieh-Yan modified teleparallel gravity}},\ \href
  {https://doi.org/10.1103/PhysRevD.104.124084} {\bibfield  {journal} {Phys.
  Rev. D\ }\textbf {\bibinfo {volume} {104}},\ \bibinfo {pages} {124084}
  (\bibinfo {year} {2021})}\BibitemShut {NoStop}%
\bibitem [{\citenamefont {Li}\ and\ \citenamefont {Zhao}(2022)}]{Li:2021mdp}%
  \BibitemOpen
  \bibfield  {author} {\bibinfo {author} {\bibfnamefont {M.}~\bibnamefont
  {Li}}\ and\ \bibinfo {author} {\bibfnamefont {D.}~\bibnamefont {Zhao}},\
  }\bibinfo {title} {{A simple parity violating model in the symmetric
  teleparallel gravity and its cosmological perturbations}},\ \href
  {https://doi.org/10.1016/j.physletb.2022.136968} {\bibfield  {journal} {Phys.
  Lett. B\ }\textbf {\bibinfo {volume} {827}},\ \bibinfo {pages} {136968}
  (\bibinfo {year} {2022})}\BibitemShut {NoStop}%
\bibitem [{\citenamefont {Battista}\ and\ \citenamefont
  {De~Falco}(2021)}]{Battista:2021rlh}%
  \BibitemOpen
  \bibfield  {author} {\bibinfo {author} {\bibfnamefont {E.}~\bibnamefont
  {Battista}}\ and\ \bibinfo {author} {\bibfnamefont {V.}~\bibnamefont
  {De~Falco}},\ }\bibinfo {title} {{First post-Newtonian generation of
  gravitational waves in Einstein-Cartan theory}},\ \href
  {https://doi.org/10.1103/PhysRevD.104.084067} {\bibfield  {journal} {Phys.
  Rev. D\ }\textbf {\bibinfo {volume} {104}},\ \bibinfo {pages} {084067}
  (\bibinfo {year} {2021})}\BibitemShut {NoStop}%
\bibitem [{\citenamefont {Li}\ {\it et~al.}(2022{\natexlab{a}})\citenamefont
  {Li}, \citenamefont {Li},\ and\ \citenamefont {Rao}}]{Li:2022mti}%
  \BibitemOpen
  \bibfield  {author} {\bibinfo {author} {\bibfnamefont {M.}~\bibnamefont
  {Li}}, \bibinfo {author} {\bibfnamefont {Z.}~\bibnamefont {Li}},\ and\
  \bibinfo {author} {\bibfnamefont {H.}~\bibnamefont {Rao}},\ }\bibinfo {title}
  {{Ghost instability in the teleparallel gravity model with parity
  violations}},\ \href {https://doi.org/10.1016/j.physletb.2022.137395}
  {\bibfield  {journal} {Phys. Lett. B\ }\textbf {\bibinfo {volume} {834}},\
  \bibinfo {pages} {137395} (\bibinfo {year} {2022}{\natexlab{a}})}\BibitemShut
  {NoStop}%
\bibitem [{\citenamefont {Li}\ {\it et~al.}(2022{\natexlab{b}})\citenamefont
  {Li}, \citenamefont {Tong},\ and\ \citenamefont {Zhao}}]{Li:2022vtn}%
  \BibitemOpen
  \bibfield  {author} {\bibinfo {author} {\bibfnamefont {M.}~\bibnamefont
  {Li}}, \bibinfo {author} {\bibfnamefont {Y.}~\bibnamefont {Tong}},\ and\
  \bibinfo {author} {\bibfnamefont {D.}~\bibnamefont {Zhao}},\ }\bibinfo
  {title} {{Possible consistent model of parity violations in the symmetric
  teleparallel gravity}},\ \href {https://doi.org/10.1103/PhysRevD.105.104002}
  {\bibfield  {journal} {Phys. Rev. D\ }\textbf {\bibinfo {volume} {105}},\
  \bibinfo {pages} {104002} (\bibinfo {year} {2022}{\natexlab{b}})}\BibitemShut
  {NoStop}%
\bibitem [{\citenamefont {Hohmann}\ and\ \citenamefont
  {Pfeifer}(2021)}]{Hohmann:2020dgy}%
  \BibitemOpen
  \bibfield  {author} {\bibinfo {author} {\bibfnamefont {M.}~\bibnamefont
  {Hohmann}}\ and\ \bibinfo {author} {\bibfnamefont {C.}~\bibnamefont
  {Pfeifer}},\ }\bibinfo {title} {{Teleparallel axions and cosmology}},\ \href
  {https://doi.org/10.1140/epjc/s10052-021-09165-x} {\bibfield  {journal} {Eur.
  Phys. J. C\ }\textbf {\bibinfo {volume} {81}},\ \bibinfo {pages} {376}
  (\bibinfo {year} {2021})}\BibitemShut {NoStop}%
\bibitem [{\citenamefont {Bombacigno}\ {\it et~al.}(2021)\citenamefont
  {Bombacigno}, \citenamefont {Boudet}, \citenamefont {Olmo},\ and\
  \citenamefont {Montani}}]{Bombacigno:2021bpk}%
  \BibitemOpen
  \bibfield  {author} {\bibinfo {author} {\bibfnamefont {F.}~\bibnamefont
  {Bombacigno}}, \bibinfo {author} {\bibfnamefont {S.}~\bibnamefont {Boudet}},
  \bibinfo {author} {\bibfnamefont {G.~J.}\ \bibnamefont {Olmo}},\ and\
  \bibinfo {author} {\bibfnamefont {G.}~\bibnamefont {Montani}},\ }\bibinfo
  {title} {{Big bounce and future time singularity resolution in Bianchi I
  cosmologies: The projective invariant Nieh-Yan case}},\ \href
  {https://doi.org/10.1103/PhysRevD.103.124031} {\bibfield  {journal} {Phys.
  Rev. D\ }\textbf {\bibinfo {volume} {103}},\ \bibinfo {pages} {124031}
  (\bibinfo {year} {2021})}\BibitemShut {NoStop}%
\bibitem [{\citenamefont {Iosifidis}\ and\ \citenamefont
  {Ravera}(2021)}]{Iosifidis:2020dck}%
  \BibitemOpen
  \bibfield  {author} {\bibinfo {author} {\bibfnamefont {D.}~\bibnamefont
  {Iosifidis}}\ and\ \bibinfo {author} {\bibfnamefont {L.}~\bibnamefont
  {Ravera}},\ }\bibinfo {title} {{Parity Violating Metric-Affine Gravity
  Theories}},\ \href {https://doi.org/10.1088/1361-6382/abde1a} {\bibfield
  {journal} {Classical Quantum Gravity\ }\textbf {\bibinfo {volume} {38}},\
  \bibinfo {pages} {115003} (\bibinfo {year} {2021})}\BibitemShut {NoStop}%
\bibitem [{\citenamefont {Hohmann}\ and\ \citenamefont
  {Pfeifer}(2022)}]{Hohmann:2022wrk}%
  \BibitemOpen
  \bibfield  {author} {\bibinfo {author} {\bibfnamefont {M.}~\bibnamefont
  {Hohmann}}\ and\ \bibinfo {author} {\bibfnamefont {C.}~\bibnamefont
  {Pfeifer}},\ }\bibinfo {title} {{Gravitational wave birefringence in
  spatially curved teleparallel cosmology}},\ \href
  {https://doi.org/10.1016/j.physletb.2022.137437} {\bibfield  {journal} {Phys.
  Lett. B\ }\textbf {\bibinfo {volume} {834}},\ \bibinfo {pages} {137437}
  (\bibinfo {year} {2022})}\BibitemShut {NoStop}%
\bibitem [{\citenamefont {Conroy}\ and\ \citenamefont
  {Koivisto}(2019)}]{Conroy:2019ibo}%
  \BibitemOpen
  \bibfield  {author} {\bibinfo {author} {\bibfnamefont {A.}~\bibnamefont
  {Conroy}}\ and\ \bibinfo {author} {\bibfnamefont {T.}~\bibnamefont
  {Koivisto}},\ }\bibinfo {title} {{Parity-Violating Gravity and GW170817 in
  Non-Riemannian Cosmology}},\ \href
  {https://doi.org/10.1088/1475-7516/2019/12/016} {J. Cosmol. Astropart. Phys.\
  \bibinfo {volume} {12}\bibfield  {year} {\bibinfo  {year} { (\textbf
  {2019})}\ }\bibinfo  {pages} {016}}\BibitemShut {NoStop}%
\bibitem [{\citenamefont {Iosifidis}(2022)}]{Iosifidis:2021bad}%
  \BibitemOpen
\bibfield  {pages} {  }\bibfield  {author} {\bibinfo {author} {\bibfnamefont
  {D.}~\bibnamefont {Iosifidis}},\ }\bibinfo {title} {{The full quadratic
  metric-affine gravity (including parity odd terms): exact solutions for the
  affine-connection}},\ \href {https://doi.org/10.1088/1361-6382/ac6058}
  {\bibfield  {journal} {Classical Quantum Gravity\ }\textbf {\bibinfo {volume}
  {39}},\ \bibinfo {pages} {095002} (\bibinfo {year} {2022})}\BibitemShut
  {NoStop}%
\bibitem [{\citenamefont {Pagani}\ and\ \citenamefont
  {Percacci}(2015)}]{Pagani:2015ema}%
  \BibitemOpen
  \bibfield  {author} {\bibinfo {author} {\bibfnamefont {C.}~\bibnamefont
  {Pagani}}\ and\ \bibinfo {author} {\bibfnamefont {R.}~\bibnamefont
  {Percacci}},\ }\bibinfo {title} {{Quantum gravity with torsion and
  non-metricity}},\ \href {https://doi.org/10.1088/0264-9381/32/19/195019}
  {\bibfield  {journal} {Classical Quantum Gravity\ }\textbf {\bibinfo {volume}
  {32}},\ \bibinfo {pages} {195019} (\bibinfo {year} {2015})}\BibitemShut
  {NoStop}%
\bibitem [{\citenamefont {Chen}\ {\it et~al.}(2023)\citenamefont {Chen},
  \citenamefont {Yu},\ and\ \citenamefont {Gao}}]{Chen:2022wtz}%
  \BibitemOpen
  \bibfield  {author} {\bibinfo {author} {\bibfnamefont {Z.}~\bibnamefont
  {Chen}}, \bibinfo {author} {\bibfnamefont {Y.}~\bibnamefont {Yu}},\ and\
  \bibinfo {author} {\bibfnamefont {X.}~\bibnamefont {Gao}},\ }\bibinfo {title}
  {{Polarized gravitational waves in the parity violating scalar-nonmetricity
  theory}},\ \href {https://doi.org/10.1088/1475-7516/2023/06/001} {J. Cosmol.
  Astropart. Phys.\ \bibinfo {volume} {06}\bibfield  {year} {\bibinfo  {year} {
  (\textbf {2023})}\ }\bibinfo  {pages} {001}}\BibitemShut {NoStop}%
\bibitem [{\citenamefont {Gialamas}\ and\ \citenamefont
  {Tamvakis}(2023)}]{Gialamas:2022xtt}%
  \BibitemOpen
\bibfield  {pages} {  }\bibfield  {author} {\bibinfo {author} {\bibfnamefont
  {I.~D.}\ \bibnamefont {Gialamas}}\ and\ \bibinfo {author} {\bibfnamefont
  {K.}~\bibnamefont {Tamvakis}},\ }\bibinfo {title} {{Inflation in
  metric-affine quadratic gravity}},\ \href
  {https://doi.org/10.1088/1475-7516/2023/03/042} {J. Cosmol. Astropart. Phys.\
  \bibinfo {volume} {03}\bibfield  {year} {\bibinfo  {year} { (\textbf
  {2023})}\ }\bibinfo  {pages} {042}}\BibitemShut {NoStop}%
\bibitem [{\citenamefont {Papanikolaou}\ {\it et~al.}(2023)\citenamefont
  {Papanikolaou}, \citenamefont {Tzerefos}, \citenamefont {Basilakos},\ and\
  \citenamefont {Saridakis}}]{Papanikolaou:2022hkg}%
  \BibitemOpen
\bibfield  {pages} {  }\bibfield  {author} {\bibinfo {author} {\bibfnamefont
  {T.}~\bibnamefont {Papanikolaou}}, \bibinfo {author} {\bibfnamefont
  {C.}~\bibnamefont {Tzerefos}}, \bibinfo {author} {\bibfnamefont
  {S.}~\bibnamefont {Basilakos}},\ and\ \bibinfo {author} {\bibfnamefont
  {E.~N.}\ \bibnamefont {Saridakis}},\ }\bibinfo {title} {{No constraints for
  f(T) gravity from gravitational waves induced from primordial black hole
  fluctuations}},\ \href {https://doi.org/10.1140/epjc/s10052-022-11157-4}
  {\bibfield  {journal} {Eur. Phys. J. C\ }\textbf {\bibinfo {volume} {83}},\
  \bibinfo {pages} {31} (\bibinfo {year} {2023})}\BibitemShut {NoStop}%
\bibitem [{\citenamefont {Salvio}(2022)}]{Salvio:2022suk}%
  \BibitemOpen
  \bibfield  {author} {\bibinfo {author} {\bibfnamefont {A.}~\bibnamefont
  {Salvio}},\ }\bibinfo {title} {{Inflating and reheating the Universe with an
  independent affine connection}},\ \href
  {https://doi.org/10.1103/PhysRevD.106.103510} {\bibfield  {journal} {Phys.
  Rev. D\ }\textbf {\bibinfo {volume} {106}},\ \bibinfo {pages} {103510}
  (\bibinfo {year} {2022})}\BibitemShut {NoStop}%
\bibitem [{\citenamefont {Gialamas}\ and\ \citenamefont
  {Veerm\"ae}(2023)}]{Gialamas:2023emn}%
  \BibitemOpen
  \bibfield  {author} {\bibinfo {author} {\bibfnamefont {I.~D.}\ \bibnamefont
  {Gialamas}}\ and\ \bibinfo {author} {\bibfnamefont {H.}~\bibnamefont
  {Veerm\"ae}},\ }\bibinfo {title} {{Electroweak vacuum decay in metric-affine
  gravity}},\ \href {https://doi.org/10.1016/j.physletb.2023.138109} {\bibfield
   {journal} {Phys. Lett. B\ }\textbf {\bibinfo {volume} {844}},\ \bibinfo
  {pages} {138109} (\bibinfo {year} {2023})}\BibitemShut {NoStop}%
\bibitem [{\citenamefont {Tzerefos}\ {\it et~al.}(2023)\citenamefont
  {Tzerefos}, \citenamefont {Papanikolaou}, \citenamefont {Saridakis},\ and\
  \citenamefont {Basilakos}}]{Tzerefos:2023mpe}%
  \BibitemOpen
  \bibfield  {author} {\bibinfo {author} {\bibfnamefont {C.}~\bibnamefont
  {Tzerefos}}, \bibinfo {author} {\bibfnamefont {T.}~\bibnamefont
  {Papanikolaou}}, \bibinfo {author} {\bibfnamefont {E.~N.}\ \bibnamefont
  {Saridakis}},\ and\ \bibinfo {author} {\bibfnamefont {S.}~\bibnamefont
  {Basilakos}},\ }\bibinfo {title} {{Scalar induced gravitational waves in
  modified teleparallel gravity theories}},\ \href
  {https://doi.org/10.1103/PhysRevD.107.124019} {\bibfield  {journal} {Phys.
  Rev. D\ }\textbf {\bibinfo {volume} {107}},\ \bibinfo {pages} {124019}
  (\bibinfo {year} {2023})}\BibitemShut {NoStop}%
\bibitem [{\citenamefont {De~Falco}\ {\it et~al.}(2024)\citenamefont
  {De~Falco}, \citenamefont {Battista}, \citenamefont {Usseglio},\ and\
  \citenamefont {Capozziello}}]{DeFalco:2024ojf}%
  \BibitemOpen
  \bibfield  {author} {\bibinfo {author} {\bibfnamefont {V.}~\bibnamefont
  {De~Falco}}, \bibinfo {author} {\bibfnamefont {E.}~\bibnamefont {Battista}},
  \bibinfo {author} {\bibfnamefont {D.}~\bibnamefont {Usseglio}},\ and\
  \bibinfo {author} {\bibfnamefont {S.}~\bibnamefont {Capozziello}},\ }\bibinfo
  {title} {{Radiative losses and radiation-reaction effects at the first
  post-Newtonian order in Einstein\textendash{}Cartan theory}},\ \href
  {https://doi.org/10.1140/epjc/s10052-024-12476-4} {\bibfield  {journal} {Eur.
  Phys. J. C\ }\textbf {\bibinfo {volume} {84}},\ \bibinfo {pages} {137}
  (\bibinfo {year} {2024})}\BibitemShut {NoStop}%
\bibitem [{\citenamefont {Yu}\ {\it et~al.}(2024)\citenamefont {Yu},
  \citenamefont {Chen},\ and\ \citenamefont {Gao}}]{Yu:2024drx}%
  \BibitemOpen
  \bibfield  {author} {\bibinfo {author} {\bibfnamefont {Y.}~\bibnamefont
  {Yu}}, \bibinfo {author} {\bibfnamefont {Z.}~\bibnamefont {Chen}},\ and\
  \bibinfo {author} {\bibfnamefont {X.}~\bibnamefont {Gao}},\ }\bibinfo {title}
  {{Spatially covariant gravity with nonmetricity}},\ \href
  {https://doi.org/10.1140/epjc/s10052-024-12893-5} {\bibfield  {journal} {Eur.
  Phys. J. C\ }\textbf {\bibinfo {volume} {84}},\ \bibinfo {pages} {549}
  (\bibinfo {year} {2024})}\BibitemShut {NoStop}%
\bibitem [{\citenamefont {Li}\ {\it et~al.}(2020)\citenamefont {Li},
  \citenamefont {Rao},\ and\ \citenamefont {Zhao}}]{Li:2020xjt}%
  \BibitemOpen
  \bibfield  {author} {\bibinfo {author} {\bibfnamefont {M.}~\bibnamefont
  {Li}}, \bibinfo {author} {\bibfnamefont {H.}~\bibnamefont {Rao}},\ and\
  \bibinfo {author} {\bibfnamefont {D.}~\bibnamefont {Zhao}},\ }\bibinfo
  {title} {{A simple parity violating gravity model without ghost
  instability}},\ \href {https://doi.org/10.1088/1475-7516/2020/11/023} {J.
  Cosmol. Astropart. Phys.\ \bibinfo {volume} {11}\bibfield  {year} {\bibinfo
  {year} { (\textbf {2020})}\ }\bibinfo  {pages} {023}}\BibitemShut {NoStop}%
\bibitem [{\citenamefont {Nieh}\ and\ \citenamefont {Yan}(1982)}]{Nieh:1981ww}%
  \BibitemOpen
\bibfield  {pages} {  }\bibfield  {author} {\bibinfo {author} {\bibfnamefont
  {H.~T.}\ \bibnamefont {Nieh}}\ and\ \bibinfo {author} {\bibfnamefont {M.~L.}\
  \bibnamefont {Yan}},\ }\bibinfo {title} {{An Identity in Riemann-cartan
  Geometry}},\ \href {https://doi.org/10.1063/1.525379} {\bibfield  {journal}
  {J. Math. Phys. (N.Y.)\ }\textbf {\bibinfo {volume} {23}},\ \bibinfo {pages}
  {373} (\bibinfo {year} {1982})}\BibitemShut {NoStop}%
\bibitem [{\citenamefont {Nieh}(2008)}]{Nieh:2008btw}%
  \BibitemOpen
  \bibfield  {author} {\bibinfo {author} {\bibfnamefont {H.~T.}\ \bibnamefont
  {Nieh}},\ }in\ \href {https://doi.org/10.1142/9789812794185_0003} {{\it
  \bibinfo {booktitle} {{Conference in Honor of C.N. Yang's 85th Birthday}:
  {Statistical Physics, High Energy, Condensed Matter and Mathematical
  Physics}}}}\ (\bibinfo {year} {2008})\ pp.\ \bibinfo {pages} {29--37},\
  \Eprint {https://arxiv.org/abs/1309.0915} {arXiv:1309.0915} \BibitemShut
  {NoStop}%
\bibitem [{\citenamefont {Nissinen}\ and\ \citenamefont
  {Volovik}(2020)}]{Nissinen:2019mkw}%
  \BibitemOpen
  \bibfield  {author} {\bibinfo {author} {\bibfnamefont {J.}~\bibnamefont
  {Nissinen}}\ and\ \bibinfo {author} {\bibfnamefont {G.~E.}\ \bibnamefont
  {Volovik}},\ }\bibinfo {title} {{Thermal Nieh-Yan anomaly in Weyl
  superfluids}},\ \href {https://doi.org/10.1103/PhysRevResearch.2.033269}
  {\bibfield  {journal} {Phys. Rev. Res.\ }\textbf {\bibinfo {volume} {2}},\
  \bibinfo {pages} {033269} (\bibinfo {year} {2020})}\BibitemShut {NoStop}%
\bibitem [{\citenamefont {Li}\ {\it et~al.}(2021)\citenamefont {Li},
  \citenamefont {Rao},\ and\ \citenamefont {Tong}}]{Li:2021wij}%
  \BibitemOpen
  \bibfield  {author} {\bibinfo {author} {\bibfnamefont {M.}~\bibnamefont
  {Li}}, \bibinfo {author} {\bibfnamefont {H.}~\bibnamefont {Rao}},\ and\
  \bibinfo {author} {\bibfnamefont {Y.}~\bibnamefont {Tong}},\ }\bibinfo
  {title} {{Revisiting a parity violating gravity model without ghost
  instability: Local Lorentz covariance}},\ \href
  {https://doi.org/10.1103/PhysRevD.104.084077} {\bibfield  {journal} {Phys.
  Rev. D\ }\textbf {\bibinfo {volume} {104}},\ \bibinfo {pages} {084077}
  (\bibinfo {year} {2021})}\BibitemShut {NoStop}%
\bibitem [{\citenamefont {Cai}\ {\it et~al.}(2022)\citenamefont {Cai},
  \citenamefont {Fu},\ and\ \citenamefont {Yu}}]{Cai:2021uup}%
  \BibitemOpen
  \bibfield  {author} {\bibinfo {author} {\bibfnamefont {R.-G.}\ \bibnamefont
  {Cai}}, \bibinfo {author} {\bibfnamefont {C.}~\bibnamefont {Fu}},\ and\
  \bibinfo {author} {\bibfnamefont {W.-W.}\ \bibnamefont {Yu}},\ }\bibinfo
  {title} {{Parity violation in stochastic gravitational wave background from
  inflation in Nieh-Yan modified teleparallel gravity}},\ \href
  {https://doi.org/10.1103/PhysRevD.105.103520} {\bibfield  {journal} {Phys.
  Rev. D\ }\textbf {\bibinfo {volume} {105}},\ \bibinfo {pages} {103520}
  (\bibinfo {year} {2022})}\BibitemShut {NoStop}%
\bibitem [{\citenamefont {Li}\ and\ \citenamefont {Rao}(2023)}]{Li:2023fto}%
  \BibitemOpen
  \bibfield  {author} {\bibinfo {author} {\bibfnamefont {M.}~\bibnamefont
  {Li}}\ and\ \bibinfo {author} {\bibfnamefont {H.}~\bibnamefont {Rao}},\
  }\bibinfo {title} {{Irregular universe in the Nieh-Yan modified teleparallel
  gravity}},\ \href {https://doi.org/10.1016/j.physletb.2023.137929} {\bibfield
   {journal} {Phys. Lett. B\ }\textbf {\bibinfo {volume} {841}},\ \bibinfo
  {pages} {137929} (\bibinfo {year} {2023})}\BibitemShut {NoStop}%
\bibitem [{\citenamefont {Zhang}\ {\it et~al.}(2023)\citenamefont {Zhang},
  \citenamefont {Feng},\ and\ \citenamefont {Gao}}]{Zhang:2023scq}%
  \BibitemOpen
  \bibfield  {author} {\bibinfo {author} {\bibfnamefont {F.}~\bibnamefont
  {Zhang}}, \bibinfo {author} {\bibfnamefont {J.-X.}\ \bibnamefont {Feng}},\
  and\ \bibinfo {author} {\bibfnamefont {X.}~\bibnamefont {Gao}},\ }\bibinfo
  {title} {{Scalar induced gravitational waves in symmetric teleparallel
  gravity with a parity-violating term}},\ \href
  {https://doi.org/10.1103/PhysRevD.108.063513} {\bibfield  {journal} {Phys.
  Rev. D\ }\textbf {\bibinfo {volume} {108}},\ \bibinfo {pages} {063513}
  (\bibinfo {year} {2023})}\BibitemShut {NoStop}%
\bibitem [{\citenamefont {Blas}\ {\it et~al.}(2009)\citenamefont {Blas},
  \citenamefont {Pujolas},\ and\ \citenamefont {Sibiryakov}}]{Blas:2009yd}%
  \BibitemOpen
  \bibfield  {author} {\bibinfo {author} {\bibfnamefont {D.}~\bibnamefont
  {Blas}}, \bibinfo {author} {\bibfnamefont {O.}~\bibnamefont {Pujolas}},\ and\
  \bibinfo {author} {\bibfnamefont {S.}~\bibnamefont {Sibiryakov}},\ }\bibinfo
  {title} {{On the Extra Mode and Inconsistency of Horava Gravity}},\ \href
  {https://doi.org/10.1088/1126-6708/2009/10/029} {J. High Energ. Phys.\
  \bibinfo {volume} {10}\bibfield  {year} {\bibinfo  {year} { (\textbf
  {2009})}\ }\bibinfo  {pages} {029}}\BibitemShut {NoStop}%
\bibitem [{\citenamefont {Charmousis}\ {\it et~al.}(2009)\citenamefont
  {Charmousis}, \citenamefont {Niz}, \citenamefont {Padilla},\ and\
  \citenamefont {Saffin}}]{Charmousis:2009tc}%
  \BibitemOpen
\bibfield  {pages} {  }\bibfield  {author} {\bibinfo {author} {\bibfnamefont
  {C.}~\bibnamefont {Charmousis}}, \bibinfo {author} {\bibfnamefont
  {G.}~\bibnamefont {Niz}}, \bibinfo {author} {\bibfnamefont {A.}~\bibnamefont
  {Padilla}},\ and\ \bibinfo {author} {\bibfnamefont {P.~M.}\ \bibnamefont
  {Saffin}},\ }\bibinfo {title} {{Strong coupling in Horava gravity}},\ \href
  {https://doi.org/10.1088/1126-6708/2009/08/070} {J. High Energ. Phys.\
  \bibinfo {volume} {08}\bibfield  {year} {\bibinfo  {year} { (\textbf
  {2009})}\ }\bibinfo  {pages} {070}}\BibitemShut {NoStop}%
\bibitem [{\citenamefont {Blas}\ {\it et~al.}(2010{\natexlab{a}})\citenamefont
  {Blas}, \citenamefont {Pujolas},\ and\ \citenamefont
  {Sibiryakov}}]{Blas:2009qj}%
  \BibitemOpen
\bibfield  {pages} {  }\bibfield  {author} {\bibinfo {author} {\bibfnamefont
  {D.}~\bibnamefont {Blas}}, \bibinfo {author} {\bibfnamefont {O.}~\bibnamefont
  {Pujolas}},\ and\ \bibinfo {author} {\bibfnamefont {S.}~\bibnamefont
  {Sibiryakov}},\ }\bibinfo {title} {{Consistent Extension of Horava
  Gravity}},\ \href {https://doi.org/10.1103/PhysRevLett.104.181302} {\bibfield
   {journal} {Phys. Rev. Lett.\ }\textbf {\bibinfo {volume} {104}},\ \bibinfo
  {pages} {181302} (\bibinfo {year} {2010}{\natexlab{a}})}\BibitemShut
  {NoStop}%
\bibitem [{\citenamefont {Papazoglou}\ and\ \citenamefont
  {Sotiriou}(2010)}]{Papazoglou:2009fj}%
  \BibitemOpen
  \bibfield  {author} {\bibinfo {author} {\bibfnamefont {A.}~\bibnamefont
  {Papazoglou}}\ and\ \bibinfo {author} {\bibfnamefont {T.~P.}\ \bibnamefont
  {Sotiriou}},\ }\bibinfo {title} {{Strong coupling in extended Horava-Lifshitz
  gravity}},\ \href {https://doi.org/10.1016/j.physletb.2010.01.054} {\bibfield
   {journal} {Phys. Lett. B\ }\textbf {\bibinfo {volume} {685}},\ \bibinfo
  {pages} {197} (\bibinfo {year} {2010})}\BibitemShut {NoStop}%
\bibitem [{\citenamefont {Blas}\ {\it et~al.}(2010{\natexlab{b}})\citenamefont
  {Blas}, \citenamefont {Pujolas},\ and\ \citenamefont
  {Sibiryakov}}]{Blas:2009ck}%
  \BibitemOpen
  \bibfield  {author} {\bibinfo {author} {\bibfnamefont {D.}~\bibnamefont
  {Blas}}, \bibinfo {author} {\bibfnamefont {O.}~\bibnamefont {Pujolas}},\ and\
  \bibinfo {author} {\bibfnamefont {S.}~\bibnamefont {Sibiryakov}},\ }\bibinfo
  {title} {{Comment on `Strong coupling in extended Horava-Lifshitz
  gravity'}},\ \href {https://doi.org/10.1016/j.physletb.2010.03.073}
  {\bibfield  {journal} {Phys. Lett. B\ }\textbf {\bibinfo {volume} {688}},\
  \bibinfo {pages} {350} (\bibinfo {year} {2010}{\natexlab{b}})}\BibitemShut
  {NoStop}%
\bibitem [{\citenamefont {Ferraro}\ and\ \citenamefont
  {Guzm\'an}(2018)}]{Ferraro:2018tpu}%
  \BibitemOpen
  \bibfield  {author} {\bibinfo {author} {\bibfnamefont {R.}~\bibnamefont
  {Ferraro}}\ and\ \bibinfo {author} {\bibfnamefont {M.~J.}\ \bibnamefont
  {Guzm\'an}},\ }\bibinfo {title} {{Hamiltonian formalism for f(T) gravity}},\
  \href {https://doi.org/10.1103/PhysRevD.97.104028} {\bibfield  {journal}
  {Phys. Rev. D\ }\textbf {\bibinfo {volume} {97}},\ \bibinfo {pages} {104028}
  (\bibinfo {year} {2018})}\BibitemShut {NoStop}%
\bibitem [{\citenamefont {Li}\ {\it et~al.}(2011)\citenamefont {Li},
  \citenamefont {Miao},\ and\ \citenamefont {Miao}}]{Li:2011rn}%
  \BibitemOpen
  \bibfield  {author} {\bibinfo {author} {\bibfnamefont {M.}~\bibnamefont
  {Li}}, \bibinfo {author} {\bibfnamefont {R.-X.}\ \bibnamefont {Miao}},\ and\
  \bibinfo {author} {\bibfnamefont {Y.-G.}\ \bibnamefont {Miao}},\ }\bibinfo
  {title} {{Degrees of freedom of $f(T)$ gravity}},\ \href
  {https://doi.org/10.1007/JHEP07(2011)108} {J. High Energ. Phys.\ \bibinfo
  {volume} {07}\bibfield  {year} {\bibinfo  {year} { (\textbf {2011})}\
  }\bibinfo  {pages} {108}}\BibitemShut {NoStop}%
\bibitem [{\citenamefont {Izumi}\ and\ \citenamefont
  {Ong}(2013)}]{Izumi:2012qj}%
  \BibitemOpen
\bibfield  {pages} {  }\bibfield  {author} {\bibinfo {author} {\bibfnamefont
  {K.}~\bibnamefont {Izumi}}\ and\ \bibinfo {author} {\bibfnamefont {Y.~C.}\
  \bibnamefont {Ong}},\ }\bibinfo {title} {{Cosmological Perturbation in f(T)
  Gravity Revisited}},\ \href {https://doi.org/10.1088/1475-7516/2013/06/029}
  {J. Cosmol. Astropart. Phys.\ \bibinfo {volume} {06}\bibfield  {year}
  {\bibinfo  {year} { (\textbf {2013})}\ }\bibinfo  {pages} {029}}\BibitemShut
  {NoStop}%
\bibitem [{\citenamefont {Golovnev}\ and\ \citenamefont
  {Guzm\'an}(2021)}]{Golovnev:2020zpv}%
  \BibitemOpen
\bibfield  {pages} {  }\bibfield  {author} {\bibinfo {author} {\bibfnamefont
  {A.}~\bibnamefont {Golovnev}}\ and\ \bibinfo {author} {\bibfnamefont {M.-J.}\
  \bibnamefont {Guzm\'an}},\ }\bibinfo {title} {{Foundational issues in f(T)
  gravity theory}},\ \href {https://doi.org/10.1142/S0219887821400077}
  {\bibfield  {journal} {Int. J. Geom. Meth. Mod. Phys.\ }\textbf {\bibinfo
  {volume} {18}},\ \bibinfo {pages} {2140007} (\bibinfo {year}
  {2021})}\BibitemShut {NoStop}%
\bibitem [{\citenamefont {Hu}\ {\it et~al.}(2023)\citenamefont {Hu},
  \citenamefont {Zhao}, \citenamefont {Ren}, \citenamefont {Wang},
  \citenamefont {Saridakis},\ and\ \citenamefont {Cai}}]{Hu:2023juh}%
  \BibitemOpen
  \bibfield  {author} {\bibinfo {author} {\bibfnamefont {Y.-M.}\ \bibnamefont
  {Hu}}, \bibinfo {author} {\bibfnamefont {Y.}~\bibnamefont {Zhao}}, \bibinfo
  {author} {\bibfnamefont {X.}~\bibnamefont {Ren}}, \bibinfo {author}
  {\bibfnamefont {B.}~\bibnamefont {Wang}}, \bibinfo {author} {\bibfnamefont
  {E.~N.}\ \bibnamefont {Saridakis}},\ and\ \bibinfo {author} {\bibfnamefont
  {Y.-F.}\ \bibnamefont {Cai}},\ }\bibinfo {title} {{The effective field theory
  approach to the strong coupling issue in f(T) gravity}},\ \href
  {https://doi.org/10.1088/1475-7516/2023/07/060} {J. Cosmol. Astropart. Phys.\
  \bibinfo {volume} {07}\bibfield  {year} {\bibinfo  {year} { (\textbf
  {2023})}\ }\bibinfo  {pages} {060}}\BibitemShut {NoStop}%
\bibitem [{\citenamefont {Hu}\ {\it et~al.}(2024)\citenamefont {Hu},
  \citenamefont {Yu}, \citenamefont {Cai},\ and\ \citenamefont
  {Gao}}]{Hu:2023xcf}%
  \BibitemOpen
\bibfield  {pages} {  }\bibfield  {author} {\bibinfo {author} {\bibfnamefont
  {Y.-M.}\ \bibnamefont {Hu}}, \bibinfo {author} {\bibfnamefont
  {Y.}~\bibnamefont {Yu}}, \bibinfo {author} {\bibfnamefont {Y.-F.}\
  \bibnamefont {Cai}},\ and\ \bibinfo {author} {\bibfnamefont {X.}~\bibnamefont
  {Gao}},\ }\bibinfo {title} {{The effective field theory approach to the
  strong coupling issue in f(T) gravity with a non-minimally coupled scalar
  field}},\ \href {https://doi.org/10.1088/1475-7516/2024/03/025} {J. Cosmol.
  Astropart. Phys.\ \bibinfo {volume} {03}\bibfield  {year} {\bibinfo  {year} {
  (\textbf {2024})}\ }\bibinfo  {pages} {025}}\BibitemShut {NoStop}%
\bibitem [{\citenamefont {Bahamonde}\ {\it et~al.}(2023)\citenamefont
  {Bahamonde}, \citenamefont {Dialektopoulos}, \citenamefont
  {Escamilla-Rivera}, \citenamefont {Farrugia}, \citenamefont {Gakis},
  \citenamefont {Hendry}, \citenamefont {Hohmann}, \citenamefont {Levi~Said},
  \citenamefont {Mifsud},\ and\ \citenamefont
  {Di~Valentino}}]{Bahamonde:2021gfp}%
  \BibitemOpen
\bibfield  {pages} {  }\bibfield  {author} {\bibinfo {author} {\bibfnamefont
  {S.}~\bibnamefont {Bahamonde}}, \bibinfo {author} {\bibfnamefont {K.~F.}\
  \bibnamefont {Dialektopoulos}}, \bibinfo {author} {\bibfnamefont
  {C.}~\bibnamefont {Escamilla-Rivera}}, \bibinfo {author} {\bibfnamefont
  {G.}~\bibnamefont {Farrugia}}, \bibinfo {author} {\bibfnamefont
  {V.}~\bibnamefont {Gakis}}, \bibinfo {author} {\bibfnamefont
  {M.}~\bibnamefont {Hendry}}, \bibinfo {author} {\bibfnamefont
  {M.}~\bibnamefont {Hohmann}}, \bibinfo {author} {\bibfnamefont
  {J.}~\bibnamefont {Levi~Said}}, \bibinfo {author} {\bibfnamefont
  {J.}~\bibnamefont {Mifsud}},\ and\ \bibinfo {author} {\bibfnamefont
  {E.}~\bibnamefont {Di~Valentino}},\ }\bibinfo {title} {{Teleparallel gravity:
  from theory to cosmology}},\ \href {https://doi.org/10.1088/1361-6633/ac9cef}
  {\bibfield  {journal} {Rept. Prog. Phys.\ }\textbf {\bibinfo {volume} {86}},\
  \bibinfo {pages} {026901} (\bibinfo {year} {2023})}\BibitemShut {NoStop}%
\bibitem [{\citenamefont {Beltr\'an~Jim\'enez}\ {\it et~al.}(2018)\citenamefont
  {Beltr\'an~Jim\'enez}, \citenamefont {Heisenberg},\ and\ \citenamefont
  {Koivisto}}]{BeltranJimenez:2017tkd}%
  \BibitemOpen
  \bibfield  {author} {\bibinfo {author} {\bibfnamefont {J.}~\bibnamefont
  {Beltr\'an~Jim\'enez}}, \bibinfo {author} {\bibfnamefont {L.}~\bibnamefont
  {Heisenberg}},\ and\ \bibinfo {author} {\bibfnamefont {T.}~\bibnamefont
  {Koivisto}},\ }\bibinfo {title} {{Coincident General Relativity}},\ \href
  {https://doi.org/10.1103/PhysRevD.98.044048} {\bibfield  {journal} {Phys.
  Rev. D\ }\textbf {\bibinfo {volume} {98}},\ \bibinfo {pages} {044048}
  (\bibinfo {year} {2018})}\BibitemShut {NoStop}%
\bibitem [{\citenamefont {Fu}\ {\it et~al.}(2024)\citenamefont {Fu},
  \citenamefont {Liu}, \citenamefont {Yang}, \citenamefont {Yu},\ and\
  \citenamefont {Zhang}}]{Fu:2023aab}%
  \BibitemOpen
  \bibfield  {author} {\bibinfo {author} {\bibfnamefont {C.}~\bibnamefont
  {Fu}}, \bibinfo {author} {\bibfnamefont {J.}~\bibnamefont {Liu}}, \bibinfo
  {author} {\bibfnamefont {X.-Y.}\ \bibnamefont {Yang}}, \bibinfo {author}
  {\bibfnamefont {W.-W.}\ \bibnamefont {Yu}},\ and\ \bibinfo {author}
  {\bibfnamefont {Y.}~\bibnamefont {Zhang}},\ }\bibinfo {title} {{Explaining
  pulsar timing array observations with primordial gravitational waves in
  parity-violating gravity}},\ \href
  {https://doi.org/10.1103/PhysRevD.109.063526} {\bibfield  {journal} {Phys.
  Rev. D\ }\textbf {\bibinfo {volume} {109}},\ \bibinfo {pages} {063526}
  (\bibinfo {year} {2024})}\BibitemShut {NoStop}%
\bibitem [{\citenamefont {Golovnev}\ and\ \citenamefont
  {Koivisto}(2018)}]{Golovnev:2018wbh}%
  \BibitemOpen
  \bibfield  {author} {\bibinfo {author} {\bibfnamefont {A.}~\bibnamefont
  {Golovnev}}\ and\ \bibinfo {author} {\bibfnamefont {T.}~\bibnamefont
  {Koivisto}},\ }\bibinfo {title} {{Cosmological perturbations in modified
  teleparallel gravity models}},\ \href
  {https://doi.org/10.1088/1475-7516/2018/11/012} {J. Cosmol. Astropart. Phys.\
  \bibinfo {volume} {11}\bibfield  {year} {\bibinfo  {year} { (\textbf
  {2018})}\ }\bibinfo  {pages} {012}}\BibitemShut {NoStop}%
\bibitem [{\citenamefont {Li}\ {\it et~al.}(2018)\citenamefont {Li},
  \citenamefont {Cai}, \citenamefont {Cai},\ and\ \citenamefont
  {Saridakis}}]{Li:2018ixg}%
  \BibitemOpen
\bibfield  {pages} {  }\bibfield  {author} {\bibinfo {author} {\bibfnamefont
  {C.}~\bibnamefont {Li}}, \bibinfo {author} {\bibfnamefont {Y.}~\bibnamefont
  {Cai}}, \bibinfo {author} {\bibfnamefont {Y.-F.}\ \bibnamefont {Cai}},\ and\
  \bibinfo {author} {\bibfnamefont {E.~N.}\ \bibnamefont {Saridakis}},\
  }\bibinfo {title} {{The effective field theory approach of teleparallel
  gravity, $f(T)$ gravity and beyond}},\ \href
  {https://doi.org/10.1088/1475-7516/2018/10/001} {J. Cosmol. Astropart. Phys.\
  \bibinfo {volume} {10}\bibfield  {year} {\bibinfo  {year} { (\textbf
  {2018})}\ }\bibinfo  {pages} {001}}\BibitemShut {NoStop}%
\bibitem [{\citenamefont {Abbott}\ {\it
  et~al.}(2017{\natexlab{e}})\citenamefont {Abbott} {\it
  et~al.}}]{LIGOScientific:2017vwq}%
  \BibitemOpen
\bibfield  {pages} {  }\bibfield  {author} {\bibinfo {author} {\bibfnamefont
  {B.~P.}\ \bibnamefont {Abbott}} {\it et~al.} (\bibinfo {collaboration} {LIGO
  Scientific, Virgo}),\ }\bibinfo {title} {{GW170817: Observation of
  Gravitational Waves from a Binary Neutron Star Inspiral}},\ \href
  {https://doi.org/10.1103/PhysRevLett.119.161101} {\bibfield  {journal} {Phys.
  Rev. Lett.\ }\textbf {\bibinfo {volume} {119}},\ \bibinfo {pages} {161101}
  (\bibinfo {year} {2017}{\natexlab{e}})}\BibitemShut {NoStop}%
\bibitem [{\citenamefont {Abbott}\ {\it
  et~al.}(2017{\natexlab{f}})\citenamefont {Abbott} {\it
  et~al.}}]{LIGOScientific:2017zic}%
  \BibitemOpen
  \bibfield  {author} {\bibinfo {author} {\bibfnamefont {B.~P.}\ \bibnamefont
  {Abbott}} {\it et~al.} (\bibinfo {collaboration} {LIGO Scientific, Virgo,
  Fermi-GBM, INTEGRAL}),\ }\bibinfo {title} {{Gravitational Waves and
  Gamma-rays from a Binary Neutron Star Merger: GW170817 and GRB 170817A}},\
  \href {https://doi.org/10.3847/2041-8213/aa920c} {\bibfield  {journal}
  {Astrophys. J. Lett.\ }\textbf {\bibinfo {volume} {848}},\ \bibinfo {pages}
  {L13} (\bibinfo {year} {2017}{\natexlab{f}})}\BibitemShut {NoStop}%
\bibitem [{\citenamefont {Cai}\ {\it et~al.}(2019)\citenamefont {Cai},
  \citenamefont {Pi},\ and\ \citenamefont {Sasaki}}]{Cai:2018dig}%
  \BibitemOpen
  \bibfield  {author} {\bibinfo {author} {\bibfnamefont {R.-g.}\ \bibnamefont
  {Cai}}, \bibinfo {author} {\bibfnamefont {S.}~\bibnamefont {Pi}},\ and\
  \bibinfo {author} {\bibfnamefont {M.}~\bibnamefont {Sasaki}},\ }\bibinfo
  {title} {{Gravitational Waves Induced by non-Gaussian Scalar
  Perturbations}},\ \href {https://doi.org/10.1103/PhysRevLett.122.201101}
  {\bibfield  {journal} {Phys. Rev. Lett.\ }\textbf {\bibinfo {volume} {122}},\
  \bibinfo {pages} {201101} (\bibinfo {year} {2019})}\BibitemShut {NoStop}%
\bibitem [{\citenamefont {Unal}(2019)}]{Unal:2018yaa}%
  \BibitemOpen
  \bibfield  {author} {\bibinfo {author} {\bibfnamefont {C.}~\bibnamefont
  {Unal}},\ }\bibinfo {title} {{Imprints of Primordial Non-Gaussianity on
  Gravitational Wave Spectrum}},\ \href
  {https://doi.org/10.1103/PhysRevD.99.041301} {\bibfield  {journal} {Phys.
  Rev. D\ }\textbf {\bibinfo {volume} {99}},\ \bibinfo {pages} {041301}
  (\bibinfo {year} {2019})}\BibitemShut {NoStop}%
\bibitem [{\citenamefont {Adshead}\ {\it et~al.}(2021)\citenamefont {Adshead},
  \citenamefont {Lozanov},\ and\ \citenamefont {Weiner}}]{Adshead:2021hnm}%
  \BibitemOpen
  \bibfield  {author} {\bibinfo {author} {\bibfnamefont {P.}~\bibnamefont
  {Adshead}}, \bibinfo {author} {\bibfnamefont {K.~D.}\ \bibnamefont
  {Lozanov}},\ and\ \bibinfo {author} {\bibfnamefont {Z.~J.}\ \bibnamefont
  {Weiner}},\ }\bibinfo {title} {{Non-Gaussianity and the induced gravitational
  wave background}},\ \href {https://doi.org/10.1088/1475-7516/2021/10/080} {J.
  Cosmol. Astropart. Phys.\ \bibinfo {volume} {10}\bibfield  {year} {\bibinfo
  {year} { (\textbf {2021})}\ }\bibinfo  {pages} {080}}\BibitemShut {NoStop}%
\bibitem [{\citenamefont {Garcia-Saenz}\ {\it
  et~al.}(2023{\natexlab{b}})\citenamefont {Garcia-Saenz}, \citenamefont
  {Pinol}, \citenamefont {Renaux-Petel},\ and\ \citenamefont
  {Werth}}]{Garcia-Saenz:2022tzu}%
  \BibitemOpen
\bibfield  {pages} {  }\bibfield  {author} {\bibinfo {author} {\bibfnamefont
  {S.}~\bibnamefont {Garcia-Saenz}}, \bibinfo {author} {\bibfnamefont
  {L.}~\bibnamefont {Pinol}}, \bibinfo {author} {\bibfnamefont
  {S.}~\bibnamefont {Renaux-Petel}},\ and\ \bibinfo {author} {\bibfnamefont
  {D.}~\bibnamefont {Werth}},\ }\bibinfo {title} {{No-go theorem for
  scalar-trispectrum-induced gravitational waves}},\ \href
  {https://doi.org/10.1088/1475-7516/2023/03/057} {J. Cosmol. Astropart. Phys.\
  \bibinfo {volume} {03}\bibfield  {year} {\bibinfo  {year} { (\textbf
  {2023})}\ }\bibinfo  {pages} {057}}\BibitemShut {NoStop}%
\bibitem [{\citenamefont {Sato-Polito}\ {\it et~al.}(2019)\citenamefont
  {Sato-Polito}, \citenamefont {Kovetz},\ and\ \citenamefont
  {Kamionkowski}}]{Sato-Polito:2019hws}%
  \BibitemOpen
\bibfield  {pages} {  }\bibfield  {author} {\bibinfo {author} {\bibfnamefont
  {G.}~\bibnamefont {Sato-Polito}}, \bibinfo {author} {\bibfnamefont {E.~D.}\
  \bibnamefont {Kovetz}},\ and\ \bibinfo {author} {\bibfnamefont
  {M.}~\bibnamefont {Kamionkowski}},\ }\bibinfo {title} {{Constraints on the
  primordial curvature power spectrum from primordial black holes}},\ \href
  {https://doi.org/10.1103/PhysRevD.100.063521} {\bibfield  {journal} {Phys.
  Rev. D\ }\textbf {\bibinfo {volume} {100}},\ \bibinfo {pages} {063521}
  (\bibinfo {year} {2019})}\BibitemShut {NoStop}%
\end{thebibliography}

%

\end{document}